\documentclass[%
reprint,
amsmath,amssymb,
aps,
prd,
]{revtex4-1}
\usepackage{graphicx}% Include figure files
\usepackage{dcolumn}% Align table columns on decimal point
\usepackage{bm}% bold math
\graphicspath{{Figures/}}

\PassOptionsToPackage{breaklinks}{hyperref}
\usepackage{xurl}
\begin{document}
\preprint{APS/123-QED}

\title{Quantum entanglement of final particle states in the resonant trident pair production in a strong electromagnetic wave}

\author{S.P. Roshchupkin}
\email{serg9rsp@gmail.com}
\author{M.V. Shakhov}%
\email{shahov.mv@edu.spbstu.ru}
\affiliation{Peter the Great St. Petersburg Polytechnic University, Polytechnicheskaya 29,
	195251 Saint Petersburg, Russia}
%


%\affiliation{
%\affiliation{
%\affiliation{%Authors' institution and/or address\\
	% This line break forced with \textbackslash\textbackslash}%


\date{\today}% It is always \today, today,

\begin{abstract}
	The resonant trident pair production process in the collision of ultrarelativistic electrons with a strong electromagnetic wave is theoretically studied. Under resonant conditions, the intermediate virtual gamma-quantum becomes real. As a result, the original resonant trident pair production process effectively splits into two first-order processes by the fine structure constant: the electromagnetic field-stimulated Compton-effect and the electromagnetic field-stimulated Breit-Wheeler process. The kinematics of the resonant trident pair production process are studied in detail. It is shown that there are two different cases for the energies and outgoing angles of final particles (an electron and an electron-positron pair) in which their quantum entanglement is realized. In the first case, energies and outgoing angles of final ultrarelativistic particles are uniquely determined by the parameters of the electromagnetic field-stimulated Compton-effect (the outgoing angle of the final electron and the quantum parameter of the Compton effect). In the second case, energies and outgoing angles of final particles are uniquely determined by the electromagnetic field-stimulated Breit-Wheeler process (the electron-positron pair outgoing angle and the Breit-Wheeler quantum parameter). It is shown that in a sufficiently wide range of frequencies and intensities of a strong electromagnetic wave, and in the case of ultrarelativistic initial electrons, the differential probability of the resonant trident pair production process with simultaneous registration of the outgoing angles of final particles can significantly (by several orders of magnitude) exceed the total probability of the electromagnetic field-stimulated Compton-effect.
	%\begin{description}
	%\end{description}
\end{abstract}

\maketitle

%\tableofcontents

\section{Introduction}

With the advent of high-intensity lasers, as well as the source of high-energy particles and high-energy gamma-quanta \cite{1,2,3,4,5,6,7,8,9,10,11,12,13,14,15,16,17,18,19,20,21,22}, the processes of quantum electrodynamics (QED) in external electromagnetic fields have become widely studied in recent decades (see, for example, reviews \cite{23,24,25,26,27,28,29,30,31,32,33,34,35}, monographs \cite{36,37,38}, and articles \cite{39,40,41,42,43,44,45,46,47,48,49,50,51,52,53,54,55,56,57,58,59,60,61,62,63,64,65,66,67,68,69,70,71,72,73,74,75,76,77,78,79,80,81,82,83,84,85,86,87,88,89,90,91,92,93,94,95,96,97,98,99,100,101,102,103,104,105}). At the same time, much attention is paid to the first-order QED processes (by the fine structure constant), which occur only in an external electromagnetic field. Such processes include the electromagnetic field-stimulated Compton-effect (EFSCE) \cite{39,40,41,42,43,44,45,46,47,48,49,50,51,52,53,54}, as well as the electromagnetic field-stimulated Breit-Wheeler (EFSBW) process \cite{55,56,57,58,59,60,61,62}. It is important to note that higher-order processes in an external electromagnetic field can proceed in both resonant and non-resonant ways. The resonant flow of higher-order QED processes is associated with intermediate virtual particle reaching the mass surface. Such resonances were first considered by Oleynik \cite{63,64}. As a result, higher-order processes effectively break up into several consecutive lower-order processes (see the reviews \cite{26,31}, monographs \cite{36,37} and recent ones \cite{65,66,67,68,69}). At the same time, the probability of resonant processes can significantly exceed the corresponding probabilities of non-resonant processes. 

It is important to note that when a relativistic electron beam collides with an electromagnetic wave, along with the EFSCE, a second-order process takes place, namely, electron scattering with simultaneous generation of the electron-positron pair. This is the so-called trident pair production process. This process also attracts attention (see, for example, articles \cite{65,70,71,72,73,74,75,76}). It should be emphasized that the trident pair production process can proceed in a resonant way, when the intermediate gamma-quantum becomes real. In this case, the original resonant trident pair production (RTPP) process effectively splits into two first-order processes: the EFSCE and EFSBW processes. Note that in the case of a weak plane electromagnetic wave the RTPP process was considered in articel \cite{65}.

In this paper, we consider the RTPP process in the collision of ultrarelativistic electrons with a strong monochromatic electromagnetic wave. We will study the conditions under which the probability of the RTPP process can significantly exceed the corresponding probability of the EFSCE.

In the considered problem, the main parameter is the classical relativistic-invariant parameter
\begin{equation} \label{ZEqnNum778375} 
	\eta =\frac{eF{\mathchar'26\mkern-10mu\lambda} {\rm \; }}{mc^{2} } {\rm \mathop{>}\limits_\sim }1 ,
\end{equation} 
which is numerically equal to the ratio of the field work at the wavelength to the rest energy of the electron ($e$ and $m$ are the charge and mass of the electron, $F$ and ${\mathchar'26\mkern-10mu\lambda} ={c\mathord{\left/ {\vphantom {c \omega }} \right. \kern-\nulldelimiterspace} \omega } $ are the electric field strength and wavelength, and $\omega $ is the wave frequency). In addition, in this problem, the characteristic quantum parameters of the Compton effect $\left(\varepsilon _{iC} \right)$ and the Breit-Wheeler process $\left(\varepsilon _{iBW} \right)$ arise, which are equal to the ratio of the energy of the initial electron to the characteristic energies of the process:
\begin{equation} \label{ZEqnNum148652} 
	\varepsilon _{iC} =\frac{E_{i} }{\hbar \omega _{C} } ,\quad \varepsilon _{iBW} =\frac{E_{i} }{\hbar \omega _{BW} } . 
\end{equation} 
Here $E_{i}$ is the energy of the initial electrons, $\hbar \omega _{C} $ and $\hbar \omega _{BW} $ are the characteristic quantum energies of the Compton effect  and the Breit-Wheeler process \cite{66,67,68,69}:
\begin{eqnarray} \label{ZEqnNum291894} 
	\hbar \omega _{C} =\frac{\left(m_{*} c^{2} \right)^{2} }{4\left(\hbar \omega \right)\sin ^{2} \left({\theta _{i} \mathord{\left/ {\vphantom {\theta _{i}  2}} \right. \kern-\nulldelimiterspace} 2} \right)} =\frac{\left(mc^{2} \right)^{2} \left(1+\eta ^{2} \right)}{4\left(\hbar \omega \right)\sin ^{2} \left({\theta _{i} \mathord{\left/ {\vphantom {\theta _{i}  2}} \right. \kern-\nulldelimiterspace} 2} \right)} ,\\ \nonumber
	\quad \omega _{BW} =4\omega _{C} . 
\end{eqnarray} 

Here $m_{*} $ is the effective mass of an electron in the field of a circularly polarized wave \eqref{ZEqnNum579777}, $\theta _{i} $ is the angle between the momentum of the initial electron and the direction of propagation of the wave \eqref{ZEqnNum348638}. Note that the characteristic energies \eqref{ZEqnNum291894} are inversely proportional to the frequency  $\left(\omega \right)$, directly proportional to the intensity of the external electromagnetic wave $\left(I\sim \eta ^{2} \right)$ and dependent on the angle between the wave and the initial particles momenta. 

In this paper, we will show that the resonant energies of final particles, as well as the resonant differential probabilities, significantly depend on the value of these quantum parameters \eqref{ZEqnNum148652}, i.e., on the ratio of the initial electron energy to the corresponding characteristic energy. 

Later in the article, the relativistic system of units is used: $\hbar =c=1$.

\section{Process amplitude}
Let's choose the 4-potential of the external electromagnetic field in the form of a plane circularly polarized monochromatic wave propagating along the z axis:
\begin{eqnarray} \label{ZEqnNum190066} 
	A\left(\varphi \right)=\frac{F}{\omega } \cdot \left(e_{x} \cos \varphi +\delta e_{y} \sin \varphi \right),\\ \nonumber
	\quad \varphi =\left(kx\right)=\omega (t-z). \nonumber
\end{eqnarray} 
Here $\delta =\pm 1$, $e_{x}^{} ,e_{y}^{} $- are the polarization 4-vectors of the external field, which have the following properties: $e_{x} =(0,\pmb{e}_{x} ),\quad e_{y} =(0,\pmb{e}_{y} ),\quad e_{x} e_{y} =0,\quad (e_{x} )^{2} =(e_{y} )^{2} =-1.$ The trident pair production process stimulated by an external field is described by two Feynman diagrams (Fig. 1). 
\begin{figure}[h]
	\includegraphics{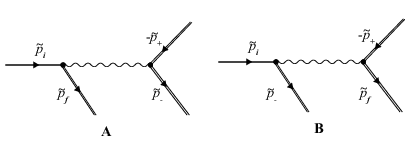}% Here is how to import EPS art
	\caption{\label{fig 1} Feynman diagrams for the trident pair production process in the collision of an electron with a strong electromagnetic field. The outer double lines correspond to the Volkov functions for an electron or positron, while the wavy lines correspond to the Green functions of an intermediate gamma-quantum.}
\end{figure}

The amplitude of the process under consideration can be represented in the following form
\begin{widetext}
	\begin{equation} \label{ZEqnNum114924} 
		S=-ie^{2} \int d^{4} x_{1} d^{4} x_{2} \overline{\psi }_{p_{f} } (x_{1} \left|A\right. )\gamma ^{\mu } \psi _{p_{i} } (x_{1} \left|A\right. )D_{\mu \nu } (x_{2} -x_{1} )\overline{\psi }_{p_{-} } (x_{2} \left|A\right. )\gamma ^{\nu } \psi _{-p_{+} } (x_{2} \left|A\right. ) -(p_{f} \leftrightarrow p_{-} ), 
	\end{equation} 
\end{widetext}
where $p_{j } =(E_{j }, \pmb{p}_{j } )$ - are the 4-momenta of final and initial electrons and positrons ($j=i,f,\pm$). Here $\psi _{p_{i} } (x_{1} \left|A\right. )$, $\psi _{-p_{+} } (x_{2} \left|A\right. )$, $\overline{\psi }_{p_{-} } (x_{2} \left|A\right. )$, $\overline{\psi }_{p_{f} } (x_{1} \left|A\right. )$ are the Volkov functions in the plane electromagnetic wave field \cite{106,108}, and $D_{\mu \nu } (x_{2} -x_{1} )$ is the propagator of the intermediate gamma-quantum \cite{36,37}. 

After performing the appropriate integrations, the amplitude \eqref{ZEqnNum114924} can be represented as:
\begin{equation} \label{ZEqnNum648653} 
	S =\sum _{l=-\infty }^{+\infty }S_{l}  , 
\end{equation} 
where the partial amplitude $S_{l} $ corresponds to the absorption or emission of $|l|$ gamma-quanta of an external wave. For channel A, the partial amplitude can be represented as follows:
\begin{widetext}
	\begin{eqnarray} \label{ZEqnNum496283} 
		S_{l} =\frac{16\pi ^{5} e^{2} e^{-id_{0} } }{\sqrt{\tilde{E}_{i} \tilde{E}_{f} \tilde{E}_{-} \tilde{E}_{+} } } \sum _{l_{2=-\infty } }^{\infty }\left[\bar{u}_{p_{f} } F_{-\left(l-l_{2} \right)}^{\mu } \left(\tilde{p}_{f} ,\tilde{p}_{i} \right)u_{p_{i} } \right] \frac{1}{q^{2} } \left[\bar{u}_{p_{-} } F_{-l_{2} ,\mu }^{} \left(\tilde{p}_{-} ,-\tilde{p}_{+} \right)\nu _{p_{+} } \right] \times  \\ \nonumber
		\times \delta ^{\left(4\right)} \left(\widetilde{p}_{+} +\widetilde{p}_{-} -\widetilde{p}_{i} +\widetilde{p}_{f} -lk\right), \quad \left(l=l_{1} +l_{2} \right) ,\nonumber
	\end{eqnarray} 
\end{widetext}
where $q$ is the 4-momentum of the intermediate gamma-quantum which takes the form
\begin{equation} \label{8)} 
	q=\widetilde{p}_{+} +\widetilde{p}_{-} -l_{2} k. 
\end{equation} 
In the expression \eqref{ZEqnNum496283} $u_{p_{i} }$ is the Dirac bispinor of the initial electron, and $\bar{u}_{p_{f} }, \bar{u}_{p_{-} } ,\nu _{p_{+} } $ - are the Dirac bispinors of the final electron and electron-positron pair, $d_{0} $ is the phase independent of the summation indices, and the function $F_{-(l-l_{2}) }^{\mu } \left(\tilde{p}_{f} ,\tilde{p}_{i} \right)$ has the form:

\begin{eqnarray} \label{ZEqnNum494460} 
	F_{-\left(l-l_{2} \right)}^{\mu } \left(\tilde{p}_{f} ,\tilde{p}_{i} \right)=a^{\mu } \left(\tilde{p}_{f} ,\tilde{p}_{i} \right)L_{-\left(l-l_{2} \right)} +\\ \nonumber
	+b_{-}^{\mu } \left(\tilde{p}_{f} ,\tilde{p}_{i} \right)L_{-\left(l-l_{2} \right)-1} +b_{+}^{\mu } \left(\tilde{p}_{f} ,\tilde{p}_{i} \right)L_{-\left(l-l_{2} \right)+1} . \nonumber
\end{eqnarray}

Here the matrices $a^{\mu } $ and  $b_{\pm }^{\mu } $ have the form:
\begin{equation} \label{10)} 
	\begin{array}{cc} {a^{\mu } \left(\tilde{p}_{f} ,\tilde{p}_{i} \right)=\gamma ^{\mu } +\frac{\eta ^{2} m^{2} }{2\left(k\tilde{p}_{f} \right)\left(k\tilde{p}_{i} \right)} \widehat{k}k^{\mu } ,} & {} \end{array} 
\end{equation} 
\begin{equation} \label{11)} 
	b_{\pm }^{\mu } \left(\tilde{p}_{f} ,\tilde{p}_{i} \right)=\frac{\eta m}{4} \left[\frac{1}{\left(k\tilde{p}_{f} \right)} \widehat{\varepsilon }_{\pm } \widehat{k}\gamma ^{\mu } +\frac{1}{\left(k\tilde{p}_{i} \right)} \gamma ^{\mu } \widehat{k}\widehat{\varepsilon }_{\pm } \right], 
\end{equation} 
\begin{equation} \label{12)} 
	\varepsilon _{\pm } =e_{x} \pm ie_{y} . 
\end{equation} 
In the relations \eqref{10)} - \eqref{11)} values with caps represent scalar product of respective 4-vectors with Dirac's gamma matrices ($\widehat{k}=\gamma^\mu k_\mu=\gamma^0 k^0-\pmb{\gamma}\pmb{k}$).
In the relations \eqref{ZEqnNum494460} special functions $L_{l} $ \cite{27} in the case of circular wave polarization can be represented using Bessel functions with integer indices
\begin{equation} \label{13)} 
	L_{-\left(l-l_{2} \right)} =\exp \left(i\left(l-l_{2} \right)\chi _{\tilde{p}_{f} p_{i} } \right)J_{-\left(l-l_{2} \right)} \left(\gamma _{\tilde{p}_{f} p_{i} } \right), 
\end{equation} 
where is indicated
\begin{equation} \label{14)} 
	\gamma _{\tilde{p}_{f} \tilde{p}_{i} } =\eta m\sqrt{-Q_{\tilde{p}_{f} \tilde{p}_{i}}^2} , 
\end{equation} 
\begin{equation} \label{15)} 
	Q_{\tilde{p}_{f} \tilde{p}_{i} } =\frac{\tilde{p}_{f} }{\left(k\tilde{p}_{f} \right)} -\frac{\tilde{p}_{i} }{\left(k\tilde{p}_{i} \right)} , 
\end{equation} 
\begin{equation} \label{ZEqnNum789072} 
	\tan \chi _{\tilde{p}_{f} \tilde{p}_{i} } =\delta \frac{\left(Q_{\tilde{p}_{f} \tilde{p}_{i} } e_{y} \right)}{\left(Q_{\tilde{p}_{f} \tilde{p}_{i} } e_{x} \right)} . 
\end{equation} 
In the above expressions $\tilde{p}_{j} =\left(\tilde{E}_{j} ,\tilde{\pmb{p}}_{j} \right)$ - are the quasi-momenta of electrons and positrons, $m_{*} $- the effective mass of the electron in the wave field \eqref{ZEqnNum190066}:
\begin{equation} \label{17)} 
	\tilde{p}_{j} =p_{j} +\eta ^{2} \frac{m^{2} }{2\left(kp_{j} \right)} k,\quad i=f,\pm  .
\end{equation} 
\begin{equation} \label{ZEqnNum579777} 
	\tilde{p}_{j}^{2} =m_{*}^{2} ,\quad m_{*} =m\sqrt{1+\eta ^{2} } . 
\end{equation} 
Note that the function $F_{-l_{2} ,\mu }^{} \left(\tilde{p}_{-} ,-\tilde{p}_{+} \right)$ in the expression \eqref{ZEqnNum496283} is obtained from the corresponding expressions \eqref{ZEqnNum494460} - \eqref{ZEqnNum789072} by replacing: $\tilde{p}_{i} \to -\tilde{p}_{+} ,\; \tilde{p}_{f} \to \tilde{p}_{-} ,\; \left(l-l_{2} \right) \to l_{2}$, as well as lowering the index $\mu $.

\section{Resonant kinematics}

Under resonant conditions, the intermediate gamma-quantum becomes real (Oleinik resonances \cite{63,64,65,66,67,68,69}). As a result, the initial second-order process in the wave field effectively decays into two first-order processes: the EFSCE (at the first vertex) \cite{24,108} and the EFSBW process (at the second vertex) \cite{24}. Diagrams of processes under resonant conditions are shown in Fig. 2.
\begin{figure}[h]
	\includegraphics{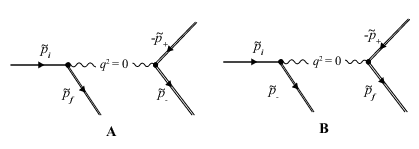}% Here is how to import EPS art
	\caption{\label{fig 2} Feynman diagrams of the resonant trident pair production process in the collision of an electron with a strong electromagnetic field, channels $A$ and $B$.}
\end{figure}

\begin{equation} \label{ZEqnNum933964} 
	q^{2} =0. 
\end{equation} 

Further consideration will be given to the resonant channel $A$. Under the conditions \eqref{ZEqnNum933964}, we write the 4-momentum conservation laws in the first and second vertices of the Feynman diagram (see Figure 2A):
\begin{equation} \label{ZEqnNum691334} 
	\widetilde{p}_{i} +l_{1} k=q+\widetilde{p}_{f} , 
\end{equation} 
\begin{equation} \label{ZEqnNum921245} 
	q+l_{2} k=\widetilde{p}_{+} +\widetilde{p}_{-} . 
\end{equation} 
Given that $\tilde{p}_{i,f}^{2} =\tilde{p}_{\pm }^{2} =m_{*}^{2} $, and $q^{2} =k^{2} =0$, the relations \eqref{ZEqnNum691334} and \eqref{ZEqnNum921245} are valid for $l_{1} \ge 1$ and $l_{2} \ge 1$.

Later in the article, we will study the case of ultrarelativistic energies of initial electrons, as well as final particles (the final electron and the electron-positron pair).
\begin{equation} \label{ZEqnNum378702} 
	E_{j} >>m,\quad j=i,f,\pm  .
\end{equation} 
Due to condition \eqref{ZEqnNum378702}, final particles will mainly fly out in a narrow cone along the momentum of the initial electron. In this case, we assume that this narrow cone of particles lies far from the direction of propagation of the external electromagnetic wave (otherwise the resonances disappear \cite{66,67,68,69}, see Figure 3). Thus, the angles between the particle momenta must satisfy the following conditions:
\begin{equation} \label{ZEqnNum886092} 
	\begin{array}{cc} {\theta _{f} \equiv \angle \left(\pmb{p}_{i} ,\pmb{p}_{f} \right)\ll 1,} & {\theta _{\pm } \equiv \angle \left(\pmb{p}_{+} ,\pmb{p}_{-} \right)\ll 1} ,\end{array} 
\end{equation} 
\begin{equation} \label{ZEqnNum348638} 
	\begin{array}{cc} {\theta _{i} \equiv \angle \left(\pmb{p}_{i} ,\pmb{k}\right)\sim 1,} & {\angle \left(\pmb{p}_{j} ,\pmb{k}\right) \approx \theta _{i}} \end{array},\quad j=f,+,- .
\end{equation} 
\begin{figure}[h]
	\includegraphics{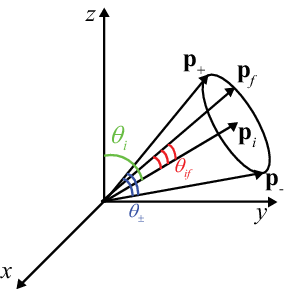}% Here is how to import EPS art
	\caption{\label{fig 3} Geometry of initial and final particles of the process.}
\end{figure}

\noindent We also assume that the classical parameter $\eta $ and the quantum parameter of the Compton Effect are bounded from above by the following conditions \cite{66,67,68,69}:
\begin{equation} \label{ZEqnNum979913} 
	\eta <<\frac{E_{j} }{m} ,\quad j=i,f,\pm \; ;\quad \varepsilon _{iC} <<\frac{E_{i} }{m} . 
\end{equation} 
Because of this, further consideration of the resonant process will be valid for sufficiently large wave intensities. However, the intensity of these fields should be less than the critical Schwinger field $F_{*} \approx 1.3\cdot 10^{16} \; {{\rm V}\mathord{\left/ {\vphantom {{\rm V} {\rm cm}}} \right. \kern-\nulldelimiterspace} {\rm cm}} $.

The resonant energy of the final electron is determined by the Compton effect stimulated by an external field. Given the law of conservation of 4-momentum at the first vertex \eqref{ZEqnNum691334}, the resonant condition \eqref{ZEqnNum933964}, and the kinematic conditions \eqref{ZEqnNum378702}\eqref{ZEqnNum378702}-\eqref{ZEqnNum348638}, after simple manipulations, we get the following expression for the resonant energy of the final electron at the first vertex:
\begin{equation} \label{ZEqnNum910925} 
	x_{f}^{} =\frac{2+l_{1} \varepsilon _{iC} \pm \sqrt{\left(l_{1} \varepsilon _{iC} \right)^{2} -4\delta _{f}^{2} } }{2\left(1+\delta _{f}^{2} +l_{1} \varepsilon _{iC} \right)} .  
\end{equation} 
Here it is indicated:
\begin{equation} \label{ZEqnNum895654} 
	\quad x_{f} =\frac{E_{f} }{E_{i} } ,\quad \delta _{f}^{2} =\frac{E_{i}^{2} \theta _{f}^{2} }{m_{*}^{2} } .  
\end{equation} 
In the expression \eqref{ZEqnNum910925}, the quantum parameter $\varepsilon _{iC} $ is defined by the relations \eqref{ZEqnNum148652}, \eqref{ZEqnNum291894}.  From the expression \eqref{ZEqnNum910925}, it follows that the energy of the final electron is determined by the number of absorbed gamma-quanta of the wave $\left(l_{1} \right)$, the quantum parameter $\varepsilon _{iC} $, and the outgoing angle of the final electron relative to the momentum of the initial electron (an ultrarelativistic parameter $\delta _{f}^{2} $). Solution \eqref{ZEqnNum910925} has two branches (upper $x_{f\left(u\right)} $ and lower $x_{f\left(d\right)} $) in accordance with the signs $\left(\pm \right)$ before the square root in the expression \eqref{ZEqnNum910925}. For fixed values $l_{1} $ and $\varepsilon _{iC} $, upper and lower branches depend only on outgoing angle of the electron $\left(\delta _{f}^{2} \right)$. It is important to emphasize that the interval of variation of the ultrarelativistic parameter $\delta _{f}^{2} $ will be different for the upper and lower branches of the electron energy values. So, from the expression \eqref{ZEqnNum910925}, it follows that the ultrarelativistic parameter for the lower branch $\left(\delta _{f\left(d\right)}^{2} \right)$ has the following interval of change:
\begin{equation} \label{ZEqnNum616240} 
	0\le \delta _{f\left(d\right)}^{2} \le {\left(l_{1} \varepsilon _{iC} \right)^{2} \mathord{\left/ {\vphantom {\left(l_{1} \varepsilon _{iC} \right)^{2}  4}} \right. \kern-\nulldelimiterspace} 4}  .
\end{equation} 
In this case, the minimum and maximum value of the resonant energy of an electron on the lower branch is obtained from the relation \eqref{ZEqnNum910925} when an electron is scattered at zero and maximum angles:
\begin{equation} \label{ZEqnNum131770} 
	x_{f\left(d\right)}^{\min } =\frac{1}{l_{1} \varepsilon _{iC} +1} ,\quad x_{f\left(d\right)}^{\max } =\frac{2}{l_{1} \varepsilon _{iC} +2} . 
\end{equation} 
At the same time, for the upper branch of the resonant energy of the electron $x_{f\left(u\right)} $\eqref{ZEqnNum910925}, the parameter change interval $\delta _{f}^{2} $ should be different. Indeed, in this case it is impossible to take the minimum value of the outgoing angle $\delta _{f\left(u\right)}^{2} =0$, since we get $x_{f\left(u\right)} =1$, something that does not make physical sense. Because of this, the desired range of change in the ultrarelativistic parameter $\delta _{f\left(u\right)}^{2} $ will be obtained from matching solutions for the resonant energies of the electron (at the first vertex) and the electron-positron pair (at the second vertex) (see \eqref{ZEqnNum669663}).

The resonant energy of the electron-positron pair is determined by the Breit-Wheeler process stimulated by an external field at the second vertex. Given the relations \eqref{ZEqnNum933964}, \eqref{ZEqnNum921245}-\eqref{ZEqnNum348638}, after simple manipulations, we obtain the expression for the energies of the electron $\left(x_{-} \right)$ and positron $\left(x_{+} \right)$ of a pair in units of the energy of the initial electron:
\begin{equation} \label{ZEqnNum981727} 
	\left(x_{+} +x_{-} \right)\left(x_{+} +x_{-} -l_{2} \varepsilon _{iBW} x_{+} x_{-} \right)+4\delta _{\pm }^{2} x_{+}^{2} x_{-}^{2} =0.  
\end{equation} 
Here, the quantum parameter $\varepsilon _{iBW} $ is defined by the relation \eqref{ZEqnNum148652}, \eqref{ZEqnNum291894}, and the following notation is used:
\begin{equation} \label{ZEqnNum539422} 
	\begin{array}{cc} {x_{\pm } =\frac{E_{\pm } }{E_{i} } ,} & {\delta _{\pm }^{2} =\frac{E_{i}^{2} \theta _{\pm }^{2} }{m_{*}^{2} } .} \end{array} 
\end{equation} 
The expression \eqref{ZEqnNum981727} implies that it is symmetric with respect to replacement $x_{+} \leftrightarrow x_{-} $. Therefore, the resonant energies of the electron and positron are equal and in the equation \eqref{ZEqnNum981727} it can be put that $x_{+} =x_{-} $. Given this, after simple transformations, we obtain the resonant energy of the positron (electron) of the pair:
\begin{equation} \label{ZEqnNum722916} 
	x_{\pm }^{} =\frac{l_{2} \varepsilon _{iBW} }{\delta _{\pm }^{2} } \left(1\pm \sqrt{1-\frac{\delta _{\pm }^{2} }{l_{2}^{2} \varepsilon _{iBW}^{2} } } \right). 
\end{equation} 
From the expression \eqref{ZEqnNum722916}, it follows that the energy of the positron (electron) pair is determined by the number of absorbed gamma-quanta of the wave $\left(l_{2} \right)$, the quantum parameter $\varepsilon _{iBW} $, and the angle between the momenta of the electron and positron of the pair (an ultrarelativistic parameter $\delta _{\pm }^{2} $). Solution \eqref{ZEqnNum722916} has two branches (upper $x_{\pm \left(u\right)} $ and lower $x_{\pm \left(d\right)} $) in accordance with the sign $\left(\pm \right)$ before the square root in the ratio \eqref{ZEqnNum722916}). For fixed values $l_{2} $ and $\varepsilon _{iBW} $, both branches depend only on their outgoing angle $\left(\delta _{\pm }^{2} \right)$. It is important to emphasize that the ultrarelativistic parameter $\delta _{\pm }^{2} $ has a different variation interval for the upper and lower branches of the resonant energy values of the electron-positron pair. So, from the expression \eqref{ZEqnNum722916}, it follows that the interval of variation of the ultrarelativistic parameter for the lower branch of the resonant energy of the pair  $\delta _{\pm \left(d\right)}^{2} $ has the form:
\begin{equation} \label{ZEqnNum278895} 
	0\le \delta _{\pm \left(d\right)}^{2} \le \left(l_{2} \varepsilon _{iBW} \right)^{2}  .
\end{equation} 
In this case, the minimum and maximum value of the resonant energy of the electron-positron pair is obtained from the ratio \eqref{ZEqnNum722916} for zero and maximum outgoing angle of the electron-positron pair:
\begin{equation} \label{ZEqnNum308607} 
	2x_{\pm \left(d\right)}^{\min } =\frac{1}{l_{2} \varepsilon _{iBW} } ,\quad 2x_{\pm \left(d\right)}^{\max } =\frac{2}{l_{2} \varepsilon _{iBW} }  .
\end{equation} 
At the same time, there is a lower limit on the number of absorbed gamma-quanta of the wave at the second vertex:
\begin{equation} \label{ZEqnNum326382} 
	l_{2} \ge l_{2\left(d\right)}^{\min } ,\quad l_{2\left(d\right)}^{\min } =\left\lceil 2\varepsilon _{iBW}^{-1} \right\rceil  .
\end{equation} 
Note that the condition \eqref{ZEqnNum326382} is necessary, but not sufficient (see the condition \eqref{ZEqnNum521789}). For the upper branch solutions $x_{\pm \left(u\right)} $\eqref{ZEqnNum722916}, you can not take the minimum value of the outgoing angle $\delta _{\pm \left(u\right)}^{2} \to 0$, since in this case we get $x_{\pm \left(u\right)} \sim \delta _{\pm \left(u\right)}^{-2} \to \infty $, which does not make physical sense. The range of change in the ultrarelativistic parameter $\delta _{\pm \left(u\right)}^{2} $ will be obtained from matching solutions for the resonant energies of the electron (at the first vertex) and the electron-positron pair (at the second vertex) (see \eqref{ZEqnNum100457}). 

Let us write down the law of conservation of energy in general for the process under consideration:
\begin{equation} \label{ZEqnNum698860} 
	x_{f} +2x_{\pm } \approx 1. 
\end{equation} 
Note that in the relation \eqref{ZEqnNum698860}, we ignored small corrections (${\left|l\right|\omega \mathord{\left/ {\vphantom {\left|l\right|\omega  E_{i} }} \right. \kern-\nulldelimiterspace} E_{i} } <<1$ and $\eta ^{2} {m^{2} \mathord{\left/ {\vphantom {m^{2}  E_{j}^{2} }} \right. \kern-\nulldelimiterspace} E_{j}^{2} } <<1$, see \eqref{ZEqnNum979913}).  The resonant energies of the electron \eqref{ZEqnNum910925} and the electron-positron pair \eqref{ZEqnNum722916} each have two possible solutions. In this case, four possible cases formally arise with the energies of final particles: each branch of the pair energy corresponds to two branches of the final electron energy. However, due to the general law of conservation of energy, only two cases that correspond to the upper (lower) energy branch of the electron-positron pair and the lower (upper) energy branch of the final electron make physical sense for the resonant process under study.  It should also be noted, that, unlike the lower branches of final particle energies, the upper branches give non-physical values and require additional investigation.  

First, we find physical solutions for the upper energy branch of the electron-positron pair using the lower energy branch of the electron and the general energy conservation law \eqref{ZEqnNum698860}, which in this case  takes the form:
\begin{equation} \label{ZEqnNum161588} 
	2x_{\pm \left(u\right)} =1-x_{f\left(d\right)}  .
\end{equation} 
Substituting in this expression the upper branch for the electron-positron pair \eqref{ZEqnNum722916} and the lower branch for the electron energy \eqref{ZEqnNum910925}, we get:
\begin{equation} \label{ZEqnNum419337} 
	\sqrt{1-\frac{\delta _{\pm \left(u\right)}^{2} }{\left(l_{2} \varepsilon _{iBW} \right)^{2} } } =\frac{\delta _{\pm \left(u\right)}^{2} }{2l_{2} \varepsilon _{iBW} } \left(1-x_{f\left(d\right)} \right)-1>0 .
\end{equation} 
Using simple manipulations, we obtain a condition for matching the outgoing angles of the electron-positron pair and the final electron: 
\begin{equation} \label{ZEqnNum100457} 
	\delta _{\pm \left(u\right)}^{2} =\frac{4}{\left(1-x_{f\left(d\right)} \right)} \left[l_{2} \varepsilon _{iBW} -\frac{1}{\left(1-x_{f\left(d\right)} \right)} \right]>0. 
\end{equation} 
The \eqref{ZEqnNum100457} relation allows us to determine the interval of variation of the ultrarelativistic parameter $\delta _{\pm \left(u\right)}^{2} $ for the upper energy branch of the electron-positron pair:
\begin{equation} \label{ZEqnNum169247} 
	\delta _{\pm \left(\min \right)}^{2} \le \delta _{\pm \left(u\right)}^{2} \le \delta _{\pm \left(\max \right)}^{2}  
\end{equation} 
where is indicated
\begin{equation} \label{ZEqnNum272695} 
	\delta _{\pm \left(\min \right)}^{2} =4\left(1+\frac{1}{l_{1} \varepsilon _{iC} } \right)\left[l_{2} \varepsilon _{iBW} -\left(1+\frac{1}{l_{1} \varepsilon _{iC} } \right)\right]>0 ,
\end{equation} 
\begin{equation} \label{ZEqnNum575279} 
	\delta _{\pm \left(\max \right)}^{2} =4\left(1+\frac{2}{l_{1} \varepsilon _{iC} } \right)\left[l_{2} \varepsilon _{iBW} -\left(1+\frac{2}{l_{1} \varepsilon _{iC} } \right)\right]>0 .
\end{equation} 
Note that the right-hand sides of the expressions \eqref{ZEqnNum419337}, \eqref{ZEqnNum272695}, and \eqref{ZEqnNum575279} must be greater than zero. This implies different conditions for the number of absorbed gamma-quanta of the wave at the second vertex. The condition under which all three equations are valid is obtained from the condition that the right-hand side of the equation \eqref{ZEqnNum419337} for $\delta _{\pm \left(u\right)}^{2} =\delta _{\pm \left(\max \right)}^{2} ,\; x_{f\left(d\right)} =x_{f\left(d\right)}^{\max }$. As a result, we obtain the following lower restrictions on the number of absorbed gamma-quanta of the wave at the second vertex for the upper energy branch of the electron-positron pair:
\begin{equation} \label{ZEqnNum521789} 
	l_{2} \ge l_{2\left(\min \right)} ,\quad l_{2\left(\min \right)} =\left\lceil \frac{2}{\varepsilon _{iBW} } \left(1+\frac{2}{l_{1} \varepsilon _{iC} } \right)\right\rceil ,\quad l_{1} \ge 1 .
\end{equation} 
Note that substituting the expression \eqref{ZEqnNum100457} into the relation for the upper branch of the resonant energy of the electron-positron pair \eqref{ZEqnNum722916} results in expression \eqref{ZEqnNum161588}, as it should be. In this case, the maximum and minimum energies of the electron-positron pair take the form:
\begin{equation} \label{ZEqnNum333964} 
	x_{\pm \left(u\right)}^{\max } \approx \frac{1}{2} \left[1-x_{f\left(d\right)}^{\min } \right]=\frac{1}{2} \left(1+\frac{1}{l_{1} \varepsilon _{iC} } \right)^{-1} , 
\end{equation} 
\begin{equation} \label{ZEqnNum238534} 
	x_{\pm \left(u\right)}^{\min } \approx \frac{1}{2} \left[1-x_{f\left(d\right)}^{\max } \right]=\frac{1}{2} \left(1+\frac{2}{l_{1} \varepsilon _{iC} } \right)^{-1} . 
\end{equation} 

It is important to emphasize that the condition for the number of absorbed gamma-quanta of the wave at the second vertex \eqref{ZEqnNum521789} ensures the validity of the energies of final particles in the entire range of outgoing angles of the final electron \eqref{ZEqnNum616240} and the electron-positron pair \eqref{ZEqnNum169247}. From the condition \eqref{ZEqnNum521789}, we can determine the range of the quantum parameter $\varepsilon _{iBW} $ values with the minimum number of gamma-quanta in the second vertex equal to one. After simple manipulations, we get:
\begin{equation} \label{ZEqnNum213828} 
	\varepsilon _{iBW} \ge \varepsilon _{*} ,\quad \varepsilon _{*} =1+\sqrt{1+\frac{1}{l_{1} } } ,\quad \left(l_{1} \ge 1,\; l_{2\min } =1\right) .
\end{equation} 

Note also that under conditions when final particles fly out at the minimum angles $\delta _{f\left(d\right)}^{2} =0$, $\delta _{\pm \left(u\right)}^{2} =\delta _{\pm \left(\min \right)}^{2} $ the number of absorbed gamma-quanta of the wave at the second vertex will be determined by a more "soft" condition, which follows from the relation \eqref{ZEqnNum272695}:
\begin{equation} \label{ZEqnNum162642} 
	l_{2} \ge l_{2\min \left(0\right)} =\left\lceil \frac{1}{\varepsilon _{iBW} } \left[1+\frac{1}{l_{1} \varepsilon _{iC} } \right]\right\rceil  .
\end{equation} 
From this, we obtain a condition for the possible values of the quantum parameter $\varepsilon _{iBW} $, under which the minimum number of gamma-quanta in the second vertex is equal to one:
\begin{equation} \label{ZEqnNum957720} 
	\varepsilon _{iBW} \ge \frac{1}{2} \varepsilon _{*} ,\quad \left(l_{1} \ge 1,\; l_{2\min \left(0\right)} =1\right) .
\end{equation} 
Figure 4a shows the dependences of the energies of final particles (an electron on the lower branch and the electron-positron pair on the upper branch) on the square of the outgoing angle of the final electron for different numbers of absorbed gamma-quanta of the wave at the first vertex. This figure shows that as the number of absorbed gamma-quanta increases, the electron energy decreases, and the energy of the electron-positron pair increases and tends to the energy of the initial electron.  At the same time, the energies of final particles for a different number of gamma-quanta of the wave at the first vertex are in different energy regions (do not intersect). Figure 4b shows the dependences of the square of the outgoing angle of the electron-positron pair (on the upper energy branch) on the square \eqref{ZEqnNum616240} for fixed values of the number of absorbed gamma-quanta of the wave at the first and second vertices. Figure 4b shows that for different numbers of absorbed gamma-quanta of the wave in the first and second vertices, each outgoing angle of the final electron uniquely corresponds to a certain outgoing angle of the electron-positron pair. Expressions \eqref{ZEqnNum161588}, \eqref{ZEqnNum100457}-\eqref{ZEqnNum238534} (see also Figure 4) completely solve the problem of resonant energies of final particles (on the lower branch of the energies of the final electron and the upper branch of the energies of the electron-positron pair). In this case, the outgoing angle of the final electron $\left(\delta _{f\left(d\right)}^{2} \right)$ in the range of angles \eqref{ZEqnNum616240} and the values of the quantum parameter of the Compton effect $\left(\varepsilon _{iC} \right)$ uniquely determine the outgoing angle of the electron-positron pair $\left(\delta _{\pm \left(u\right)}^{2} \right)$ (see the relations \eqref{ZEqnNum100457}-\eqref{ZEqnNum521789}), as well as final particle energies (see the relations \eqref{ZEqnNum910925} and \eqref{ZEqnNum161588}). Because of this, there is a quantum entanglement of states of final particles. Thus, the Compton effect stimulated by an external field, which occurs at the first vertex, completely determines the resonant states of final particles in this case.

Now we turn to the case of the upper energy branch of the final electron and the lower energy branch of the electron-positron pair. In this case energy conservation law \eqref{ZEqnNum698860} takes the form:
\begin{equation} \label{ZEqnNum277929} 
	x_{f\left(u\right)} =1-2x_{\pm \left(d\right)}  .
\end{equation} 
Substituting into this equation the corresponding energies of final particles \eqref{ZEqnNum910925}, \eqref{ZEqnNum722916}, we obtain
\begin{eqnarray} \label{ZEqnNum742010} 
	\sqrt{\left(l_{1} \varepsilon _{iC} \right)^{2} -4\delta _{f\left(u\right)}^{2} }=2\left[1-2x_{\pm \left(d\right)} \right]\times\\
	\times\left(1+l_{1} \varepsilon _{iC} +\delta _{f\left(u\right)}^{2} \right)-\left(2+l_{1} \varepsilon _{iC} \right)>0. \nonumber
\end{eqnarray} 
\begin{figure}[h]
	\includegraphics[width=0.45\textwidth]{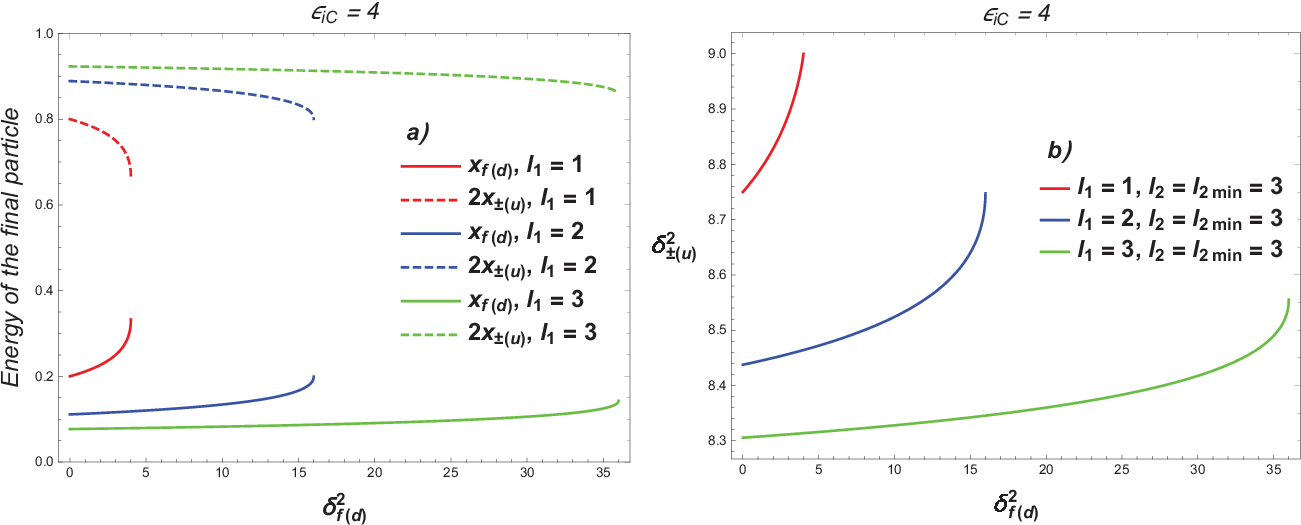}
	\caption{\label{fig 4} Dependence of the energies of final particles \eqref{ZEqnNum910925}, \eqref{ZEqnNum161588} (case a) and the square of the electron-positron pair outgoing angle \eqref{ZEqnNum100457} (case b) on the square of the outgoing angle of the final electron \eqref{ZEqnNum616240} for fixed values of the quantum parameter $\varepsilon _{iC} =4$ $\left(\varepsilon _{iBW} =1\right)$ and the number of absorbed gamma-quanta of the wave in the first and second vertices.}
\end{figure}

After simple transformations from this equation, we obtain the desired dependence of the outgoing angle of the final electron on the outgoing angle of the electron-positron pair:
\begin{equation} \label{ZEqnNum669663} 
	\delta _{f\left(u\right)}^{2} =\left[\frac{1}{2x_{\pm \left(d\right)} } -1\right]^{-1} \left\{l_{1} \varepsilon _{iC} -\left[\frac{1}{2x_{\pm \left(d\right)} } -1\right]^{-1} \right\}. 
\end{equation} 
The relation \eqref{ZEqnNum669663} allows us to determine the range of $\delta _{f\left(u\right)}^{2} $ values for the upper energy branch of the final electron:
\begin{equation} \label{ZEqnNum965893} 
	\delta _{f\left(\min \right)}^{2} \le \delta _{f\left(u\right)}^{2} \le \delta _{f\left(\max \right)}^{2}  ,
\end{equation} 
where is indicated
\begin{equation} \label{ZEqnNum237967} 
	\delta _{f\left(\min \right)}^{2} =\left(\frac{1}{l_{2} \varepsilon _{iBW} -1} \right)\left[l_{1} \varepsilon _{iC} -\left(\frac{1}{l_{2} \varepsilon _{iBW} -1} \right)\right]>0, 
\end{equation} 
\begin{equation} \label{ZEqnNum277186} 
	\delta _{f\left(\max \right)}^{2} =\left(\frac{2}{l_{2} \varepsilon _{iBW} -2} \right)\left[l_{1} \varepsilon _{iC} -\left(\frac{2}{l_{2} \varepsilon _{iBW} -2} \right)\right]>0. 
\end{equation} 
Note that the right-hand sides of the expressions \eqref{ZEqnNum742010}, \eqref{ZEqnNum237967}, and \eqref{ZEqnNum277186} must be greater than zero. This implies different conditions for the number of absorbed gamma-quanta of the wave at the second vertex. The condition under which all three equations are valid is obtained from the condition that the right-hand side of the equation \eqref{ZEqnNum742010} at $\delta _{f\left(u\right)}^{2} =\delta _{f\left(\max \right)}^{2} ,\; x_{\pm \left(d\right)} =x_{\pm \left(d\right)}^{\max } $ is positive.  This condition matches expression \eqref{ZEqnNum521789}. Note that substituting the expression \eqref{ZEqnNum669663} into the relation for the upper branch of the resonant energy of the final electron \eqref{ZEqnNum910925} results in the following expression \eqref{ZEqnNum277929}, as it should be. In this case, the maximum and minimum energies of the final electron take the form:
\begin{equation} \label{ZEqnNum666204} 
	x_{f\left(u\right)}^{\max } =1-2x_{\pm \left(d\right)}^{\min } =1-\frac{1}{l_{2} \varepsilon _{iBW} } , 
\end{equation} 
\begin{equation} \label{ZEqnNum906538} 
	x_{f\left(u\right)}^{\min } =1-2x_{\pm \left(d\right)}^{\max } =1-\frac{2}{l_{2} \varepsilon _{iBW} } . 
\end{equation} 
Figure 5a shows the dependence of the energies of final particles (the final electron on the upper branch and the electron-positron pair on the lower branch) on the square of the angle between the momenta of the electron and positron of the pair for different numbers of absorbed gamma-quanta of the wave at the second vertex, when $l_{2\min } =3$. It follows from this figure that with an increase in the number of absorbed gamma-quanta of the wave, the energy of the electron-positron pair decreases, and the energy of the final electron increases and tends to the energy of the initial electron. At the same time, the energies of final particles for a different number of gamma-quanta of the wave at the second vertex are in different energy regions (do not intersect). Figure5b shows the dependences of the square of the final electron outgoing angle (on the upper energy branch) on the square of the electron-positron pair outgoing angle for fixed values of the number of absorbed gamma-quanta of the wave at the first and second vertices. Figure 5b shows that for different numbers of absorbed gamma-quanta of the wave at the first and second vertices, each outgoing angle of the electron-positron pair uniquely corresponds to a certain outgoing angle of the final electron. Obtained relations \eqref{ZEqnNum277929}, \eqref{ZEqnNum669663}-\eqref{ZEqnNum906538} (see 5) completely solve the problem of resonant energies of final particles (on the lower branch of the electron-positron pair energies and the upper branch of the electron energies). In this case, the outgoing angle of the electron-positron pair $\left(\delta _{\pm \left(d\right)}^{2} \right)$ in \eqref{ZEqnNum278895} and the value of the quantum parameter of the Breit-Wheeler $\left(\varepsilon _{iBW} \right)$ process uniquely determine the outgoing angle of the final electron $\left(\delta _{f\left(u\right)}^{2} \right)$ (see the relations \eqref{ZEqnNum669663}-\eqref{ZEqnNum277186}), as well as final particle energies (see the relations \eqref{ZEqnNum722916} and \eqref{ZEqnNum277929}). Because of this, there is a quantum entanglement of states of final particles. Thus, the Breit-Wheeler process stimulated by an external field, which takes place at the second vertex, completely determines the resonant states of final particles in this case.
\begin{figure}[h]
	\includegraphics[width=0.45\textwidth]{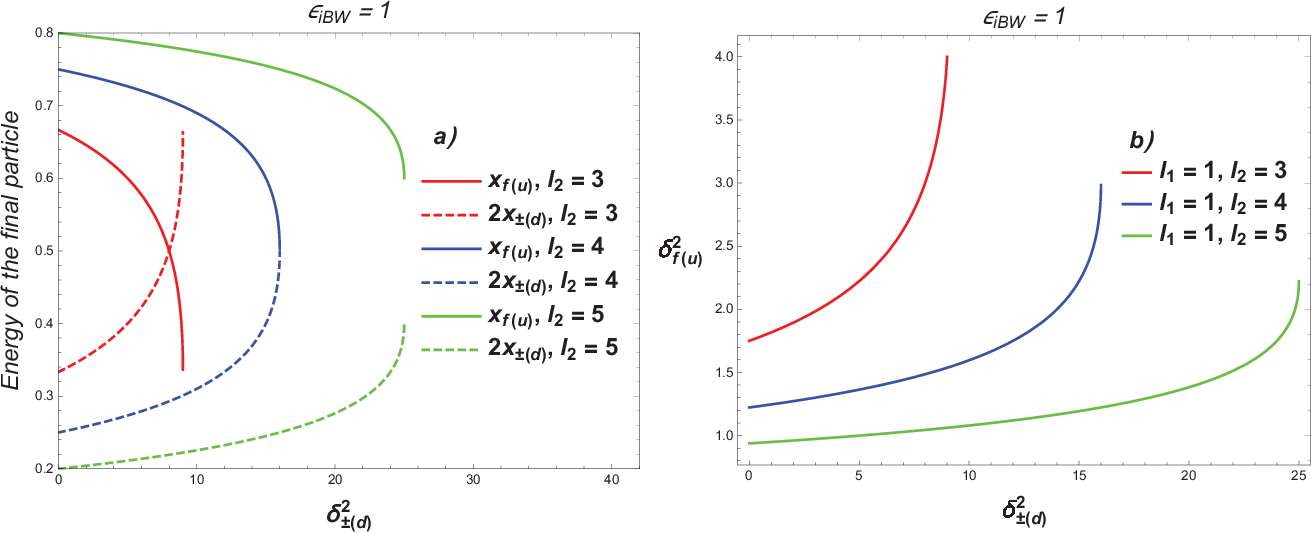}% Here is how to import EPS art
	\caption{\label{fig 5} Dependence of the energies of final particles \eqref{ZEqnNum722916}, \eqref{ZEqnNum277929} (case a) and the square of the final electron outgoing angle \eqref{ZEqnNum669663} (case b) on the square of the angle between the electron and positron of the pair momenta  \eqref{ZEqnNum278895} for fixed values of the quantum parameter $\varepsilon _{iBW} =1$$\left(\varepsilon _{iC} =4\right)$ and the number of absorbed gamma-quanta of the wave at the first and second vertices, and .}
\end{figure}

It is important to emphasize that the condition for the number of absorbed gamma-quanta of the wave at the second vertex \eqref{ZEqnNum521789} ensures the validity of final particles energies in the entire range of electron-positron pair outgoing angles \eqref{ZEqnNum278895} and the final electron outgoing angles \eqref{ZEqnNum965893}. In this case, the range of the quantum parameter $\varepsilon _{iC} $ values at which the minimum number of gamma-quanta in the second vertex is equal to one is determined by the relation \eqref{ZEqnNum213828}. When final particles fly out at the minimum angles $\delta _{\pm \left(d\right)}^{2} =0$, $\delta _{f\left(u\right)}^{2} =\delta _{f\left(\min \right)}^{2} $ the number of absorbed gamma-quanta of the wave at the second vertex is determined by a "softer" condition \eqref{ZEqnNum162642}, which follows from the relation \eqref{ZEqnNum237967}. In this case, the range of the quantum parameter $\varepsilon _{iC} $ values at which the minimum number of gamma-quanta in the second vertex is equal to one is determined by the expression \eqref{ZEqnNum957720}.

Let us consider the case of initial electron energies significantly exceeding the characteristic Compton effect energy, but simultaneously of the same order as the characteristic Breit-Wheeler energy:
\begin{equation} \label{ZEqnNum303594} 
	E_{i} >>\omega _{C} ,\quad E_{i} \sim \omega _{BW} . 
\end{equation} 
In this case, $\varepsilon _{iC} >>1$ and $\varepsilon _{iBW} \sim 1$. Therefore, the resonant energy of the final electron (lower branch, see the relations \eqref{ZEqnNum910925} and \eqref{ZEqnNum131770}) will be small compared to the initial electron energy, and the resonant energy of the electron-positron pair (upper branch, see the relations \eqref{ZEqnNum161588} and \eqref{ZEqnNum333964}, \eqref{ZEqnNum238534}) will be close to the energy of the initial electron:
\begin{equation} \label{ZEqnNum803024} 
	x_{f\left(d\right)} \sim \frac{1}{l_{1} \varepsilon _{iC} } <<1,\quad 2x_{\pm \left(u\right)} \approx 1-x_{f\left(d\right)} \approx 1 .
\end{equation} 
In addition, as follows from the relations \eqref{ZEqnNum169247}-\eqref{ZEqnNum575279}, the ultrarelativistic parameter of the pair \eqref{ZEqnNum100457}, which determines their outgoing angle, will be close to the value $\delta _{\pm \left(u\right)}^{2} \approx \delta _{\pm \left(*\right)}^{2} $:
\begin{eqnarray} \label{ZEqnNum986696} 
	&&\delta _{\pm \left(\min \right)}^{2} \approx \delta _{\pm \left(*\right)}^{2} \left(1+\frac{\xi }{l_{1} \varepsilon _{iC} } \right)\approx \delta _{\pm \left(*\right)}^{2} ,\quad \\ \nonumber
	&&\delta _{\pm \left(\max \right)}^{2} \approx \delta _{\pm \left(*\right)}^{2} \left(1+\frac{2\xi }{l_{1} \varepsilon _{iC} } \right)\approx \delta _{\pm \left(*\right)}^{2} , \nonumber
\end{eqnarray} 
\begin{equation} \label{ZEqnNum394505} 
	\delta _{\pm \left(*\right)}^{2} =4\left(l_{2} \varepsilon _{iBW} -1\right),\quad \xi =\frac{\left(l_{2} \varepsilon _{iBW} -2\right)}{\left(l_{2} \varepsilon _{iBW} -1\right)}  .
\end{equation} 
It is also important to note that the number of gamma-quanta of the wave in the second vertex \eqref{ZEqnNum521789} in this case \eqref{ZEqnNum303594} weakly depends on the process parameters in the first vertex
\begin{equation} \label{ZEqnNum560543} 
	l_{2\left(\min \right)} =\left\lceil \frac{8}{\varepsilon _{iC} } \left(1+\frac{2}{l_{1} \varepsilon _{iC} } \right)\right\rceil \approx \left\lceil \frac{8}{\varepsilon _{iC} } \right\rceil =1 .
\end{equation} 
For the second case of final particle energies, we have: $2x_{\pm \left(d\right)} \sim x_{f\left(u\right)} $ (see the relations \eqref{ZEqnNum722916}, \eqref{ZEqnNum308607}, \eqref{ZEqnNum277929}, \eqref{ZEqnNum666204}, \eqref{ZEqnNum906538}).

Now let us consider a more stringent condition for initial electrons energies than \eqref{ZEqnNum303594}, when they significantly exceeds the characteristic Breit-Wheeler energy
\begin{equation} \label{ZEqnNum157133} 
	E_{i} >>\omega _{BW}  .
\end{equation} 
In this case, the characteristic parameters at the first and second vertices are significantly greater than unity: $\varepsilon _{iC} >>1,\; \varepsilon _{iBW} >>1$. Therefore, only the just described case of resonant energies of final particles is realized here, when the energy of the initial electron mainly turns into the energy of the electron-positron pair (see the relations \eqref{ZEqnNum803024}-\eqref{ZEqnNum394505}). At the same time,
\begin{equation} \label{63)} 
	\xi \approx 1,\quad \delta _{\pm \left(*\right)}^{2} \approx 4l_{2} \varepsilon _{iBW} . 
\end{equation} 
In addition, for the second case of final particle energies, the energy of the electron-positron pair (lower branch, see relations \eqref{ZEqnNum722916}, \eqref{ZEqnNum308607}) will be small compared to the energy of the initial electrons, and the resonant energy of the final electron (upper branch, see relations \eqref{ZEqnNum277929}, \eqref{ZEqnNum666204}, \eqref{ZEqnNum906538}) will be close to the energy of the initial electrons:
\begin{equation} \label{ZEqnNum173271} 
	2x_{\pm \left(d\right)}^{\min } \sim \frac{1}{l_{2} \varepsilon _{iBW} } <<1,\quad x_{f\left(u\right)} =1-2x_{\pm \left(d\right)}^{\min } \approx 1 .
\end{equation} 
In addition, as follows from the relations \eqref{ZEqnNum965893}-\eqref{ZEqnNum277186}, the ultrarelativistic parameter  of the final electron \eqref{ZEqnNum669663}, which defines its outgoing angle, will be enclosed in the interval:
\begin{equation} \label{ZEqnNum225692} 
	\delta _{f\left(*\right)}^{2} \le \delta _{f\left(u\right)}^{2} \le 2\delta _{f\left(*\right)}^{2} ,\quad \delta _{f\left(*\right)}^{2} =\frac{l_{1} }{4l_{2} }  .
\end{equation} 
It is also important to note that the number of gamma-quanta of the wave in the second vertex \eqref{ZEqnNum722916} in this case \eqref{ZEqnNum157133} \eqref{ZEqnNum157133} satisfies the relation $l_{2} \ge 1$.

\section{Resonant differential probability}

It was shown in the previous section that, due to resonant kinematics, in the wave field the initial second-order process effectively decays into two first-order processes: at the first vertex, we have the Compton effect stimulated by an external field, and at the second vertex, the Breit-Wheeler process stimulated by an external field. In this case, the energy of the final electron (on the lower branch) is determined by the first-order process at the first vertex, and the energy of the electron-positron pair (on the lower branch) is given by the first-order process at the second vertex. It is important to note that the two final electrons (in the first and second vertices) have different energies, since their energies are determined by different first-order processes in the first and second vertices. Because of this, the final electron and the pair electron are distinguishable (see Fig.2) and as a result, channels A and B do not interfere and are topologically identical.  Therefore, in the future we will consider only channel A. It is important to emphasize that for the resonant channel A, the energies of final particles for different numbers of absorbed gamma-quanta of the wave are also different (see Fig. 4 and Fig. 5). Therefore, the amplitudes of the process with different numbers of gamma-quanta of the wave $\left(l_{1} ,l_{2} \right)$ also do not interfere with each other. Given the expression for \eqref{ZEqnNum648653}-\eqref{ZEqnNum579777} and the resonant conditions \eqref{ZEqnNum933964}-\eqref{ZEqnNum921245},  it is easy to get partial resonant differential probability of the process (per unit time) for unpolarized initial and final particles \eqref{ZEqnNum378702}-\eqref{ZEqnNum348638}:
\begin{eqnarray} \label{ZEqnNum305958} 
	dw_{l_{1} l_{2} } =\frac{4\pi \alpha ^{2} m^{4} }{\tilde{E}_{i} \tilde{E}_{f} \tilde{E}_{-} \tilde{E}_{+} } \frac{1}{\left|q^{2} \right|^{2} } K_{l_{1} } \left(u_{f} ,z_{f} \right)P_{l_{2} } \left(u_{\pm } ,z_{\pm } \right)\times \\ \nonumber
	\times \delta ^{\left(4\right)} \left[\widetilde{p}_{+} +\widetilde{p}_{-} -\widetilde{p}_{i} +\widetilde{p}_{f} -\left(l_{1} +l_{2} \right)k\right]d^{3} \widetilde{p}_{f} d^{3} \widetilde{p}_{+} d^{3} \widetilde{p}_{-} .\nonumber
\end{eqnarray} 
In the expression \eqref{ZEqnNum305958}, the function $K_{l_{1} } $ defines the probability of the Compton effect stimulated by an external field, and the function $P_{l_{2} } $ -- defines the probability of the Breit-Wheeler process stimulated by an external field \cite{24}. 
\begin{widetext}
	\begin{equation} \label{ZEqnNum302085} 
		K_{l_{1} } \left(u_{f} ,z_{f} \right)=-4J_{l_{1} }^{2} \left(z_{f} \right)+\eta ^{2} \left(2+\frac{u_{f}^{2} }{1+u_{f} } \right)\left[J_{l_{1} -1}^{2} \left(z_{f} \right)+J_{l_{1} +1}^{2} \left(z_{f} \right)-2J_{l_{1} }^{2} \left(z_{f} \right)\right], 
	\end{equation} 
	\begin{equation} \label{ZEqnNum395097} 
		P_{l_{2} } \left(u_{\pm } ,z_{\pm } \right)=J_{l_{2} }^{2} \left(z_{\pm } \right)+\eta ^{2} \left(2u_{\pm } -1\right)\left[\left(\frac{l_{2}^{2} }{z_{\pm }^{2} } -1\right)J_{l_{2} }^{2} \left(z_{\pm } \right)+J_{l_{2} }^{'2} \left(z_{\pm } \right)\right]. 
	\end{equation} 
\end{widetext}
Here it is indicated:
\begin{equation} \label{ZEqnNum514236} 
	\begin{array}{cc} {z_{j} =2l_{j} \frac{\eta }{\sqrt{1+\eta ^{2} } } \sqrt{\frac{u_{j} }{v_{j} } \left(1-\frac{u_{j} }{v_{j} } \right)} ,} & {\left(j=f,\pm \right)} \end{array}, 
\end{equation} 
\begin{equation} \label{ZEqnNum905734} 
	\quad u_{f} =\frac{\left(kq\right)}{\left(kp_{f} \right)} \approx \frac{1}{x_{f} } -1,\quad \quad v_{f} =2l_{1} \frac{\left(kp_{i} \right)}{m_{*}^{2} } \approx l_{1} \varepsilon _{iC} , 
\end{equation} 
\begin{equation} \label{ZEqnNum203124} 
	\quad u_{\pm } =\frac{\left(kq\right)^{2} }{4\left(kp_{-} \right)\left(kp_{+} \right)} \approx \frac{1}{4} \left(\frac{x_{+} }{x_{-} } +\frac{x_{-} }{x_{+} } +2\right)=1, 
\end{equation} 
\begin{equation} \label{ZEqnNum860452} 
	\quad v_{\pm } =l_{2} \frac{\left(kq\right)}{2m_{*}^{2} } \approx \left(x_{+} +x_{-} \right)l_{2} \varepsilon _{iBW} =2x_{\pm } l_{2} \varepsilon _{iBW} . 
\end{equation} 
Substituting \eqref{ZEqnNum905734} - \eqref{ZEqnNum860452} in \eqref{ZEqnNum514236}, taking into account the relations \eqref{ZEqnNum910925} and \eqref{ZEqnNum722916}, we get:
\begin{equation} \label{ZEqnNum806316} 
	z_{f} \approx \frac{2\eta }{\sqrt{1+\eta ^{2} } } \frac{\sqrt{\delta _{f}^{2} } }{\varepsilon _{iC} } ,\quad z_{\pm } \approx \frac{\eta }{\sqrt{1+\eta ^{2} } } \frac{\sqrt{\delta _{\pm }^{2} } }{\varepsilon _{iBW} } . 
\end{equation} 
At the same time, if we take into account the recurrent formulas for Bessel functions, then after simple transformations, the function \eqref{ZEqnNum395097}, taking into account the value $u_{2} $ \eqref{ZEqnNum203124}, will look like:
\begin{eqnarray} \label{ZEqnNum633873} 
	P_{l_{2} } \left(z_{\pm } \right)=J_{l_{2} }^{2} \left(z_{\pm } \right)+\\ \nonumber
	+\frac{1}{2} \eta ^{2} \left[J_{l_{2} +1}^{2} \left(z_{\pm } \right)+J_{l_{2} -1}^{2} \left(z_{\pm } \right)-2J_{l_{2} }^{2} \left(z_{\pm } \right)\right]. \nonumber
\end{eqnarray} 
The resonant energies of the electron and positron pair are the same, i.e. there is a symmetry between the particles of the electron-positron pair. For this reason, integration over $d^{3} \tilde{p}_{+} $ or $d^{3} \tilde{p}_{-} $ can be done with three-dimensional delta function in resonant probability \eqref{ZEqnNum305958}. In addition, it is easy to integrate along the azimuthal angles of departure of the final electron and one of the particles of the pair. Given this, and the ultrarelativistic kinematics \eqref{ZEqnNum378702}-\eqref{ZEqnNum979913}, the resonant differential probability \eqref{ZEqnNum305958} will look like this:
\begin{eqnarray} \label{ZEqnNum588448} 
	\frac{dw_{l_{1} l_{2} } }{d\delta _{f}^{2} d\delta _{\pm }^{2} } =\frac{4\pi ^{3} \alpha ^{2} m^{4} m_{*}^{4} }{E_{i}^{3} \left|q^{2} \right|^{2} } K_{l_{1} } \left(u_{f} ,z_{f} \right)P_{l_{2} } \left(z_{\pm } \right) \times\\ \nonumber
	\times\delta \left(1-x_{f} -x_{-} -x_{+} \right)\frac{x_{f} x_{\pm } }{x_{\mp } } dx_{f} dx_{\pm } . \nonumber
\end{eqnarray} 
The delta function in the relation \eqref{ZEqnNum588448} makes it easy to integrate over the energy of the final electron or one of the particles of the pair. At the same time, two possible variants of the final particle energy should be considered. When the energy of the final electron lies on the lower branch $x_{f\left(d\right)} $, in the relation \eqref{ZEqnNum588448}, integration should be performed with respect to the energy of the particle pair. If the energy of the pair particle lies on the lower branch, then in the relation \eqref{ZEqnNum588448}, integration should be performed with respect to the energy of the final electron. Given this, the expression \eqref{ZEqnNum588448} looks like:
\begin{eqnarray} \label{ZEqnNum852103} 
	\frac{dw_{l_{1} l_{2} }^{\left(f\right)} }{d\delta _{\pm \left(u\right)}^{2} dx_{f\left(d\right)} } =\frac{4\pi ^{3} \alpha ^{2} m^{4} }{E_{i}^{3} } \frac{m_{*}^{4} x_{f\left(d\right)} }{\left|q^{2} \left(x_{f\left(d\right)} \right)\right|^{2} }\times \\ \nonumber
	\times K_{l_{1} } \left(u_{f\left(d\right)} ,z_{f\left(d\right)} \right)P_{l_{2} } \left(z_{\pm \left(u\right)} \right)d\delta _{f}^{2}, \nonumber  
\end{eqnarray} 
\begin{eqnarray} \label{ZEqnNum358334} 
	\frac{dw_{l_{1} l_{2} }^{\left(\pm \right)} }{d\delta _{f\left(u\right)}^{2} dx_{\pm \left(d\right)} } =\frac{4\pi ^{3} \alpha ^{2} m^{4} }{E_{i}^{3} } \frac{m_{*}^{4} \left(1-2x_{\pm \left(d\right)} \right)}{\left|q^{2} \left(x_{\pm \left(d\right)} \right)\right|^{2} }\times \\ \nonumber
	\times K_{l_{1} } \left(u_{f\left(u\right)} ,z_{f\left(u\right)} \right)P_{l_{2} } \left(z_{\pm \left(d\right)} \right)d\delta _{\pm }^{2} . \nonumber
\end{eqnarray} 
Here, functions $K_{l_{1} } $ and $P_{l_{2} }$ are defined by the expressions \eqref{ZEqnNum302085} and \eqref{ZEqnNum633873}, in which the parameters $u_{f} $ \eqref{ZEqnNum905734} and $z_{f} ,z_{\pm } $ \eqref{ZEqnNum806316} are taken on the lower and upper energy branches of the final electron and electron-positron pair (see section 3).

Elimination of resonant infinity in the expressions \eqref{ZEqnNum852103}, \eqref{ZEqnNum358334} is performed by the Breit-Wigner procedure \cite{24,107}:
\begin{equation} \label{ZEqnNum431316} 
	E_{i} \to E_{i} -i{\Gamma _{BW} \mathord{\left/ {\vphantom {\Gamma _{BW}  2}} \right. \kern-\nulldelimiterspace} 2} ,\quad E_{j} \to E_{j} +i{\Gamma _{BW} \mathord{\left/ {\vphantom {\Gamma _{BW}  2}} \right. \kern-\nulldelimiterspace} 2}  , 
\end{equation}
\begin{eqnarray*}
	j=f,\pm ;\quad \Gamma _{BW} =\frac{1}{2}W_{BW} .
\end{eqnarray*} 
where $W_{BW} $ is the total probability (per unit time) of the Braith-Wheeler process stimulated by an external field \cite{24} 
\begin{equation} \label{ZEqnNum178508} 
	W_{BW} \left(\eta ,\varepsilon _{iC} \right)=\frac{\alpha m^{2} }{8\pi q_{0} } {\rm P}\left(\eta ,\varepsilon _{iC} \right), 
\end{equation} 
\begin{equation} \label{ZEqnNum372123} 
	{\rm P}\left(\eta ,\varepsilon _{iC} \right)=\sum _{k=k_{\min } =\left\lceil \varepsilon _{iC} {}^{-1} \right\rceil }^{\infty }{\rm P}_{k} \left(\eta ,\varepsilon _{iC} \right)  , 
\end{equation} 
\begin{eqnarray*}
	{\rm P}_{k} \left(\varepsilon _{iC} \right)=\int _{1}^{k\varepsilon _{iC} }\frac{du}{u\sqrt{u\left(u-1\right)} } P\left(u,\eta ,k\varepsilon _{iC} \right)
\end{eqnarray*}
\begin{eqnarray} \label{81)} 
	P\left(\eta ,u,k\varepsilon _{iC} \right)=J_{r}^{2} \left(z\right)+\eta ^{2} \left(u-\frac{1}{2} \right)[J_{l_{2} +1}^{2} \left(z\right)+\\ \nonumber
	+J_{l_{2} -1}^{2} \left(z\right)-2J_{l_{2} }^{2} \left(z\right)], \nonumber 
\end{eqnarray} 
\begin{equation} \label{ZEqnNum942506} 
	z=2k\frac{\eta }{\sqrt{1+\eta ^{2} } } \sqrt{\frac{u}{k\varepsilon _{iC} } \left(1-\frac{u}{k\varepsilon _{iC} } \right)} . 
\end{equation} 
Due to relations \eqref{ZEqnNum431316} the square of the 4-momentum of the intermediate gamma-quantum obtain an imaginary additive:
\begin{equation} \label{83)} 
	q_{0} \to q_{0} -i\Gamma _{BW} ,\quad q^{2} \to q^{2} -2iq_{0} \Gamma _{BW}  .
\end{equation} 
Taking this into account, the resonant differential probabilities \eqref{ZEqnNum852103} and \eqref{ZEqnNum358334} will take the following form:
\begin{eqnarray} \label{ZEqnNum269941} 
	\frac{dw_{l_{1} l_{2} }^{\left(f\right)} }{d\delta _{\pm \left(u\right)}^{2} dx_{f\left(d\right)} } =\frac{4\pi ^{3} \alpha ^{2} m^{4} }{E_{i}^{3} x_{f\left(d\right)} }K_{l_{1} } \left(u_{f\left(d\right)} ,z_{f\left(d\right)} \right) \times \\ \nonumber
	\times P_{l_{2} } \left(z_{\pm \left(u\right)} \right)\int _{0}^{\infty }\frac{d\delta _{f}^{2} }{\left[\left(\delta _{f}^{2} -\delta _{f\left(d\right)}^{2} \right)^{2} +\Upsilon _{f}^{2} \right]}  , \nonumber
\end{eqnarray} 
\begin{eqnarray} \label{ZEqnNum439593} 
	&&\frac{dw_{l_{1} l_{2} }^{\left(\pm \right)} }{d\delta _{f\left(u\right)}^{2} dx_{\pm \left(d\right)} } =\\ \nonumber
	&&=\frac{4\pi ^{3} \alpha ^{2} m^{4} }{E_{i}^{3} }  K_{l_{1} } \left(u_{f\left(u\right)} ,z_{f\left(u\right)} \right) P_{l_{2} } \left(z_{\pm \left(d\right)} \right)\times \\ \nonumber
	&&\times \frac{\left(1-2x_{\pm \left(d\right)} \right)}{x_{\pm \left(d\right)}^{4} }\int _{0}^{\infty }\frac{d\delta _{\pm }^{2} }{\left[\left(\delta _{\pm }^{2} -\delta _{\pm \left(d\right)}^{2} \right)^{2} +\Upsilon _{\pm }^{2} \right]}  . 
\end{eqnarray} 
Here, the ultrarelativistic parameters $\delta _{f\left(d\right)}^{2} $ and $\delta _{\pm \left(d\right)}^{2} $ determine resonant energies of the final electron and electron-positron pair (see expressions \eqref{ZEqnNum910925} and \eqref{ZEqnNum722916}), and the parameters $\delta _{f}^{2} $ and $\delta _{\pm }^{2} $ can take arbitrary values that do not depend on the energy of final particles and, therefore, can be integrated according to these parameters. In this case, the angular resonant widths are determined by the following expressions:
\begin{equation} \label{86)} 
	\Upsilon _{f} =\frac{\alpha m^{2} }{8\pi m_{*}^{2} x_{f\left(d\right)} } P\left(\eta ,\varepsilon _{iBW} \right), 
\end{equation} 
\begin{equation} \label{87)} 
	\Upsilon _{\pm } =\frac{\alpha m^{2} }{8\pi m_{*}^{2} x_{\pm \left(d\right)}^{2} } P\left(\eta ,\varepsilon _{iBW} \right). 
\end{equation} 
Note that the expressions \eqref{ZEqnNum269941} and \eqref{ZEqnNum439593} have a characteristic Breit-Wigner resonant structure and take the maximum value at $\delta _{f}^{2} =\delta _{f\left(d\right)}^{2} $ and $\delta _{\pm }^{2} =\delta _{\pm \left(d\right)}^{2} $, respectively. Let us integrate expressions \eqref{ZEqnNum269941} and \eqref{ZEqnNum439593} with respect to ultrarelativistic parameters $\delta _{f}^{2} $ and $\delta _{\pm }^{2} $ near their respective resonant values $\delta _{f\left(d\right)}^{2} $ and $\delta _{\pm \left(d\right)}^{2} $. It should be taken into account that the resonant width in these expressions has a dominant value only in the resonant denominators (in all other expressions, this width can be ignored). Therefore, this integration is reduced to an integral of only the resonant denominator. For example, for the expression \eqref{ZEqnNum269941}, we get:
\begin{equation} \label{88)} 
	\int _{-\infty }^{\infty }\frac{dy}{\left[y^{2} +\Upsilon _{f}^{2} \right]}  =\frac{\pi }{\Upsilon _{f}^{} } \quad \left(y=\delta _{f}^{2} -\delta _{f\left(d\right)}^{2} \right) .
\end{equation} 
Here, the limits of integration are extended to $\pm \infty $ due to the fast convergence of the integral. Taking this into account, the expressions for the resonant probabilities \eqref{ZEqnNum269941} and \eqref{ZEqnNum439593} will take the form:
\begin{equation} \label{ZEqnNum870558} 
	\frac{dw_{l_{1} l_{2} }^{\left(f\right)} }{d\delta _{\pm \left(u\right)}^{2} dx_{f\left(d\right)} } =a_{i} \Psi _{l_{1} l_{2} }^{\left(f\right)} \left(x_{f\left(d\right)} ,\delta _{\pm \left(u\right)}^{2} \right), 
\end{equation} 
\begin{equation} \label{ZEqnNum833334} 
	\frac{dw_{l_{1} l_{2} }^{\left(\pm \right)} }{d\delta _{f\left(u\right)}^{2} dx_{\pm \left(d\right)} } =a_{i} \Psi _{l_{1} l_{2} }^{\left(\pm \right)} \left(x_{\pm \left(d\right)} ,\delta _{f\left(u\right)}^{2} \right) .
\end{equation} 
Here the function $a_{i} $ is defined by the initial installation parameters 
\begin{equation} \label{ZEqnNum268191} 
	a_{i} =\frac{32\pi ^{5} \alpha m^{2} m_{*}^{2} }{E_{i}^{3} P\left(\eta ,\varepsilon _{iBW} \right)} ,  
\end{equation} 
and the functions $\Psi _{l_{1} l_{2} }^{\left(f\right)} $ and $\Psi _{l_{1} l_{2} }^{\left(\pm \right)} $ determine the spectral-angular distribution of final particles:
\begin{equation} \label{ZEqnNum677244} 
	\Psi _{l_{1} l_{2} }^{\left(f\right)} \left(x_{f\left(d\right)} ,\delta _{\pm \left(u\right)}^{2} \right)=K_{l_{1} } \left(u_{f\left(d\right)} ,z_{f\left(d\right)} \right)P_{l_{2} } \left(z_{\pm \left(u\right)} \right), 
\end{equation} 
\begin{eqnarray} \label{ZEqnNum238301} 
	&&\Psi _{l_{1} l_{2} }^{\left(\pm \right)} \left(x_{\pm \left(d\right)} ,\delta _{f\left(u\right)}^{2} \right)=\\ \nonumber 
	&&=\frac{\left(1-2x_{\pm \left(d\right)} \right)}{x_{\pm \left(d\right)}^{2} } K_{l_{1} } \left(u_{f\left(u\right)} ,z_{f\left(u\right)} \right)P_{l_{2} } \left(z_{\pm \left(d\right)} \right). \nonumber
\end{eqnarray} 
Note that expressions \eqref{ZEqnNum870558}, \eqref{ZEqnNum268191}, \eqref{ZEqnNum677244} are determined by the resonant differential probability of the RTPP process with simultaneous registration of the energy of the final electron and the angle between the electron and positron of the pair momenta. In this case, the energy of the final electron uniquely determines the outgoing angle of the pair (see the \eqref{ZEqnNum100457} relation). Expressions \eqref{ZEqnNum833334}, \eqref{ZEqnNum268191}, \eqref{ZEqnNum238301} determine the differential probability of the RTPP process with simultaneous registration of the pair energy and the final electron outgoing angle. In this case, the energy of the pair uniquely determines the outgoing angle of the final electron (see the relation \eqref{ZEqnNum669663}). Thus, the differential probabilities \eqref{ZEqnNum870558}, \eqref{ZEqnNum833334} together with the relations \eqref{ZEqnNum100457} and \eqref{ZEqnNum669663} reflect the spectral-angular entanglement of final particles.  

Given the relations for the energies of the electron \eqref{ZEqnNum910925} and the electron-positron pair \eqref{ZEqnNum722916} on their lower branches, we can obtain the relation of differential $dx_{f\left(d\right)} $ with $d\delta _{f\left(d\right)}^{2} $ as well as the relation of differential $dx_{\pm \left(d\right)} $ with $d\delta _{\pm \left(d\right)}^{2} $. Thus, we obtain the corresponding differential probabilities of the RTPP process with angular entanglement of final particles:
\begin{equation} \label{ZEqnNum786463} 
	\frac{dw_{l_{1} l_{2} }^{\left(f\right)} }{d\delta _{\pm \left(u\right)}^{2} d\delta _{f\left(d\right)}^{2} } =a_{i} \Phi _{l_{1} l_{2} }^{\left(f\right)} \left(\delta _{f\left(d\right)}^{2} ,\delta _{\pm \left(u\right)}^{2} \right), 
\end{equation} 
\begin{equation} \label{ZEqnNum978706} 
	\frac{dw_{l_{1} l_{2} }^{\left(\pm \right)} }{d\delta _{f\left(u\right)}^{2} d\delta _{\pm \left(d\right)}^{2} } =a_{i} \Phi _{l_{1} l_{2} }^{\left(\pm \right)} \left(\delta _{\pm \left(d\right)}^{2} ,\delta _{f\left(u\right)}^{2} \right) .
\end{equation} 
Here, the function $\Phi _{l_{1} l_{2} }^{\left(f\right)} $ determines the probability of departure of the final electron by an angle $\delta _{f\left(d\right)}^{2} $ (on the lower branch of the electron energy) and the electron-positron pair by an angle $\delta _{\pm \left(u\right)}^{2} $ (on the upper branch of the pair energy). At the same time, these angles are uniquely related by the \eqref{ZEqnNum100457} relation. 
\begin{widetext}
	\begin{equation} \label{96)} 
		\Phi _{l_{1} l_{2} }^{\left(f\right)} \left(\delta _{f\left(d\right)}^{2} ,\delta _{\pm \left(u\right)}^{2} \right)=\frac{2\sqrt{\left(l_{1} \varepsilon _{iC} \right)^{2} -4\delta _{f\left(d\right)}^{2} } \left(1+\delta _{f\left(d\right)}^{2} +l_{1} \varepsilon _{iC} \right)^{2} K_{l_{1} } \left(u_{f\left(d\right)} ,z_{f\left(d\right)} \right)P_{l_{2} } \left(z_{\pm \left(u\right)} \right)}{2\left(1-\delta _{f\left(d\right)}^{2} \right)+\left(2+l_{1} \varepsilon _{iC} \right)\left(l_{1} \varepsilon _{iC} -\sqrt{\left(l_{1} \varepsilon _{iC} \right)^{2} -4\delta _{f\left(d\right)}^{2} } \right)}  .
	\end{equation} 
\end{widetext}
And the function $\Phi _{l_{1} l_{2} }^{\left(\pm \right)} $ determines the probability of the departure of the electron-positron pair by an angle $\delta _{\pm \left(d\right)}^{2} $ (on the lower branch of the pair energy) and the final electron by an angle $\delta _{f\left(u\right)}^{2} $ (on the upper branch of the electron energy). At the same time, these angles are uniquely related by the \eqref{ZEqnNum669663} relation. 
\begin{widetext}
	\begin{equation} \label{97)} 
		\Phi _{l_{1} l_{2} }^{\left(\pm \right)} \left(\delta _{f\left(u\right)}^{2} ,\delta _{\pm \left(d\right)}^{2} \right)=\frac{\delta _{\pm \left(d\right)}^{2} \left(1-2x_{\pm \left(d\right)} \right)\sqrt{1-{\delta _{\pm \left(d\right)}^{2} \mathord{\left/ {\vphantom {\delta _{\pm \left(d\right)}^{2}  \left(l_{2}^{2} \varepsilon _{iBW}^{2} \right)}} \right. \kern-\nulldelimiterspace} \left(l_{2}^{2} \varepsilon _{iBW}^{2} \right)} } }{x_{\pm \left(d\right)}^{2} \left[x_{\pm \left(d\right)} -{1\mathord{\left/ {\vphantom {1 \left(2l_{2} \varepsilon _{iBW} \right)}} \right. \kern-\nulldelimiterspace} \left(2l_{2} \varepsilon _{iBW} \right)} \right]} K_{l_{1} } \left(u_{f\left(u\right)} ,z_{f\left(u\right)} \right)P_{l_{2} } \left(z_{\pm \left(d\right)} \right) 
	\end{equation} 
\end{widetext}
Note that when an ultrarelativistic electron beam collides with an electromagnetic wave, first-order process takes place (the Compton effect stimulated by an external field). At the same time, the second-order resonant process studied in this paper also takes place. In order, to evaluate the efficiency of this resonant process, it makes sense to consider the resonant differential probability in units of the total probability (per unit time) of the Compton effect stimulated by an external field.
\begin{equation} \label{ZEqnNum958945} 
	W_{C} \left(\eta ,\varepsilon _{iC} \right)=\frac{\alpha m^{2} }{4\pi E_{i} } {\rm K} \left(\eta ,\varepsilon _{iC} \right), 
\end{equation} 
where
\begin{equation} \label{ZEqnNum132969} 
	{\rm K} \left(\eta ,\varepsilon _{iC} \right)=\sum _{n=1}^{\infty }\int _{0}^{n\varepsilon _{iC} }\frac{du}{\left(1+u\right)^{2} } K_{n} \left(\eta ,u,n\varepsilon _{iC} \right) .  
\end{equation} 
Here, functions $K_{n} \left(\eta ,u,n\varepsilon _{iC} \right)$ are defined by the expression: 
\begin{eqnarray} \label{100)} 
	K_{n} \left(\eta ,u,n\varepsilon _{iC} \right)=-4J_{n}^{2} \left(z'\right)+ \\ \nonumber
	+\eta ^{2} \left(2+\frac{u^{2} }{1+u} \right)\left(J_{n+1}^{2} +J_{n-1}^{2} -2J_{n}^{2} \right),  \nonumber
\end{eqnarray} 
\begin{equation} \label{ZEqnNum573036} 
	z'=2n\frac{\eta }{\sqrt{1+\eta ^{2} } } \sqrt{\frac{u}{n\varepsilon _{iC} } \left(1-\frac{u}{n\varepsilon _{iC} } \right)} .  
\end{equation} 
Dividing the relations \eqref{ZEqnNum870558}, \eqref{ZEqnNum833334} and \eqref{ZEqnNum786463}, \eqref{ZEqnNum978706} by the expression \eqref{ZEqnNum958945}, we obtain the relative resonant differential probabilities (in units of the total probability of stimulated by an external field Compton effect):

\begin{enumerate}
	\item  If the final electron energy on the lower branch $\left(x_{f\left(d\right)} \right)$ and the angle between the electron and positron of the pair momenta on the upper branch are simultaneously recorded $\left(\delta _{\pm \left(u\right)}^{2} \right)$:
	\begin{eqnarray} \label{ZEqnNum593059} 
		R_{l_{1} l_{2} }^{\left(f\right)} \left(x_{f\left(d\right)} ,\delta _{\pm \left(u\right)}^{2} \right)=\frac{dw_{l_{1} l_{2} }^{\left(f\right)} }{d\delta _{\pm \left(u\right)}^{2} dx_{f\left(d\right)} } \times\\ \nonumber 
		\times\frac{1}{W_{C} \left(\eta ,\varepsilon _{iC} \right)}=b_{i} \Psi _{l_{1} l_{2} }^{\left(f\right)} \left(x_{f\left(d\right)} ,\delta _{\pm \left(u\right)}^{2} \right). \nonumber
	\end{eqnarray} 
	
	\item  If the outgoing angles of the final electron on the lower branch $\left(\delta _{f\left(d\right)}^{2} \right)$ and the electron-positron pair on the upper branch are simultaneously recorded $\left(\delta _{\pm \left(u\right)}^{2} \right)$:
	\begin{eqnarray} \label{ZEqnNum609209} 
		{\rm H} _{l_{1} l_{2} }^{\left(f\right)} \left(\delta _{f\left(d\right)}^{2} ,\delta _{\pm \left(u\right)}^{2} \right)=\frac{dw_{l_{1} l_{2} }^{\left(f\right)} }{d\delta _{\pm \left(u\right)}^{2} d\delta _{f\left(d\right)}^{2} } \times\\ \nonumber 
		\times\frac{1}{W_{C} \left(\eta ,\varepsilon _{iC} \right)}=b_{i} \Phi _{l_{1} l_{2} }^{\left(f\right)} \left(\delta _{f\left(d\right)}^{2} ,\delta _{\pm \left(u\right)}^{2} \right). \nonumber
	\end{eqnarray} 
	
	\item  If the energy of the electron-positron pair on the lower branch $\left(x_{\pm \left(d\right)} \right)$ and the outgoing angle of the elecron on the upper branch are simultaneously recorded $\left(\delta _{f\left(u\right)}^{2} \right)$:
	\begin{eqnarray} \label{ZEqnNum548453} 
		R_{l_{1} l_{2} }^{\left(\pm \right)} \left(x_{\pm \left(d\right)} ,\delta _{f\left(u\right)}^{2} \right)=\frac{dw_{l_{1} l_{2} }^{\left(\pm \right)} }{d\delta _{f\left(u\right)}^{2} dx_{\pm \left(d\right)} } \times\\ \nonumber 
		\times\frac{1}{W_{C} \left(\eta ,\varepsilon _{iC} \right)} =b_{i} \Psi _{l_{1} l_{2} }^{\left(\pm \right)} \left(x_{\pm \left(d\right)} ,\delta _{f\left(u\right)}^{2} \right). \nonumber
	\end{eqnarray} 
	
	\item  If the outgoing angles of the electron-positron pair on the lower branch $\left(\delta _{\pm \left(d\right)}^{2} \right)$ and the final electron on the upper branch are simultaneously recorded $\left(\delta _{f\left(u\right)}^{2} \right)$:
	\begin{eqnarray} \label{ZEqnNum842068} 
		{\rm H} _{l_{1} l_{2} }^{\left(\pm \right)} \left(\delta _{\pm \left(d\right)}^{2} ,\delta _{f\left(u\right)}^{2} \right)=\frac{dw_{l_{1} l_{2} }^{\left(\pm \right)} }{d\delta _{f\left(u\right)}^{2} d\delta _{\pm \left(d\right)}^{2} } \times\\ \nonumber 
		\times\frac{1}{W_{C} \left(\eta ,\varepsilon _{iC} \right)} =b_{i} \Phi _{l_{1} l_{2} }^{\left(\pm \right)} \left(\delta _{\pm \left(d\right)}^{2} ,\delta _{f\left(u\right)}^{2} \right). \nonumber
	\end{eqnarray} 
\end{enumerate}
Here the dimensionless function $b_{i} $ is determined by the initial setup parameters 
\begin{equation} \label{ZEqnNum905934} 
	b_{i} =\frac{2\left(4\pi ^{2} \right)^{3} \left(1+\eta ^{2} \right)}{P\left(\eta ,\varepsilon _{iBW} \right){\rm K} \left(\eta ,\varepsilon _{iC} \right)\varepsilon _{iC}^{2} } \left(\frac{m}{\omega _{C} } \right)^{2} . 
\end{equation} 

It is important to emphasize that the initial installation $\eta $ \eqref{ZEqnNum778375}, $\omega _{C} $ \eqref{ZEqnNum291894}, $\varepsilon _{iC} $ \eqref{ZEqnNum148652}, as well as the number of absorbed gamma-quanta of the wave at the first vertex $l_{1} \ge 1$ and the outgoing angle of the final electron $\delta _{f\left(d\right)}^{2} $ \eqref{ZEqnNum895654}, \eqref{ZEqnNum616240} uniquely determine the resonant relative probabilities \eqref{ZEqnNum593059}, \eqref{ZEqnNum609209}. At the same time, the parameters that define stimulated by an external field Breit-Wheeler process in the second vertex (see the functions $P_{l_{2} } \left(z_{\pm \left(u\right)} \right)$ \eqref{ZEqnNum633873}) are determined by the relations: $\varepsilon _{iBW} ={\varepsilon _{iC} \mathord{\left/ {\vphantom {\varepsilon _{iC}  4}} \right. \kern-\nulldelimiterspace} 4} $, $l_{2\left(\min \right)} $ \eqref{ZEqnNum521789},  $\delta _{\pm \left(u\right)}^{2} $ \eqref{ZEqnNum100457}. On the other hand, the initial setup $\eta $ \eqref{ZEqnNum778375}, $\omega _{BW} $ \eqref{ZEqnNum291894}, $\varepsilon _{iBW} $ \eqref{ZEqnNum148652}, as well as number of absorbed gamma-quanta of the wave in the second vertex $l_{2} \ge l_{2\left(\min \right)} $ \eqref{ZEqnNum521789} and the outgoing angle of the electron-positron pair $\delta _{\pm \left(d\right)}^{2} $ \eqref{ZEqnNum539422}, \eqref{ZEqnNum278895}, uniquely determine the resonant relative probabilities \eqref{ZEqnNum548453}, \eqref{ZEqnNum842068}. At the same time, the parameters that determine the Compton effect stimulated by an external field at the first vertex (see the functions $K_{l_{1} } \left(u_{f\left(u\right)} ,z_{f\left(u\right)} \right)$ \eqref{ZEqnNum302085}) are determined by the relations: $\varepsilon _{iC} =4\varepsilon _{iBW} $, $l_{1} \ge 1$ \eqref{ZEqnNum521789},  $\delta _{f\left(u\right)}^{2} $ \eqref{ZEqnNum669663}. Taking this into account, we can plot the corresponding relative probabilities multiplied by the parameter  $\left({\omega _{C} \mathord{\left/ {\vphantom {\omega _{C}  m}} \right. \kern-\nulldelimiterspace} m} \right)^{2} $ as a function of the outgoing angles of final particles at different values of the initial parameters $\eta $ and $\varepsilon _{iC} \left(\varepsilon _{iBW} \right)$, which determine the parameters of the plane electromagnetic wave and the initial electron energy. In order for the resonant probability to take maximum values, we will assume that the characteristic quantum parameter of the Breit-Wheeler process is chosen from the condition $\varepsilon _{iBW} \ge \varepsilon _{*} $ $\left(\varepsilon _{iC} \ge 4\varepsilon _{*} \right)$. This condition allows us to select the minimum number of absorbed gamma-quanta of the wave at the second vertex equal to one (see the relation \eqref{ZEqnNum213828}). As was shown earlier (see the relations \eqref{ZEqnNum303594} and \eqref{ZEqnNum157133}), for large values of quantum parameters $\varepsilon _{iC} \left(\varepsilon _{iBW} \right)$, the energy of the initial electron mainly goes into the electron-positron pair energy or the final electron energy. It is important to emphasize that for a fixed energy of the initial electron, the quantum parameter $\varepsilon _{iC} \left(\varepsilon _{iBW} \right)$ increases as the corresponding characteristic energy decreases $\omega _{C} \left(\omega _{BW} \right)$. If we fix the characteristic energy of the process, then the quantum parameter $\varepsilon _{iC} \left(\varepsilon _{iBW} \right)$ increases with increasing energy of the initial electron. 

\noindent 
\begin{figure}[h!]
	\includegraphics[width=0.45\textwidth]{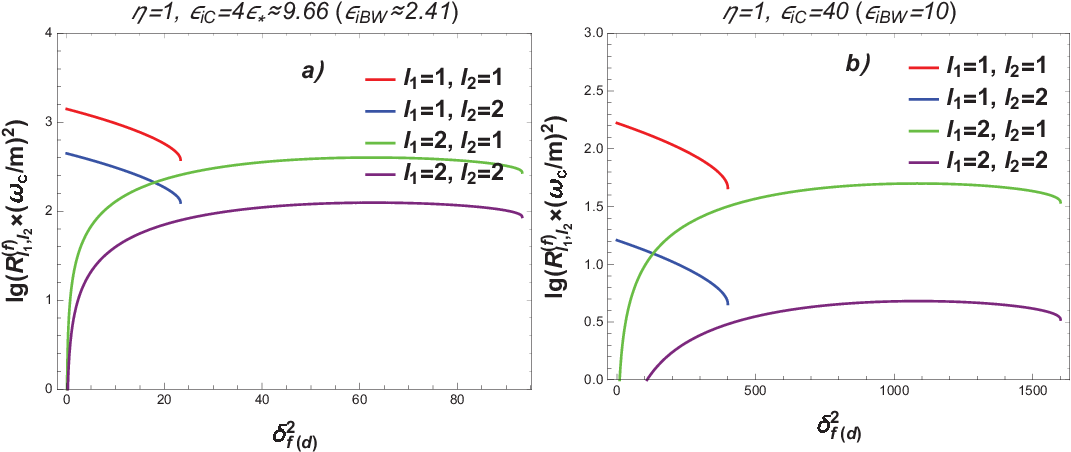}
	\caption{\label{fig 6} Dependence of the resonant relative probability $R_{l_{1} l_{2} }^{\left(f\right)} \times \left({\omega _{C} \mathord{\left/ {\vphantom {\omega _{C}  m}} \right. \kern-\nulldelimiterspace} m} \right)^{2} $ \eqref{ZEqnNum593059} on the square of the final electron outgoing angle \eqref{ZEqnNum616240} for different numbers of absorbed gamma-quanta of a wave in the first and second vertexes with fixed initial parameters.}
\end{figure}

\begin{table*}%The best place to locate the table environment is directly after its first reference in text
	\caption{\label{table1}%
		The maximum values of the resonant relative probability $R_{l_{1} l_{2} \left(\max \right)}^{\left(f\right)} \times \left({\omega _{C} \mathord{\left/ {\vphantom {\omega _{C}  m}} \right. \kern-\nulldelimiterspace} m} \right)^{2} $ \eqref{ZEqnNum593059} and the corresponding values of the energies and outgoing angles of final particles for different numbers of absorbed gamma-quanta of the wave at the first and second vertices and fixed quantum parameters of the Breit-Wheeler process and the Compton effect.
	}
	\begin{ruledtabular}
		\begin{tabular}{ccccccc}
			& $l_{1} {\rm ,\; }l_{2} $ & $\delta _{f\left(d\right)}^{2*} $ &  $\delta _{\pm \left(u\right)}^{2*} $ & $x_{f\left(d\right)} $ & $2x_{\pm \left(u\right)} $ & $R_{l_{1} l_{2} \left(\max \right)}^{\left(f\right)} \times \left(\frac{\omega _{C} }{m} \right)^{2} $ \\ \hline
			$\eta =1$& 1, 1 & 0 & 5.79 & 0.09 & 0.91 & $1.41\times 10^{3} $\\
			$\varepsilon _{iBW} \approx 2.41$& 1, 2 & 0 & 16.44 & 0.09 & 0.91 & $4.45\times 10^{2} $ \\
			$\varepsilon _{iC} \approx 9.66$& 2, 1 & 62.40 & 5.75 & 0.06 & 0.94 & $4.0\times 10^{2} $ \\
			& 2, 2 & 62.84 & 16.04 & 0.06 & 0.94 & $125$ \\ \hline
			$\eta =1$& 1, 1 & 0 & 36.80 & 0.02 & 0.98 & $167$ \\ 
			$\varepsilon _{iBW} \approx 10$& 1, 2 & 0 & 77.80 & 0.02 & 0.98 & $16.15$ \\
			$\varepsilon _{iC} \approx 40$& 2, 1 & 1083 & 36.51 & 0.02 & 0.98 & $50$\\
			& 2, 2 & 1086 & 77.15 & 0.02 & 0.98 & $4.81$ \\
		\end{tabular}
	\end{ruledtabular}
\end{table*}

\begin{figure}[h!]
	\includegraphics[width=0.45\textwidth]{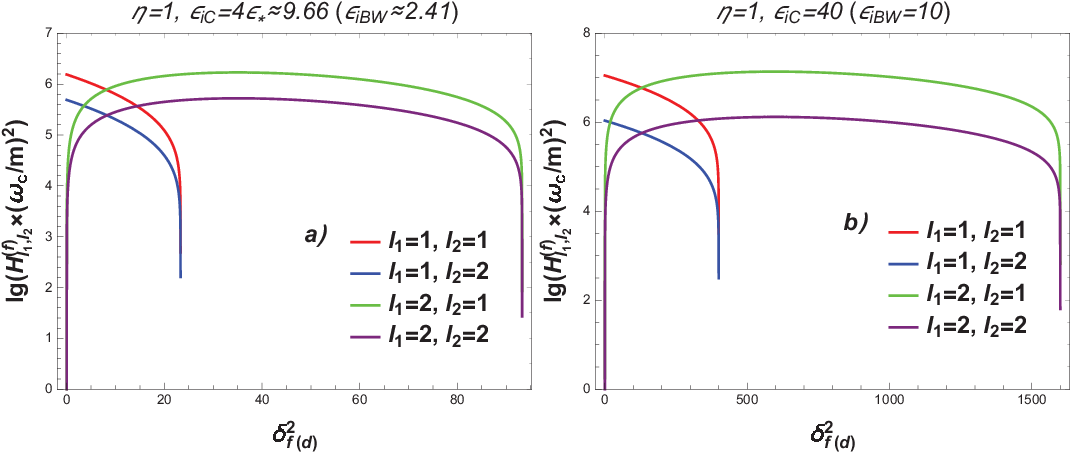}
	\caption{\label{fig 7} Dependence of the resonant relative probability ${\rm H} _{l_{1} l_{2} }^{\left(f\right)} \times \left({\omega _{C} \mathord{\left/ {\vphantom {\omega _{C}  m}} \right. \kern-\nulldelimiterspace} m} \right)^{2} $ \eqref{ZEqnNum609209} on the square of the final electron outgoing angle \eqref{ZEqnNum616240} for different numbers of absorbed gamma-quanta of the wave in the first and second vertexes with fixed initial parameters.}
\end{figure}

\begin{table*}%The best place to locate the table environment is directly after its first reference in text
	\caption{\label{table2}%
		The maximum values of the resonant relative probability $H_{l_{1} l_{2} \left(\max \right)}^{\left(f\right)} \times \left({\omega _{C} \mathord{\left/ {\vphantom {\omega _{C}  m}} \right. \kern-\nulldelimiterspace} m} \right)^{2} $ \eqref{ZEqnNum609209} and the corresponding values of the energies and outgoing angles of final particles for different numbers of absorbed gamma-quanta of the wave at the first and second vertices and fixed quantum parameters of the Breit-Wheeler process and the Compton effect.
	}
	\begin{ruledtabular}
		\begin{tabular}{ccccccc}
			& $l_{1} {\rm ,\; }l_{2} $ & $\delta _{f\left(d\right)}^{2*} $ & $\delta _{\pm \left(u\right)}^{2*} $ & $x_{f\left(d\right)} $ & $2x_{\pm \left(u\right)} $ & $H_{l_{1} l_{2} \left(\max \right)}^{\left(f\right)} \times \left(\frac{\omega _{C} }{m} \right)^{2} $  \\ \hline
			$\eta =1$& 1, 1 & 0 & 5.79 & 0.09 & 0.91 & $1.545\times 10^{6} $ \\
			$\varepsilon _{iBW} \approx 2.41$& 1, 2 & 0 & 16.44 & 0.09 & 0.91 & $4.90\times 10^{5} $ \\
			$\varepsilon _{iC} \approx 9.66$& 2, 1 & 34.95 & 5.74 & 0.05 & 0.95 & $1.67\times 10^{6} $ \\
			& 2, 2 & 35.05 & 15.95 & 0.05 & 0.95 & $5.25\times 10^{5} $ \\ \hline
			$\eta =1$& 1, 1 & 0 & 36.80 & 0.02 & 0.98 & $1.125\times 10^{7} $ \\ 
			$\varepsilon _{iBW} \approx 10$	& 1, 2 & 0 & 77.80 & 0.02 & 0.98 & $1.09\times 10^{6} $ \\
			$\varepsilon _{iC} \approx 40$& 2, 1 & 597.61 & 36.45 & 0.01 & 0.99 & $1.38\times 10^{7} $ \\
			& 2, 2 & 598.34 & 77.00 & 0.01 & 0.99 & $1.32\times 10^{6} $ \\
		\end{tabular}
	\end{ruledtabular}
\end{table*}

\begin{figure}[h]
	\includegraphics[width=0.45\textwidth]{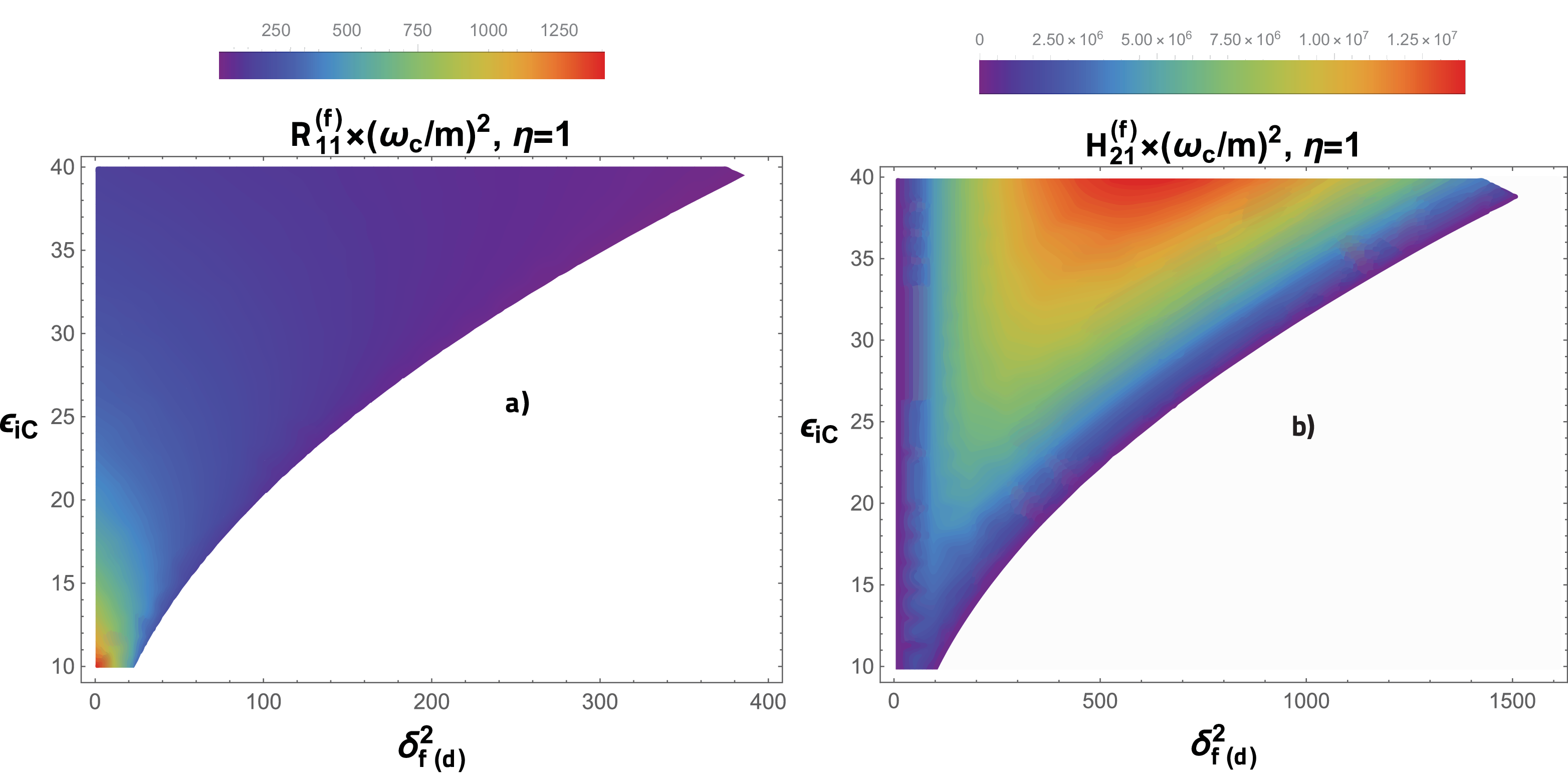}
	\caption{\label{fig 8} Dependence of the relative probabilities $R_{11}^{\left(f\right)} \times \left({\omega _{C} \mathord{\left/ {\vphantom {\omega _{C}  m}} \right. \kern-\nulldelimiterspace} m} \right)^{2} $ \eqref{ZEqnNum593059} and $H_{21}^{\left(f\right)} \times \left({\omega _{C} \mathord{\left/ {\vphantom {\omega _{C}  m}} \right. \kern-\nulldelimiterspace} m} \right)^{2} $ \eqref{ZEqnNum609209} on the square of the final electron outgoing angle and the quantum parameter of the Compton effect.}
\end{figure}

\noindent Figures 6 and 7 show graphs of the relative probabilities $R_{l_{1} l_{2} }^{\left(f\right)} $ \eqref{ZEqnNum593059} and ${\rm H} _{l_{1} l_{2} }^{\left(f\right)} $ \eqref{ZEqnNum609209} (multiplied by a factor $\left({\omega _{C} \mathord{\left/ {\vphantom {\omega _{C}  m}} \right. \kern-\nulldelimiterspace} m} \right)^{2} $) depending on the outgoing angle of the final electron (on lower branch of its energies) for fixed initial setup parameters and different numbers of absorbed gamma-quanta of the wave. Tables 1 and 2 show the corresponding maximum probabilities for different numbers of absorbed gamma-quanta of the wave (at points $\delta _{f\left(d\right)}^{2*} $ corresponding to the maxima of the relative probability distributions in Fig. 6 and Fig. 7), as well as the energies of the final electron and electron-positron pair (in units of the energy of the initial electron). It can be seen from the figures and tables that for $l_{1} =1$ and $l_{2} =1,2,...$, the maximum of relative probabilities occurs when an electron is scattered at a zero angle $\left(\delta _{f\left(d\right)}^{2*} =0\right)$. If $l_{1} =2,3,...$(for any values $l_{2} $), then the maximum of resonant probabilities is shifted to the right, in the region of non-zero values of the electron outgoing angles. Figure 6 and Table 1 show that the maximum relative probability $R_{l_{1} l_{2} \left(\max \right)}^{\left(f\right)} $ holds for the minimum number of absorbed gamma-quanta of the wave $l_{1} =l_{2} =1$. As the number of absorbed gamma-quanta increases, the relative probability decreases. In addition, when the quantum parameter of the Compton effect increases by a factor of 4, the maximum relative probability decreases by one order of magnitude: ${\left(R_{11\left(\max \right)}^{\left(f\right)} \left|{}_{\varepsilon _{iC=40} } \right. \right)\mathord{\left/ {\vphantom {\left(R_{11\left(\max \right)}^{\left(f\right)} \left|{}_{\varepsilon _{iC=40} } \right. \right) \left(R_{11\left(\max \right)}^{\left(f\right)} \left|{}_{\varepsilon _{iC=9.66} } \right. \right)}} \right. \kern-\nulldelimiterspace} \left(R_{11\left(\max \right)}^{\left(f\right)} \left|{}_{\varepsilon _{iC=9.66} } \right. \right)} \approx 0.09$. Figure 7 and Table 2 show that the maximum relative probability $H_{l_{1} l_{2} \left(\max \right)}^{\left(f\right)} $ holds for the number of absorbed gamma-quanta of the wave $l_{1} =2,\; l_{2} =1$. At the same time, $H_{11\left(\max \right)}^{\left(f\right)} {\rm \mathop{<}\limits_\sim }H_{21\left(\max \right)}^{\left(f\right)} $. As the number of absorbed gamma-quanta increases, the relative probability decreases. In addition, when the quantum parameter of the Compton effect is increased by a factor of 4, the maximum relative probability $H_{21\left(\max \right)}^{\left(f\right)} $ is reduced by one order of magnitude: ${\left(H_{21\left(\max \right)}^{\left(f\right)} \left|{}_{\varepsilon _{iC=40} } \right. \right)\mathord{\left/ {\vphantom {\left(H_{21\left(\max \right)}^{\left(f\right)} \left|{}_{\varepsilon _{iC=40} } \right. \right) \left(H_{21\left(\max \right)}^{\left(f\right)} \left|{}_{\varepsilon _{iC=9.66} } \right. \right)}} \right. \kern-\nulldelimiterspace} \left(H_{21\left(\max \right)}^{\left(f\right)} \left|{}_{\varepsilon _{iC=9.66} } \right. \right)} \approx 8.33$. 

Figure 8 shows the dependence of the maximum relative probabilities $R_{11\left(\max \right)}^{\left(f\right)} $ \eqref{ZEqnNum593059} and $H_{21\left(\max \right)}^{\left(f\right)} $ \eqref{ZEqnNum609209} on the final electron outgoing angle and the quantum parameter of the Compton effect at optimal number of absorbed gamma-quanta in the first and second vertices. Figure 8a shows that the maximum relative probability with simultaneous registration of the electron outgoing angle and the electron-positron pair $R_{11\left(\max \right)}^{\left(f\right)} $ energy has a maximum value at zero outgoing angle of the final electron and when the quantum parameter $\varepsilon _{iC} \approx 4\varepsilon _{*} $. As the value of this quantum parameter increases, the relative probability $R_{11\left(\max \right)}^{\left(f\right)} $ decreases rather rapidly. On the other hand, Figure 8b shows that the maximum relative probability with simultaneous registration of the final electron and electron-positron pair outgoing angles $H_{21\left(\max \right)}^{\left(f\right)} $ increases quite rapidly with increasing quantum parameter $\varepsilon _{iC} $, taking maximum values for sufficiently large outgoing angles of the final electron. It is important to note that the final electron outgoing angle uniquely determines the outgoing angle of the electron-positron pair (see the relations \eqref{ZEqnNum100457}, \eqref{ZEqnNum986696}), and also the energies of final particles (see the relations \eqref{ZEqnNum910925}, \eqref{ZEqnNum161588}), i.e. quantum entanglement of final particles takes place. At the same time, as the quantum parameter of the Compton effect increases $\varepsilon _{iC} $ , the energy of the electron-positron pair tends to the energy of the initial electrons \eqref{ZEqnNum803024}.

\begin{figure}[h]
	\includegraphics[width=0.45\textwidth]{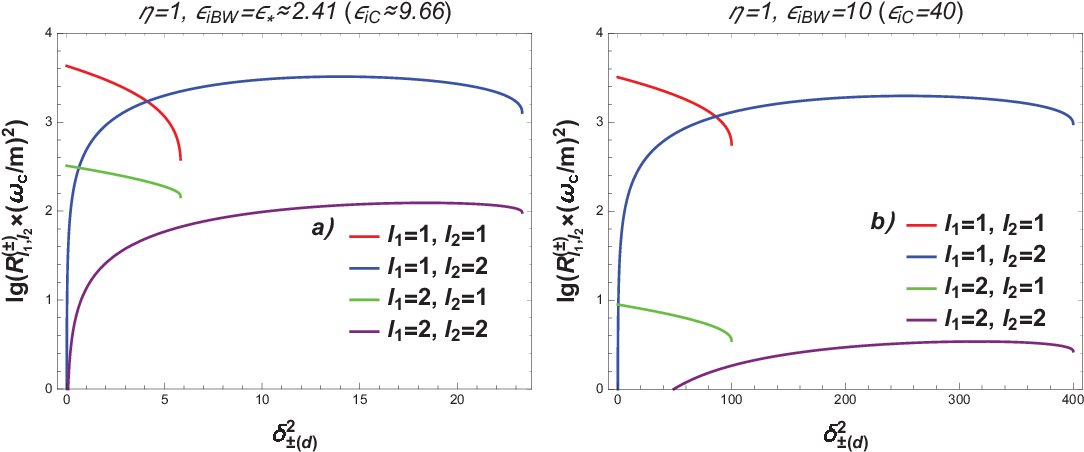}
	\caption{\label{fig 9} Dependence of the resonant relative probability $R_{l_{1} l_{2} }^{\left(\pm \right)} \times \left({\omega _{C} \mathord{\left/ {\vphantom {\omega _{C}  m}} \right. \kern-\nulldelimiterspace} m} \right)^{2} $  \eqref{ZEqnNum548453} on the square of the electron-positron pair outgoing angle \eqref{ZEqnNum278895} for different numbers of absorbed gamma-quanta of the wave in the first and second vertex with fixed initial parameters.}
\end{figure}
\begin{table*}%The best place to locate the table environment is directly after its first reference in text
	\caption{\label{table3}%
		The maximum values of the resonant relative probability $R_{l_{1} l_{2} \left(\max \right)}^{\left(\pm \right)} \times \left({\omega _{C} \mathord{\left/ {\vphantom {\omega _{C}  m}} \right. \kern-\nulldelimiterspace} m} \right)^{2} $\eqref{ZEqnNum548453} and the corresponding values of the energies and outgoing angles of final particles for different numbers of absorbed gamma-quanta of the wave at the first and second vertices and fixed quantum parameters of the Breit-Wheeler process and the Compton effect.
	}
	\begin{ruledtabular}
		\begin{tabular}{ccccccc}
			& $l_{1} {\rm ,\; }l_{2} $ & $\delta _{f\left(u\right)}^{2*} $ & $\delta _{\pm \left(d\right)}^{2*} $ & $x_{f\left(u\right)} $ & $2x_{\pm \left(d\right)} $ & $R_{l_{1} l_{2} \left(\max \right)}^{\left(\pm \right)} \times \left(\frac{\omega _{C}}{m} \right)^{2} $ \\ \hline
			$\eta =1$ & 1, 1 & 6.33 & 0 & 0.59 & 0.41 & $4.28\times 10^{3} $ \\
			$\varepsilon _{iBW} \approx 2.41$& 1, 2 & 3.17 & 13.98 & 0.75 & 0.25 & $3.23\times 10^{3} $ \\
			$\varepsilon _{iC} \approx 9.66$& 2, 1 & 13.16 & 0 & 0.59 & 0.41 & $3.25\times 10^{2} $ \\
			& 2, 2 & 7.48 & 18.34 & 0.72 & 0.28 & $123.75$ \\ \hline
			$\eta =1$& 1, 1 & 4.43 & 0 & 0.90 & 0.10 & $3.21\times 10^{3} $ \\ 
			$\varepsilon _{iBW} \approx 10$& 1, 2 & 2.65 & 253.18 & 0.94 & 0.06 & $1.98\times 10^{3} $ \\
			$\varepsilon _{iC} \approx 40$& 2, 1 & 8.88 & 0 & 0.90 & 0.10 & $8.925$ \\
			& 2, 2 & 5.88 & 315.62 & 0.93 & 0,07 & $3.430$ \\
		\end{tabular}
	\end{ruledtabular}
\end{table*}

\begin{figure}[h]
	\includegraphics[width=0.45\textwidth]{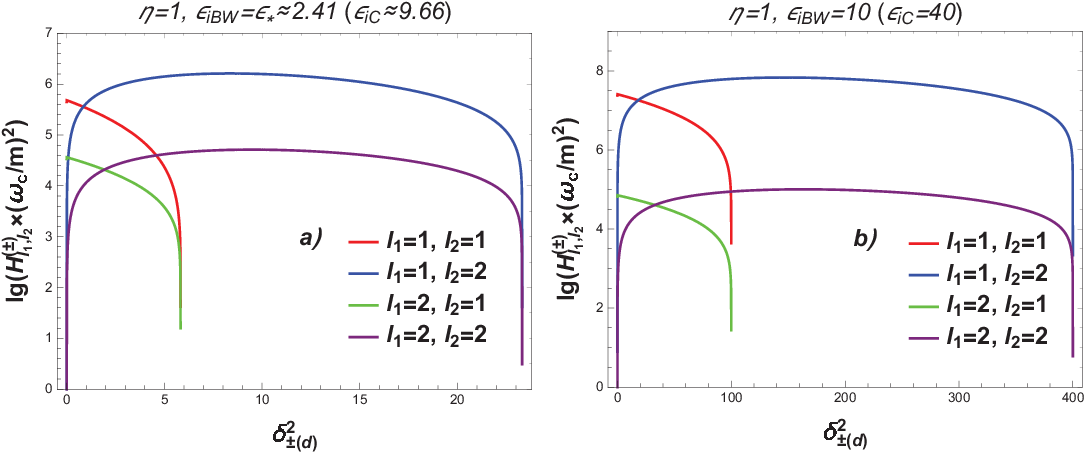}
	\caption{\label{fig 10} Dependence of the resonant relative probability $H_{l_{1} l_{2} }^{\left(\pm \right)} \times \left({\omega _{C} \mathord{\left/ {\vphantom {\omega _{C}  m}} \right. \kern-\nulldelimiterspace} m} \right)^{2} $ \eqref{ZEqnNum842068} on the square of the electron-positron pair outgoing angle  \eqref{ZEqnNum278895} for different numbers of absorbed electrons. gamma-quanta of the first wave and the second vertex with fixed initial parameters.}
\end{figure}

\begin{table*}%The best place to locate the table environment is directly after its first reference in text
	\caption{\label{table4}%
		The maximum values of the resonant relative probability $H_{l_{1} l_{2} \left(\max \right)}^{\left(\pm \right)} \times \left({\omega _{C} \mathord{\left/ {\vphantom {\omega _{C}  m}} \right. \kern-\nulldelimiterspace} m} \right)^{2} $ \eqref{ZEqnNum842068} and the corresponding values of the energies and outgoing angles of final particles for different numbers of absorbed gamma-quanta of the wave at the first and second vertices and fixed quantum parameters of the Breit-Wheeler process and the Compton effect.
	}
	\begin{ruledtabular}
		\begin{tabular}{ccccccc}
			& $l_{1} {\rm ,\; }l_{2} $ & $\delta _{f\left(u\right)}^{2*} $ & $\delta _{\pm \left(d\right)}^{2*} $ & $x_{f\left(u\right)} $ & $2x_{\pm \left(d\right)} $ & $H_{l_{1} l_{2} \left(\max \right)}^{\left(\pm \right)} \times \left(\frac{\omega _{C} }{m} \right)^{2} $ \\ \hline
			$\eta =1$ & 1, 1 & 6.33 & 0 & 0.59 & 0.41 & $5.00\times 10^{5} $ \\
			$\varepsilon _{iBW} \approx 2.41$& 1, 2 & 2.79 & 8.35 & 0.77 & 0.23 & $1.62\times 10^{6} $ \\
			$\varepsilon _{iC} \approx 9.66$& 2, 1 & 13.16 & 0 & 0.59 & 0.41 & $3.80\times 10^{4} $ \\
			& 2, 2 & 5.80 & 9.47 & 0.77 & 0.23 & $5.0\times 10^{4} $ \\ \hline
			$\eta =1$& 1, 1 & 4.43 & 0 & 0.90 & 0.10 & $2.60\times 10^{7} $ \\ 
			$\varepsilon _{iBW} \approx 10$& 1, 2 & 2.36 & 146.94 & 0.94 & 0.06 & $6.80\times 10^{7} $ \\
			$\varepsilon _{iC} \approx 40$& 2, 1 & 8.88 & 0 & 0.90 & 0.10 & $7.0\times 10^{4} $ \\
			& 2, 2 & 4.79 & 162.89 & 0.94 & 0.06 & $1.0\times 10^{5} $\\
		\end{tabular}
	\end{ruledtabular}
\end{table*}

\begin{figure}[h]
	\includegraphics[width=0.45\textwidth]{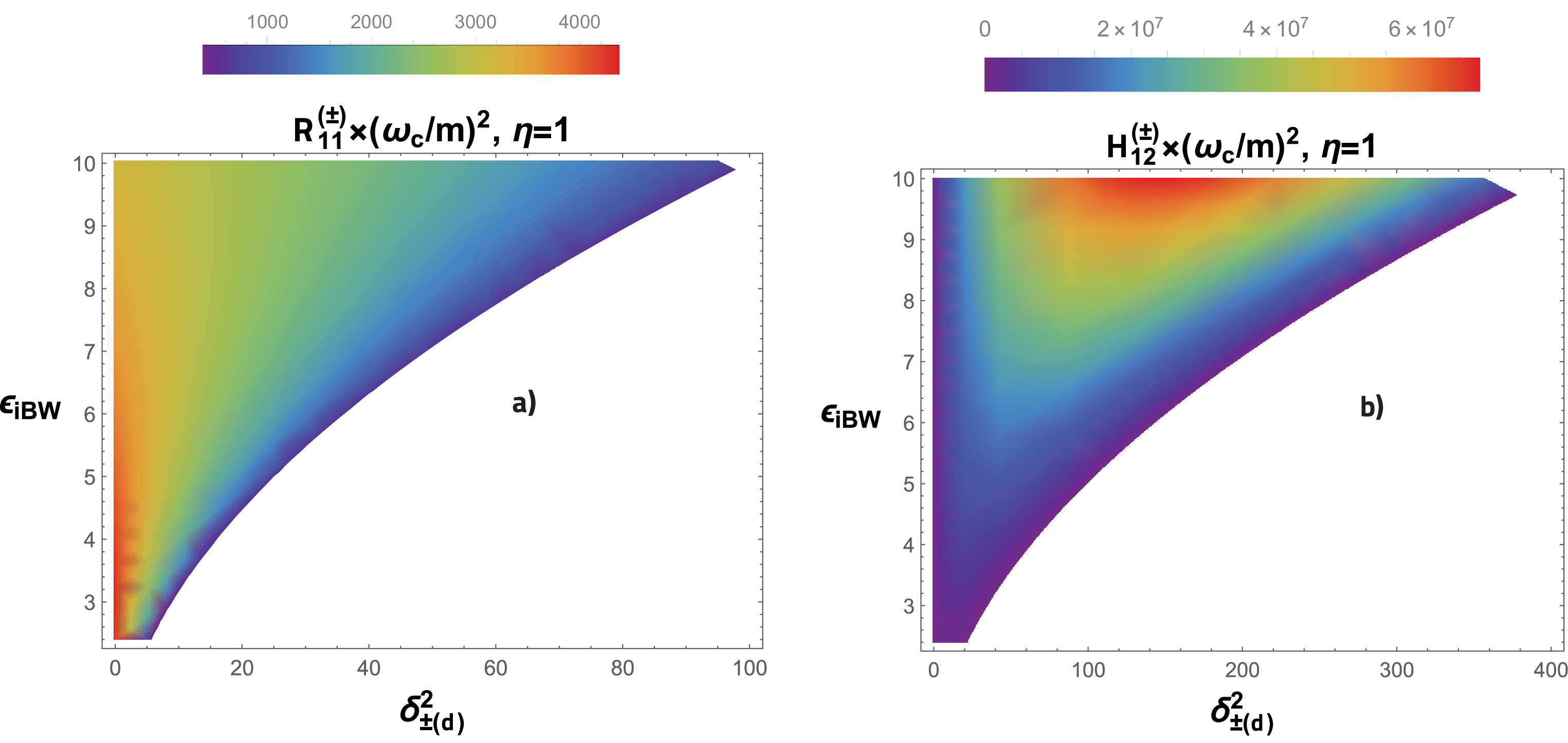}
	\caption{\label{fig 11} Dependence of the relative probabilities $R_{11}^{\left(\pm \right)} \times \left({\omega _{C} \mathord{\left/ {\vphantom {\omega _{C}  m}} \right. \kern-\nulldelimiterspace} m} \right)^{2} $ \eqref{ZEqnNum548453} and $H_{12}^{\left(\pm \right)} \times \left({\omega _{C} \mathord{\left/ {\vphantom {\omega _{C}  m}} \right. \kern-\nulldelimiterspace} m} \right)^{2} $ on the square of the electron-positron pair outgoing angle \eqref{ZEqnNum842068}  and the quantum parameter of the Breit-Wheeler process.}
\end{figure}

Figures 9 and 10 show graphs of the relative probabilities $R_{l_{1} l_{2} }^{\left(\pm \right)} $ \eqref{ZEqnNum548453}\eqref{ZEqnNum548453} and ${\rm H} _{l_{1} l_{2} }^{\left(\pm \right)} $ \eqref{ZEqnNum842068} (multiplied by a factor $\left({\omega _{C} \mathord{\left/ {\vphantom {\omega _{C}  m}} \right. \kern-\nulldelimiterspace} m} \right)^{2} $) as a function of the angle between momenta of electron and positron pairs on the lower branch of their energies for fixed initial setup parameters and different numbers of absorbed gamma-quanta of the wave. Tables 3 and 4 show the corresponding maximum probabilities for different numbers of absorbed gamma-quanta of the wave (at points $\delta _{\pm \left(d\right)}^{2*} $ corresponding to the maxima of the relative probability distributions in Fig. 9 and Fig.10), as well as the energy of the final electron and the electron-positron pair (in units of the energy of the initial electron). It can be seen from the figures and tables that for $l_{2} =1$ and $l_{1} =1,2,...$, the maximum of relative probabilities occurs at zero outgoing angle of the pair $\left(\delta _{\pm \left(d\right)}^{2*} =0\right)$. If $l_{2} =2,3,...$ (for any values $l_{1} $), then the maximum of resonant probabilities is shifted to the right, in the region of non-zero values of the ultrarelativistic parameter $\delta _{\pm \left(d\right)}^{2*} $. Figure 9 and Table 3 show that the maximum relative probability $R_{l_{1} l_{2} \left(\max \right)}^{\left(\pm \right)} $ holds for the minimum number of absorbed gamma-quanta of the wave $l_{1} =l_{2} =1$. As the number of absorbed gamma-quanta increases, the relative probability decreases. In addition, when the quantum parameter of the Breit-Wheeler process increases by about 4 times, the maximum relative probability $R_{l_{1} l_{2} \left(\max \right)}^{\left(\pm \right)} $ decreases slightly: ${\left(R_{11\left(\max \right)}^{\left(\pm \right)} \left|{}_{\varepsilon _{iBW=10} } \right. \right)\mathord{\left/ {\vphantom {\left(R_{11\left(\max \right)}^{\left(\pm \right)} \left|{}_{\varepsilon _{iBW=10} } \right. \right) \left(R_{11\left(\max \right)}^{\left(\pm \right)} \left|{}_{\varepsilon _{iBW=2.41} } \right. \right)}} \right. \kern-\nulldelimiterspace} \left(R_{11\left(\max \right)}^{\left(\pm \right)} \left|{}_{\varepsilon _{iBW=2.41} } \right. \right)} \approx 0.75$. Figure 10 and Table 4 show that the maximum relative probability $H_{l_{1} l_{2} \left(\max \right)}^{\left(\pm \right)} $ holds for the optimal number of absorbed gamma-quanta of the wave $l_{1} =1,{\rm \; }\; l_{2} =2$. At the same time, $H_{11\left(\max \right)}^{\left(\pm \right)} {\rm \mathop{<}\limits_\sim }H_{12\left(\max \right)}^{\left(\pm \right)} $. As the number of absorbed gamma-quanta increases, the relative probability decreases. In addition, when the quantum parameter of the Breit-Wheeler process increases by about 4 times, the maximum relative probability $H_{12\left(\max \right)}^{\left(f\right)} $ increases by 42 times: ${\left(H_{12\left(\max \right)}^{\left(\pm \right)} \left|{}_{\varepsilon _{iBW=10} } \right. \right)\mathord{\left/ {\vphantom {\left(H_{12\left(\max \right)}^{\left(\pm \right)} \left|{}_{\varepsilon _{iBW=10} } \right. \right) \left(H_{12\left(\max \right)}^{\left(\pm \right)} \left|{}_{\varepsilon _{iBW=2.41} } \right. \right)}} \right. \kern-\nulldelimiterspace} \left(H_{12\left(\max \right)}^{\left(\pm \right)} \left|{}_{\varepsilon _{iBW=2.41} } \right. \right)} \approx 42$. 

Figure 11 shows the dependence of the maximum relative probabilities $R_{11\left(\max \right)}^{\left(\pm \right)} $ \eqref{ZEqnNum548453} and $H_{12\left(\max \right)}^{\left(\pm \right)} $ \eqref{ZEqnNum842068} on the electron-positron pair outgoing angle and the quantum parameter of the Breit-Wheeler process at the optimal number absorbed gamma-quanta at the first and second vertices. Figure 11a shows that the maximum relative probability with simultaneous registration of the electron-positron pair outgoing angle and the energy of the final electron $R_{11\left(\max \right)}^{\left(\pm \right)} $ has a maximum value when the electron-positron pair outgoing angle is zero and the quantum parameter $\varepsilon _{iBW} \approx \varepsilon _{*} $. As the value of this quantum parameter increases, the relative probability $R_{11\left(\max \right)}^{\left(\pm \right)} $ decreases. On the other hand, Figure 11b shows that the maximum relative probability with simultaneous registration of the electron and electron-positron pair outgoing angles $H_{12\left(\max \right)}^{\left(\pm \right)} $ increases significantly with an increase in the quantum parameter $\varepsilon _{iBW} $. So, if the quantum parameter $\varepsilon _{iBW} $ increases four times, then the corresponding probability increases by two orders of magnitude. In this case, the outgoing angles of the electron-positron pair increase as $\delta _{\pm \left(d\right)}^{2*} \sim \varepsilon _{iBW}^{2} $. It is important to note that the outgoing angle of the electron-positron pair uniquely determines the outgoing angle of the final electron (see the relation \eqref{ZEqnNum669663}), as well as the energies of final particles (see the relations \eqref{ZEqnNum722916}, \eqref{ZEqnNum277929}), i.e. quantum entanglement of final particles occurs.  At the same time, as the quantum parameter of the Breit-Wheeler process increases $\varepsilon _{iBW} $, the energy of the final electron tends to the energy of the initial electrons \eqref{ZEqnNum173271}. 

The relations for relative probabilities \eqref{ZEqnNum593059}-\eqref{ZEqnNum842068} are significantly simplified under conditions when final particles fly out at minimal angles. Moreover, if $\delta _{f\left(d\right)}^{2} =0$, $\delta _{\pm \left(u\right)}^{2} =\delta _{\pm \left(\min \right)}^{2} $, then for the functions $R_{l_{1} l_{2} }^{\left(f\right)} \left(x_{f\left(d\right)}^{\min } ,\delta _{\pm \left(\min \right)}^{2} \right)$ \eqref{ZEqnNum593059} and ${\rm H} _{l_{1} l_{2} }^{\left(f\right)} \left(0,\delta _{\pm \left(\min \right)}^{2} \right)$ \eqref{ZEqnNum609209}, the main contribution to the probability of the process is made by the process with the absorption of one gamma-quantum of the wave in the first top (for the Compton effect \eqref{ZEqnNum302085}, \eqref{ZEqnNum806316} argument of the Bessel functions $z_{f} =0$ and $l_{1} =1$).  In this case, we get:
\begin{equation} \label{ZEqnNum591731} 
	R_{1l_{2} }^{\left(f\right)} \left(x_{f\left(d\right)}^{\min } ,\delta _{\pm \left(\min \right)}^{2} \right)=b'_{i} P_{l_{2} } \left(z_{\pm \left(\min \right)} \right), 
\end{equation} 
\begin{equation} \label{ZEqnNum668878} 
	{\rm H} _{1l_{2} }^{\left(f\right)} \left(0,\delta _{\pm \left(\min \right)}^{2} \right)=\varepsilon _{iC} \left(1+\varepsilon _{iC} \right)^{2} b_{i} {}^{{'} } P_{l_{2} } \left(z_{\pm \left(\min \right)} \right).  
\end{equation} 
Here it is indicated:
\begin{equation} \label{109)} 
	b'_{i} =\eta ^{2} \left[\left(1+\varepsilon _{iC} \right)+\frac{1}{\left(1+\varepsilon _{iC} \right)} \right]b_{i}  .
\end{equation} 

In the expressions \eqref{ZEqnNum591731}, \eqref{ZEqnNum668878}, the functions $P_{l_{2} } \left(z_{\pm \left(\min \right)} \right)$ have the form \eqref{ZEqnNum633873} with the argument of the Bessel functions equal to
\begin{equation} \label{ZEqnNum116595} 
	z_{\pm \left(\min \right)} \approx \frac{4\eta }{\varepsilon _{iC} \sqrt{1+\eta ^{2} } } \sqrt{\left(1+\varepsilon _{iC} \right)\left[l_{2} -\frac{4}{\varepsilon _{iC} } \left(1+\frac{1}{\varepsilon _{iC} } \right)\right]} . 
\end{equation} 
We emphasize that in the relations \eqref{ZEqnNum591731}, \eqref{ZEqnNum668878}, \eqref{ZEqnNum116595} the number of gamma-quanta absorbed at the second vertex $l_{2} $ is determined by the expression \eqref{ZEqnNum162642}. The relations \eqref{ZEqnNum591731}, \eqref{ZEqnNum668878} show that these relative probabilities have the same dependence on the classical parameter $\eta $, but different dependence on the quantum parameter Compton-the effect. Given this, we can obtain the ratio of these relative probabilities, which depends only on the quantum parameter $\varepsilon _{iC} $:
\begin{equation} \label{ZEqnNum624305} 
	\gamma ^{\left(f\right)} \left(\varepsilon _{iC} \right)=\frac{{\rm H} _{11}^{\left(f\right)} \left(0,\delta _{\pm \left(\min \right)}^{2} \right)}{R_{11}^{\left(f\right)} \left(x_{f\left(d\right)}^{\min } ,\delta _{\pm \left(\min \right)}^{2} \right)} =\varepsilon _{iC} \left(1+\varepsilon _{iC} \right)^{2}  .
\end{equation} 
At the same time, the relative probability of the RTPP process with simultaneous registration of the electron and pair outgoing angles at minimum angles exceeds the corresponding probability with simultaneous registration of the electron outgoing angle and the pair energy. Moreover, this excess will be significant $\gamma ^{\left(f\right)} \approx \varepsilon _{iC}^{3} >>1$ for large quantum parameters of the Compton effect. 

Now consider the case when $\delta _{\pm \left(d\right)}^{2} =0$, $\delta _{f\left(u\right)}^{2} =\delta _{f\left(\min \right)}^{2} $. In this case, for the functions $R_{l_{1} l_{2} }^{\left(\pm \right)} \left(x_{\pm \left(d\right)}^{\min } ,\delta _{f\left(\min \right)}^{2} \right)$ \eqref{ZEqnNum548453} and ${\rm H} _{l_{1} l_{2} }^{\left(\pm \right)} \left(0,\delta _{f\left(\min \right)}^{2} \right)$ \eqref{ZEqnNum842068}, the main contribution to the relative probability is made by the process with the absorption of one gamma-quantum of the wave at the second vertex, provided condition \eqref{ZEqnNum957720} is met (for an external field-stimulated Breit-Wheeler \eqref{ZEqnNum395097}, \eqref{ZEqnNum806316} argument of the Bessel functions $z_{2} =0$ and $l_{2} =1$).  In this case, we get:
\begin{eqnarray} \label{ZEqnNum451141} 
	R_{l_{1} 1}^{\left(\pm \right)} \left(x_{\pm \left(d\right)}^{\min } ,\delta _{f\left(\min \right)}^{2} \right)=2b_{i} \eta ^{2} \varepsilon _{iBW} \left(\varepsilon _{iBW} -1\right)\times \\ \nonumber
	\times K_{l_{1} } \left(u_{f\left(\min \right)} ,z_{f\left(\min \right)} \right), 
\end{eqnarray} 
\begin{eqnarray} \label{ZEqnNum352099} 
	{\rm H} _{l_{1} 1}^{\left(\pm \right)} \left(0,\delta _{f\left(\min \right)}^{2} \right)=32b_{i} \eta ^{2} \varepsilon _{iBW}^{4} \left(\varepsilon _{iBW} -1\right)\times \\ \nonumber
	\times K_{l_{1} } \left(u_{f\left(\min \right)} ,z_{f\left(\min \right)} \right). 
\end{eqnarray} 
Note that when obtaining the expression \eqref{ZEqnNum352099} at $\delta _{\pm \left(d\right)}^{2} \to 0$ energy of the electron-positron pair on the lower branch \eqref{ZEqnNum722916} was put into Taylor series up to second-order terms. In the expression \eqref{ZEqnNum352099}, the function $K_{l_{1} } \left(u_{f\left(\min \right)} ,z_{f\left(\min \right)} \right)$ is defined by the relations \eqref{ZEqnNum302085}, \eqref{ZEqnNum905734}, where $u_{f\left(\min \right)} $ and $z_{f\left(\min \right)} $ are defined as:
\begin{eqnarray} \label{ZEqnNum994392} 
	&&u_{f\left(\min \right)} =\frac{1}{\left(\varepsilon _{iBW} -1\right)} ,\\ \nonumber 
	&&z_{f\left(\min \right)} \approx \frac{\eta }{\sqrt{1+\eta ^{2} } } \frac{\sqrt{4l_{1} \varepsilon _{iBW} \left(\varepsilon _{iW} -1\right)-1} }{4\varepsilon _{iBW} \left(\varepsilon _{iBW} -1\right)} ,\\ \nonumber
	&&l_{1} \ge 1 \nonumber.
\end{eqnarray} 
Note that in the expressions \eqref{ZEqnNum451141}, \eqref{ZEqnNum352099}, \eqref{ZEqnNum994392} the value of the quantum parameter $\varepsilon _{iBW} $ satisfies the condition \eqref{ZEqnNum957720}. The relations \eqref{ZEqnNum451141} and \eqref{ZEqnNum352099} show that these relative probabilities have the same dependence on the classical parameter $\eta $, but different dependence on the quantum parameter of the Breit-Wheeler process. Given this, we can obtain the ratio of these relative probabilities, which depends only on the quantum parameter $\varepsilon _{iBW} $:
\begin{equation} \label{ZEqnNum375588} 
	\gamma ^{\left(\pm \right)} \left(\varepsilon _{iBW} \right)=\frac{{\rm H} _{l_{1} 1}^{\left(\pm \right)} \left(0,\delta _{f\left(\min \right)}^{2} \right)}{R_{l_{1} 1}^{\left(\pm \right)} \left(x_{\pm \left(d\right)}^{\min } ,\delta _{f\left(\min \right)}^{2} \right)} =16\varepsilon _{iBW}^{3}  .
\end{equation} 
At the same time, the relative probability of the RTPP process with simultaneous registration of the outgoing angles of the pair and the electron by the minimum angles exceeds the corresponding probability with simultaneous registration of the electron outgoing angle and the energy of the pair. Moreover, this excess will be significant for large quantum parameters of the Breit-Wheeler process. 

Expressions for resonant relative probabilities \eqref{ZEqnNum591731}, \eqref{ZEqnNum668878} and \eqref{ZEqnNum451141}, \eqref{ZEqnNum352099} are significantly simplified under the condition \eqref{ZEqnNum157133}, when the energy of the initial electrons significantly exceeds the characteristic Breit-Wheeler energy:  
\begin{equation} \label{ZEqnNum450323} 
	\varepsilon _{iBW} >>1\quad \left(\varepsilon _{iC} >>1\right) .
\end{equation} 
In this case, the resonant relative probabilities \eqref{ZEqnNum591731}, \eqref{ZEqnNum668878} will take the form:
\begin{equation} \label{ZEqnNum327588} 
	R_{1l_{2} }^{\left(f\right)} \left(x_{f\left(d\right)}^{\min } ,\delta _{\pm \left(\min \right)}^{2} \right)\approx \frac{g_{i} }{\varepsilon _{iC} } \left(\frac{m}{\omega _{C} } \right)^{2} P_{l_{2} } \left(z_{\pm \left(\min \right)} \right), 
\end{equation} 
\begin{equation} \label{ZEqnNum791332} 
	{\rm H} _{1l_{2} }^{\left(f\right)} \left(0,\delta _{\pm \left(\min \right)}^{2} \right)\approx g_{i} \varepsilon _{iC}^{2} \left(\frac{m}{\omega _{C} } \right)^{2} P_{l_{2} } \left(z_{\pm \left(\min \right)} \right) .
\end{equation} 
Here it is indicated:
\begin{equation} \label{119)} 
	g_{i} =\frac{2\left(4\pi ^{2} \right)^{3} \eta ^{2} \left(1+\eta ^{2} \right)}{P\left(\eta ,\varepsilon _{iBW} \right){\rm K} \left(\eta ,\varepsilon _{iC} \right)} \approx \frac{1.23\times 10^{5} \eta ^{2} \left(1+\eta ^{2} \right)}{P\left(\eta ,\varepsilon _{iBW} \right){\rm K} \left(\eta ,\varepsilon _{iC} \right)} . 
\end{equation} 
The argument of the functions $P_{l_{2} } \left(z_{\pm \left(\min \right)} \right)$ and the minimum angle between the electron and positron momenta of the pair take the form:
\begin{equation} \label{120)} 
	z_{\pm \left(\min \right)} \approx \frac{2\eta }{\sqrt{1+\eta ^{2} } } \sqrt{\frac{l_{2} }{\varepsilon _{iBW} } } ,\quad \delta _{\pm \left(\min \right)}^{2} \approx 4l_{2} \varepsilon _{iBW} . 
\end{equation} 
In this case, the energy of the initial electron is mainly converted into the energy of the electron-positron pair:
\begin{equation} \label{121)} 
	x_{f\left(d\right)}^{\min } \approx \frac{1}{\varepsilon _{iC} } <<1,\quad 2x_{\pm \left(u\right)}^{\max } \approx 1-\frac{1}{\varepsilon _{iC} } \approx 1. 
\end{equation} 
It is important to note that if the Breit-Wheeler quantum parameter satisfies a more stringent condition
\begin{equation} \label{ZEqnNum970786} 
	\sqrt{\varepsilon _{iBW} } >>1,  
\end{equation} 
then the argument of the functions $P_{l_{2} } \left(z_{\pm \left(\min \right)} \right)$ in the relations \eqref{ZEqnNum327588} and \eqref{ZEqnNum791332} becomes small $\left(z_{\pm \left(\min \right)} <<1\right)$. Because of this, the process with the absorption of one gamma-quantum of the wave at the second vertex becomes most likely. Given this, the relative resonant probabilities \eqref{ZEqnNum327588} and \eqref{ZEqnNum791332} within the \eqref{ZEqnNum970786} condition  takes the following form:
\begin{equation} \label{ZEqnNum973574} 
	R_{11}^{\left(f\right)} \left(x_{f\left(d\right)}^{\min } ,\delta _{\pm \left(\min \right)}^{2} \right)\approx \frac{1}{2} g_{i} \frac{\eta ^{2} }{\varepsilon _{iC} } \left(\frac{m}{\omega _{C} } \right)^{2} , 
\end{equation} 
\begin{eqnarray} \label{ZEqnNum366346} 
	{\rm H} _{11}^{\left(f\right)} \left(0,\delta _{\pm \left(\min \right)}^{2} \right)\approx \frac{1}{2} g_{i} \eta ^{2} \varepsilon _{iC}^{2} \left(\frac{m}{\omega _{C} } \right)^{2} \\ \nonumber
	\left(\delta _{\pm \left(\min \right)}^{2} \approx \varepsilon _{iC} >>1\right). \nonumber
\end{eqnarray} 
This shows that the ratio of these probabilities in accordance with the expression \eqref{ZEqnNum624305} has the form:
\begin{equation} \label{125)} 
	\gamma ^{\left(f\right)} \approx \varepsilon _{iC}^{3} >>1 .
\end{equation} 

Now let's consider the resonant relative probabilities $R_{l_{1} 1}^{\left(\pm \right)} $ \eqref{ZEqnNum451141} and $H_{l_{1} 1}^{\left(\pm \right)} $ \eqref{ZEqnNum352099} under the condition on the Breit-Wheeler quantum parameter \eqref{ZEqnNum450323}.  In this case, the main contribution to the function $K_{l_{1} } \left(u_{f\left(\min \right)} ,z_{f\left(\min \right)} \right)\approx 2\eta ^{2} $ is made by the term $l_{1} =1$ ($u_{f\left(\min \right)} \approx \varepsilon _{iBW}^{-1} <<1$, $z_{f\left(\min \right)} \sim \varepsilon _{iBW}^{-1} <<1$). Therefore, the required relative resonant probabilities take the form:
\begin{equation} \label{ZEqnNum330923} 
	R_{11}^{\left(\pm \right)} \left(x_{\pm \left(d\right)}^{\min } ,\delta _{f\left(\min \right)}^{2} \right)=\frac{1}{4} g_{i} \eta ^{2} \left(\frac{m}{\omega _{C} } \right)^{2} , 
\end{equation} 
\begin{equation} \label{ZEqnNum754352} 
	{\rm H} _{11}^{\left(\pm \right)} \left(0,\delta _{f\left(\min \right)}^{2} \right)=4g_{i} \eta ^{2} \varepsilon _{iBW}^{3} \left(\frac{m}{\omega _{C} } \right)^{2} . 
\end{equation} 
In this case the outgoing angles of final particles are given by the expressions:
\begin{equation} \label{128)} 
	\delta _{\pm \left(d\right)}^{2} =0,\quad \delta _{f\left(\min \right)}^{2} \approx 4, 
\end{equation} 
and the energy of the initial electron, basically, passes into the energy of the final electron
\begin{equation} \label{129)} 
	\quad 2x_{\pm \left(d\right)}^{\min } =\frac{1}{\varepsilon _{iBW} } <<1,\quad x_{f\left(u\right)}^{\max } \approx 1-\frac{1}{\varepsilon _{iBW} } \approx 1. 
\end{equation} 
Note that the probability ratio of \eqref{ZEqnNum451141} to \eqref{ZEqnNum330923} is \eqref{ZEqnNum375588}. 

We obtain the ratio of the relative probabilities ${\rm H} _{11}^{\left(\pm \right)} $ \eqref{ZEqnNum754352} and ${\rm H} _{11}^{\left(f\right)} $ \eqref{ZEqnNum366346} under conditions of large values of the Breit-Wheeler quantum parameters \eqref{ZEqnNum970786}:
\begin{equation} \label{ZEqnNum389568} 
	\beta =\frac{{\rm H} _{11}^{\left(\pm \right)} \left(0,\delta _{f\left(\min \right)}^{2} \right)}{{\rm H} _{11}^{\left(f\right)} \left(0,\delta _{\pm \left(\min \right)}^{2} \right)} =\frac{1}{2} \varepsilon _{iBW} >>1 .
\end{equation} 
Hence, it can be seen that for large values of the Breit-Wheeler quantum parameters, the probability of the RTPP process with simultaneous registration of the outgoing angles of final particles at minimum angles $\left(\delta _{\pm }^{2} =0,{\rm \; }\delta _{f\left(\min \right)}^{2} \right)$ and the transition of the main energy of initial electrons to the energy of final electrons $\left(E_{f\left(\max \right)} =E_{i} -\omega _{BW} ,E_{\pm \left(\min \right)} ={\omega _{BW} \mathord{\left/ {\vphantom {\omega _{BW}  2}} \right. \kern-\nulldelimiterspace} 2} \; \right)$ is several orders of magnitude higher than the corresponding probability with simultaneous registration of the outgoing angles of final particles at minimum angles $\left(\delta _{f}^{2} =0,\; \delta _{\pm \left(\min \right)}^{2} \right)$ and the transition of the main energy of initial electrons to energy of electron-positron pair $\left(2E_{\pm \left(\max \right)} =E_{i} -E_{f\left(\min \right)} ,\; E_{f\left(\min \right)} \approx {E_{i} \mathord{\left/ {\vphantom {E_{i}  \varepsilon _{iC} }} \right. \kern-\nulldelimiterspace} \varepsilon _{iC} } \right)$.

\begin{figure}[h]
	\includegraphics[width=0.45\textwidth]{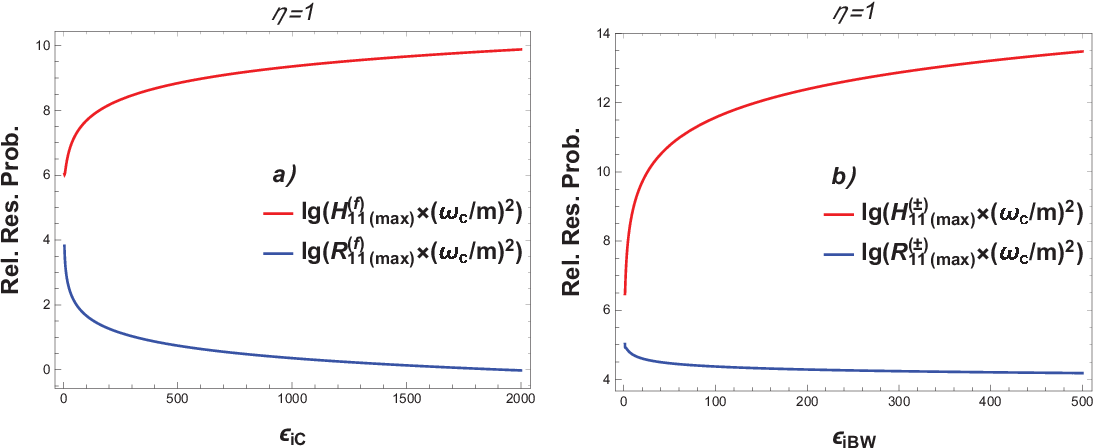}
	\caption{\label{fig 12} Dependence of the maximum relative probabilities $R_{11\left(\max \right)}^{\left(f\right)} \times \left({\omega _{C} \mathord{\left/ {\vphantom {\omega _{C}  m}} \right. \kern-\nulldelimiterspace} m} \right)^{2} $ \eqref{ZEqnNum591731}\eqref{ZEqnNum591731} and $H_{11\left(\max \right)}^{\left(f\right)} \times \left({\omega _{C} \mathord{\left/ {\vphantom {\omega _{C}  m}} \right. \kern-\nulldelimiterspace} m} \right)^{2} $ \eqref{ZEqnNum668878} (case a) and $R_{11\left(\max \right)}^{\left(\pm \right)} \times \left({\omega _{C} \mathord{\left/ {\vphantom {\omega _{C}  m}} \right. \kern-\nulldelimiterspace} m} \right)^{2} $ \eqref{ZEqnNum451141} and $H_{11\left(\max \right)}^{\left(\pm \right)} \times \left({\omega _{C} \mathord{\left/ {\vphantom {\omega _{C}  m}} \right. \kern-\nulldelimiterspace} m} \right)^{2} $ \eqref{ZEqnNum352099} (case b) on the quantum parameters $\varepsilon _{iC} $ and $\varepsilon _{iBW} $.}
\end{figure}

Figure 12a shows graphs of the maximum relative probabilities $R_{11\left(\max \right)}^{\left(f\right)} \times \left({\omega _{C} \mathord{\left/ {\vphantom {\omega _{C}  m}} \right. \kern-\nulldelimiterspace} m} \right)^{2} $ \eqref{ZEqnNum591731} and $H_{11\left(\max \right)}^{\left(f\right)} \times \left({\omega _{C} \mathord{\left/ {\vphantom {\omega _{C}  m}} \right. \kern-\nulldelimiterspace} m} \right)^{2} $ \eqref{ZEqnNum668878} on the quantum parameter of the Compton effect $\varepsilon _{iC} \ge 2\varepsilon _{*} $ (see the relation \eqref{ZEqnNum957720}) for minimum outgoing angles of the final electron $\left(\delta _{f\left(d\right)}^{2} =0\right)$ and electron-positron pair $\left(\delta _{\pm \left(\min \right)}^{2} \right)$. It can be seen from this figure that the relative probability $R_{11\left(\max \right)}^{\left(f\right)} $(with simultaneous registration of the electron outgoing angle and pair energy) decreases quite rapidly with an increase in the quantum parameter of the Compton effect (see the relations \eqref{ZEqnNum591731}, \eqref{ZEqnNum327588}, \eqref{ZEqnNum973574}). On the other hand, the relative probability $H_{11\left(\max \right)}^{\left(f\right)} $ (with simultaneous registration of the electron and electron-positron pair outgoing angles) increases rapidly with the growth of the quantum parameter of the Compton effect , and then proceeds to a smooth increase (see the relations \eqref{ZEqnNum668878}, \eqref{ZEqnNum791332}, \eqref{ZEqnNum366346}). Note that in this case, the energy of the initial electrons is mainly converted into the energy of the electron-positron pair $\left(E_{+} =E_{-} \approx {E_{i} \mathord{\left/ {\vphantom {E_{i}  2}} \right. \kern-\nulldelimiterspace} 2} \right)$. 

Figure 12b shows graphs of the maximum relative probabilities $R_{11\left(\max \right)}^{\left(\pm \right)} \times \left({\omega _{C} \mathord{\left/ {\vphantom {\omega _{C}  m}} \right. \kern-\nulldelimiterspace} m} \right)^{2} $ \eqref{ZEqnNum451141} and $H_{11\left(\max \right)}^{\left(\pm \right)} \times \left({\omega _{C} \mathord{\left/ {\vphantom {\omega _{C}  m}} \right. \kern-\nulldelimiterspace} m} \right)^{2} $ \eqref{ZEqnNum352099} on the quantum parameter of the Breit-Wheeler $\varepsilon _{iBW} \ge {\varepsilon _{*} \mathord{\left/ {\vphantom {\varepsilon _{*}  2}} \right. \kern-\nulldelimiterspace} 2} $ process (see the relation \eqref{ZEqnNum957720})  at the minimum outgoing angles of the electron-positron pair $\left(\delta _{\pm \left(d\right)}^{2} =0\right)$ of the final electron $\left(\delta _{f\left(\min \right)}^{2} \right)$. This figure shows that the relative probability $R_{11\left(\max \right)}^{\left(\pm \right)} $(with simultaneous registration of the pair outgoing angle and electron energy) gradually decreases with increasing quantum parameter of the Breit-Wheeler process (see the relations \eqref{ZEqnNum451141}, \eqref{ZEqnNum330923}). On the other hand, the relative probability $H_{11\left(\max \right)}^{\left(\pm \right)} $(with simultaneous registration of the outgoing angles of the electron-positron pair and electron) increases quite rapidly with the growth of the quantum parameter of the Breit-Wheeler process (see the relations \eqref{ZEqnNum352099}, \eqref{ZEqnNum754352}). Note that in this case, the energy of the initial electrons is mainly converted into the energy of the final electron $\left(E_{f} \approx E_{i} \right)$. 

\begin{figure}[h]
	\includegraphics[width=0.45\textwidth]{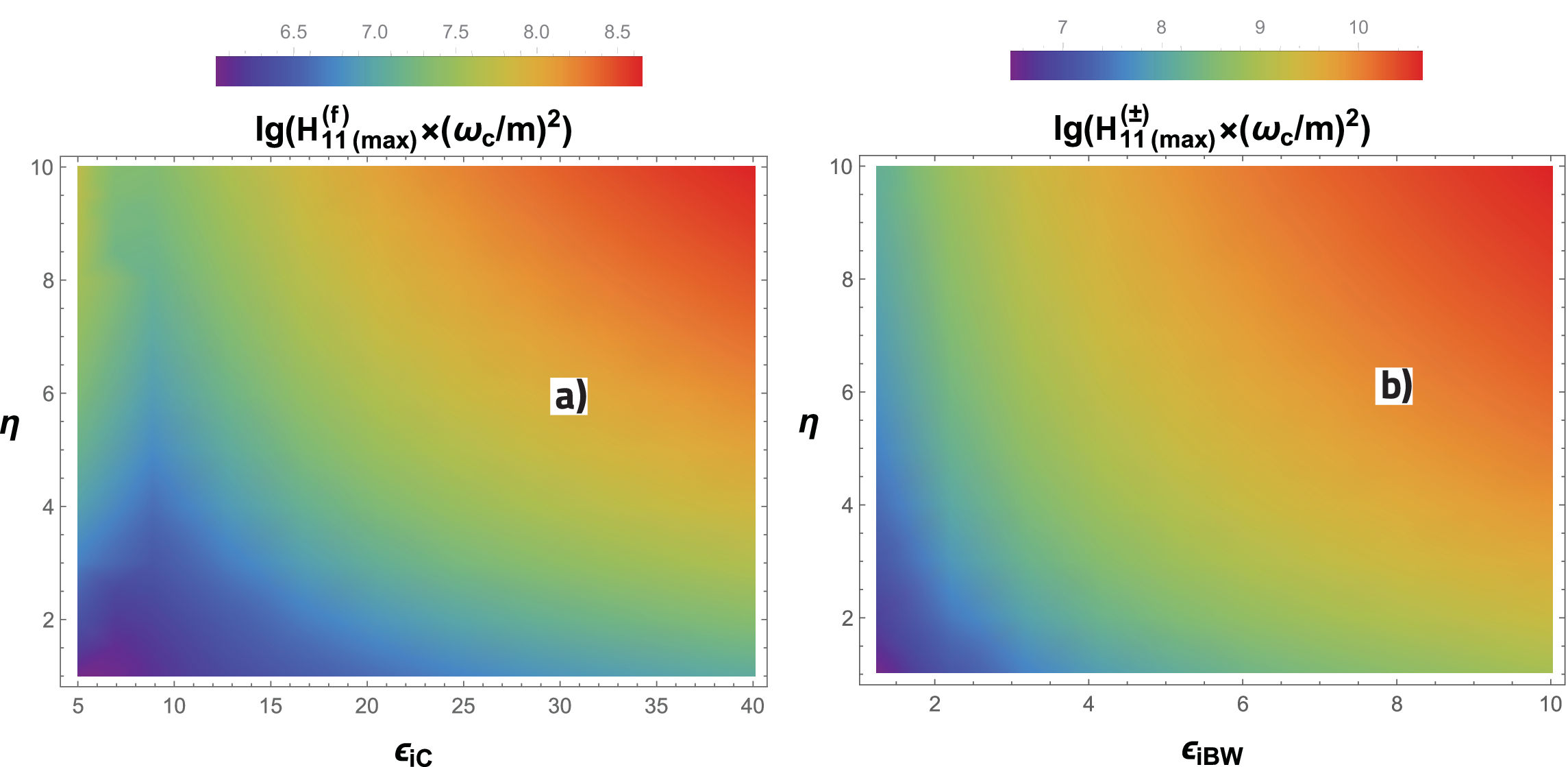}
	\caption{\label{fig 13} Dependence of the maximum relative probabilities $H_{11\left(\max \right)}^{\left(f\right)} \times \left({\omega _{C} \mathord{\left/ {\vphantom {\omega _{C}  m}} \right. \kern-\nulldelimiterspace} m} \right)^{2} $ \eqref{ZEqnNum668878} (case a) and $H_{11\left(\max \right)}^{\left(\pm \right)} \times \left({\omega _{C} \mathord{\left/ {\vphantom {\omega _{C}  m}} \right. \kern-\nulldelimiterspace} m} \right)^{2} $ \eqref{ZEqnNum352099} (case b) on the classical parameter $\eta $ \eqref{ZEqnNum778375} and quantum parameters $\varepsilon _{iC} $ and $\varepsilon _{iBW} $.}
\end{figure} 

\noindent Figure 13 shows dependence of the maximum relative probabilities $H_{11\left(\max \right)}^{\left(f\right)} \times \left({\omega _{C} \mathord{\left/ {\vphantom {\omega _{C}  m}} \right. \kern-\nulldelimiterspace} m} \right)^{2} $ \eqref{ZEqnNum668878} (case a) and $H_{11\left(\max \right)}^{\left(\pm \right)} \times \left({\omega _{C} \mathord{\left/ {\vphantom {\omega _{C}  m}} \right. \kern-\nulldelimiterspace} m} \right)^{2} $ \eqref{ZEqnNum352099} (case b) on the classical parameter $\eta $ \eqref{ZEqnNum778375} and quantum parameters of the Compton effect and Breit-Wheeler process at minimum outgoing angles of final particles. It can be seen from these graphs that as the parameters grow $\eta $, the $\varepsilon _{iC} \left(\varepsilon _{iBW} \right)$ relative resonant probabilities also increase significantly, and the function increases most strongly $H_{11\left(\max \right)}^{\left(\pm \right)} $. Indeed, we take the ratio of these functions at two points for $\eta =10,\; \varepsilon _{iC} =40\; \left(\varepsilon _{iBW} =10\right)$ and $\eta =1,\; \varepsilon _{iC} =4.84 \left(\varepsilon _{iBW} =1.21\right)$. Then from graphs 13 we get:
\begin{widetext}
	\begin{equation} \label{131)} 
		{H_{11\left(\max \right)}^{\left(f\right)} \left|{}_{\begin{array}{l} {\eta =10} \\ {\varepsilon _{iC} =40} \end{array}} \right. \mathord{\left/ {\vphantom {H_{11\left(\max \right)}^{\left(f\right)} \left|{}_{\begin{array}{l} {\eta =10} \\ {\varepsilon _{iC} =40} \end{array}} \right.  H_{11\left(\max \right)}^{\left(f\right)} \left|{}_{\begin{array}{l} {\eta =1} \\ {\varepsilon _{iC} =4.84} \end{array}} \right. }} \right. \kern-\nulldelimiterspace} H_{11\left(\max \right)}^{\left(f\right)} \left|{}_{\begin{array}{l} {\eta =1} \\ {\varepsilon _{iC} =4.84} \end{array}} \right. }\approx  2.07\times 10^{4}  ,
	\end{equation} 
	\begin{equation} \label{132)} 
		{H_{11\left(\max \right)}^{\left(\pm \right)} \left|{}_{\begin{array}{l} {\eta =10} \\ {\varepsilon _{iBW} =10} \end{array}} \right. \mathord{\left/ {\vphantom {H_{11\left(\max \right)}^{\left(\pm \right)} \left|{}_{\begin{array}{l} {\eta =10} \\ {\varepsilon _{iBW} =10} \end{array}} \right.  H_{11\left(\max \right)}^{\left(\pm \right)} \left|{}_{\begin{array}{l} {\eta =1} \\ {\varepsilon _{iBW} =1.21} \end{array}} \right. }} \right. \kern-\nulldelimiterspace} H_{11\left(\max \right)}^{\left(\pm \right)} \left|{}_{\begin{array}{l} {\eta =1} \\ {\varepsilon _{iBW} =1.21} \end{array}} \right. } \approx 0.76\times 10^{6}.
	\end{equation} 
\end{widetext}
\section{Discussion}

In the previous section, it was shown that the excess of the resonant probability of the studied process over the total probability of the stimulated by an external field Compton effect significantly depends on the quantum parameter $\varepsilon _{iBW} \left(\varepsilon _{iC} \right)$, the number of absorbed gamma-quanta of the wave at the first and second vertices, and the characteristic energy of the Compton effect (multiplier $\left({m\mathord{\left/ {\vphantom {m \omega _{C} }} \right. \kern-\nulldelimiterspace} \omega _{C} } \right)^{2} $). Let us determine the characteristic energy of the Compton effect \eqref{ZEqnNum291894} for various frequencies and intensities in case of head-on collision with external wave ($\theta _{i} =\pi $, see \eqref{ZEqnNum348638}). For the characteristic energy of the Compton effect to decrease as the frequency and intensity of a wave increase, we assume that the classical parameter $\eta $ is constant .
\begin{widetext}
	\begin{equation} \label{ZEqnNum810217} 
		\omega _{C} \approx \left\{\begin{array}{c} {43.5{\rm \; GeV,\; if\; }\omega =3{\rm \; eV,\; }I=1.675\cdot 10^{19} {\rm \; Wcm}^{-2} \left(\eta =1\right)} \\ {{\rm 130.56\; MeV,\; if\; \; }\omega {\rm \; =\; }1{\rm \; keV,\; }I=1.86\cdot {\rm }10^{24} {\rm \; Wcm}^{-2} \left(\eta =1\right)} \\ {{\rm 6.528\; MeV,\; if\; \; }\omega {\rm \; =\; 20\; keV,\; }I=7.44\cdot {\rm }10^{26} {\rm \; Wcm}^{-2} \left(\eta =1\right)} \end{array}\right.  
	\end{equation} 
\end{widetext}
Note that $\omega _{BW} =4\omega _{C} $ (see expression \eqref{ZEqnNum291894}). From the relation \eqref{ZEqnNum810217}, it follows that in the optical range of frequencies , the characteristic energy of the Compton effect is quite high (tens of GeV). Moving into the region of X-ray waves and high field strengths, the value $\omega _{C} $ decreases quite quickly. 

Given the expression \eqref{ZEqnNum810217}, as well as Figure 12, we can obtain the data presented in Tables 5 and 6 for the case when the characteristic energy of the Compton effect is fixed, and the energy of the initial electrons is chosen from the condition $E_{i} \ge 2\varepsilon _{*} \omega _{C} \approx 4.83\omega _{C} $ (see the relation \eqref{ZEqnNum957720}). At the same time, Table 5 (see Figure 12a) presents the results for the relative probability $H_{11\left(\max \right)}^{\left(f\right)} $ (at the minimum outgoing angles of the final electron $\left(\delta _{f\left(d\right)}^{2} =0\right)$ and the electron-positron pair $\left(\delta _{\pm \left(\min \right)}^{2} \right)$). And Table 6 (see Figure 12b) shows the results for the relative probability $H_{11\left(\max \right)}^{\left(\pm \right)} $ (at the minimum outgoing angles of the electron-positron pair $\left(\delta _{\pm \left(d\right)}^{2} =0\right)$ and the final electron $\left(\delta _{f\left(\min \right)}^{2} \right)$). 

In the region of optical frequencies $\left(\omega =3{\rm \; eV}\right)$ and wave intensities $\left(I=1.675\cdot 10^{19} {\rm \; Wcm}^{-2} \right)$, the characteristic energy of the Compton effect is quite large $\left(\omega _{C} \approx 43.5{\rm \; GeV}\right)$. Because of this, in a sufficiently wide range of very high energies of the initial electrons $\left(E_{i} \le 22{\rm \; TeV}\right)$, the relative probability $H_{11\left(\max \right)}^{\left(f\right)} $ is small $\left(H_{11\left(\max \right)}^{\left(f\right)} \le 0.1\right)$. And only for very large energies of the initial electrons $E_{i} =87\; {\rm TeV}$ (which is evenly distributed between the electron and positron of the pair), the probability of the RTPP process becomes of the same order with the total probability of the Compton effect stimulated by an external field  $\left(H_{11\left(\max \right)}^{\left(f\right)} \approx 1\right)$. On the other hand, in the region of optical wave frequencies, the relative probability $H_{11\left(\max \right)}^{\left(\pm \right)} $ becomes of the same order as the total probability of the Compton effect stimulated by an external field for the initial electron energies $E_{i} {\rm \mathop{>}\limits_\sim }0.25\; {\rm TeV}$. As the initial electron energy increases, this probability increases rapidly and $E_{i} =87\; {\rm TeV}$ exceeds the total probability of the Compton effect stimulated by an external field by three orders of magnitude $\left(H_{11\left(\max \right)}^{\left(\pm \right)} \approx 4.2\times 10^{3} \right)$. Thus, in the region of optical frequencies and wave intensities $I{\rm \mathop{<}\limits_\sim }10^{19} {\rm \; Wcm}^{-2} $, $E_{i} {\rm \mathop{>}\limits_\sim }1\; {\rm TeV}$ the RTPP process with the generation of the electron-positron pair at zero angle and the scattering of an electron at an angle $\delta _{f\left(\min \right)}^{2} $ will be dominant. In this case, the energy of the initial electrons will mainly convert into the energy of the final electron, and the remaining energy will be evenly distributed between the electron and the positron of the pair. 

In the region of X-ray frequencies $\left(\omega =1{\rm \; keV}\right)$ and sufficiently high wave intensities $\left(I=1.86\cdot {\rm }10^{24} {\rm \; Wcm}^{-2} \right)$, the characteristic energy of the Compton effect becomes significantly lower $\left(\omega _{C} \approx {\rm 130.56\; MeV}\right)$. Therefore, for the initial electron energies in the interval $0.63{\rm \; GeV}\le E_{i} \le 261{\rm \; GeV}$, the probability of the RTPP process significantly (by several orders of magnitude) exceeds the total probability of the Compton effect stimulated by an external field. At the same time, Tables 5 and 6 show that  $16.6\le H_{11\left(\max \right)}^{\left(f\right)} \le 1.2\times 10^{5} $ and $45.49\le H_{11\left(\max \right)}^{\left(\pm \right)} \le 4.65\times 10^{8} $. Thus, for the energy of the initial electrons $E_{i} =0.63{\rm \; GeV}$, the relative probabilities $H_{11\left(\max \right)}^{\left(f\right)} $ and $H_{11\left(\max \right)}^{\left(\pm \right)} $ differ by several times ${H_{11\left(\max \right)}^{\left(\pm \right)} \mathord{\left/ {\vphantom {H_{11\left(\max \right)}^{\left(\pm \right)}  H_{11\left(\max \right)}^{\left(f\right)} \approx 2.7}} \right. \kern-\nulldelimiterspace} H_{11\left(\max \right)}^{\left(f\right)} \approx 2.7} $. However, as the energy of the initial electrons increases, this difference quickly increases to three orders of magnitude. 

For higher X-ray frequencies $\left(\omega =20{\rm \; keV}\right)$ and wave intensities $\left(I=7.44\cdot {\rm }10^{26} {\rm \; Wcm}^{-2} \right)$, the characteristic energy of the Compton effect becomes small $\left(\omega _{C} \approx {\rm 6.528\; MeV}\right)$. Therefore, in this case, even for small initial electron energies $31{\rm \; MeV}\le E_{i} \le 13{\rm \; GeV}$, the probability of the RTPP process is much higher than the total probability of the Compton effect stimulated by an external field. So, Tables 5 and 6 show that $6.6\times 10^{3} \le H_{11\left(\max \right)}^{\left(f\right)} \le 4.7\times 10^{7} $ and $1.82\times 10^{4} \le H_{11\left(\max \right)}^{\left(\pm \right)} \le 1.86\times 10^{11} $. 

Thus, the RTPP process with the generation of the electron-positron pair with a zero angle and the scattering of an electron by an angle $\delta _{f\left(\min \right)}^{2} $ will be dominant. At the same time, under conditions when the Breit-Wheeler quantum parameter is much larger than unity, the energy of the initial electrons will mainly pass into the energy of the final electron, and the remaining part of the ultrarelativistic energy will be evenly distributed between the electron and the positron of the pair. 

\begin{table*}%The best place to locate the table environment is directly after its first reference in text
	\caption{\label{table5}%
		The maximum values of the resonant relative probability $H_{11\left(\max \right)}^{\left(f\right)} $ \eqref{ZEqnNum668878} and the corresponding values of the energies and outgoing angles of final particles for different characteristic energies of the Compton effect and energies of initial electrons.
	}
	\begin{ruledtabular}
		\begin{tabular}{cccccc}
			$\omega _{C} ,{\rm \; MeV}$ & $E_{i} ,{\rm \; GeV}$ & $\delta _{\pm \left(\min \right)}^{2} $ & $E_{\pm } ,{\rm \; GeV}$ & $E_{f} ,{\rm \; GeV}$ & $H_{11\left(\max \right)}^{\left(f\right)} $  \\ \hline
			${\rm 130.56}$& $2.175\times 10^{4} $ & $496.98$ & $1.085\times 10^{4} $ & $43.41$ & $9.55\times 10^{-2} $ \\ 
			& $8.70\times 10^{4} $ & $1997.00$ & $4.348\times 10^{4} $ & $43.48$ & $1.050$ \\
			& $0.63$ & $0.0022$ & $0.261$ & $0.108$ & $16.571$ \\  
			& $65.28$ & $496.98$ & $32.58$ & $0.1303$ & $1.061\times 10^{4} $ \\  \hline
			${\rm 6.528}$ & $261.12$ & $1997.00$ & $130.50$ & $0.1305$ & $1.165\times 10^{5} $ \\  
			& $3.153\times 10^{-2} $ & $0.0022$ & $1.306\times 10^{-2} $ & $5.408\times 10^{-3} $ & $6.630\times 10^{3} $ \\  
			& $3.264$ & $496.98$ & $1.629$ & $6.515\times 10^{-3} $ & $4.245\times 10^{6} $ \\ 
			& $13.06$ & $1997.00$ & $6.525$ & $6.525\times 10^{-3} $ & $4.660\times 10^{7} $ \\  
		\end{tabular}
	\end{ruledtabular}
\end{table*} 
\begin{table*}%The best place to locate the table environment is directly after its first reference in text
	\caption{\label{table6}%
		The maximum values of the resonant relative probability $H_{11\left(\max \right)}^{\left(\pm \right)} $ \eqref{ZEqnNum352099} and the corresponding values of the energies and outgoing angles of final particles for different characteristic energies of the Compton effect and energies of initial electrons.
	}
	\begin{ruledtabular}
		\begin{tabular}{cccccc}
			$\omega _{C} ,{\rm \; MeV}$ & $E_{i} ,{\rm \; GeV}$ & $\delta _{f\left(\min \right)}^{2} $ & $E_{f} ,{\rm \; GeV}$ & $E_{\pm } ,{\rm \; GeV}$ & $H_{11\left(\max \right)}^{\left(\pm \right)} $ \\ \hline 
			$43.5\times 10^{3} $ & $2.101\times 10^{2} $ & $0.0517$ & $36.11$ & $87.0$ & $4.098\times 10^{-4} $ \\ 
			& $2.480\times 10^{2} $ & $7.8754$ & $73.95$ & $87.0$ & $1.001$ \\  
			& $2.175\times 10^{4} $ & $4.032$ & $2.158\times 10^{4} $ & $87.0$ & $65.012$ \\  
			& $8.70\times 10^{4} $ & $4.008$ & $8.68\times 10^{4} $ & $87.0$ & $4.191\times 10^{3} $ \\ \hline 
			${\rm 130.56}$ & $0.63$ & $0.0517$ & $0.108$ & $0.261$ & $45.490$ \\ 
			& $65.28$ & $4.032$ & $64.76$ & $0.261$ & $1.055\times 10^{7} $ \\ 
			& $261.12$ & $4.008$ & $260.60$ & $0.261$ & $4.653\times 10^{8} $ \\ \hline 
			${\rm 6.528}$ & $3.153\times 10^{-2} $ & $0.0517$ & $5.418\times 10^{-3} $ & $1.306\times 10^{-2} $ & $1.820\times 10^{4} $ \\  
			& $3.264$ & $4.032$ & $3.238$ & $1.306\times 10^{-2} $ & $4.219\times 10^{9} $ \\ 
			& $13.06$ & $4.008$ & $13.03$ & $1.306\times 10^{-2} $ & $1.861\times 10^{11} $ \\  
		\end{tabular}
	\end{ruledtabular}
\end{table*}
Figure 14 shows dependence of the relative probabilities $H_{1l_{2\min } \left(\max \right)}^{\left(f\right)} $ \eqref{ZEqnNum668878} (case a) and $H_{11\left(\max \right)}^{\left(\pm \right)} $ \eqref{ZEqnNum352099} (case b) on the characteristic energy of the Compton effect $\omega _{C} $ \eqref{ZEqnNum291894} for three values of the initial electron energy. Note that for a constant value of the parameter $\eta =1$, the characteristic energy of the Compton effect increases $\omega _{C} $ with a symmetric decrease in the frequency and intensity of the electromagnetic wave. It was assumed that $\omega _{C} >>m$ (see the inequality \eqref{ZEqnNum979913} for the quantum parameter $\varepsilon _{iC} $). The interval of change in the characteristic energy of the Compton effect was chosen from value $\omega _{C} {\rm =5\; MeV}$ $\left(\omega \approx 26{\rm \; keV,}\; I\approx 1.3\times 10^{27} {\rm \; Wcm}^{-2} \right)$ to value $\omega _{C} {\rm =7.4\; GeV}$ $\left(\omega \approx 18{\rm \; eV,}\; I\approx 5.85\times 10^{20} {\rm \; Wcm}^{-2} \right)$. It can be seen from these figures that for the case when $l_{1} =l_{2} =1$ these dependences are approximately linear on a logarithmic scale. Moreover, for the minimum values of the characteristic energy of the Compton effect (maximum frequencies and intensities of the electromagnetic wave), the relative probabilities $H_{11\left(\max \right)}^{\left(f\right)} $ and $H_{11\left(\max \right)}^{\left(\pm \right)} $ take maximum values. With an increase in the characteristic energy of the Compton effect (a decrease in the frequency and intensity of the wave), the corresponding relative probabilities decrease approximately linearly. Except for the energy case $E_{i} =50\; {\rm MeV}$(see Figure 14a), where there are 5 characteristic regions of relative probability variation $H_{1l_{2\min } \left(\max \right)}^{\left(f\right)} $ that correspond to the following values $l_{2\min } =1,2,3,4,5$. At the same time, with an increase in the number of absorbed gamma-quanta of the wave, the relative probability $H_{1l_{2\min } \left(\max \right)}^{\left(f\right)} $ decreases sharply. For energies $E_{i} ={\rm 5}\; {\rm GeV,}\; {\rm 500}\; {\rm GeV}$, the number of gamma-quanta in the wave $l_{2\min } =1$. Figure 14 shows that for each energy of initial electrons, there is a region of change in the characteristic energy of the Compton effect, in which the corresponding probabilities of the RTPP process exceed the total probability of the Compton effect stimulated by an external field.  Moreover, with an increase in the energy of the initial electrons, this region of change $\omega _{C} $, as well $H_{1l_{2\min } \left(\max \right)}^{\left(f\right)} $ as the relative $H_{11\left(\max \right)}^{\left(\pm \right)} $ probabilities, increases. So, for $\omega _{C} {\rm =5\; MeV}$ and the energies of the initial electrons $E_{i} =50\; {\rm MeV,}\; {\rm 5}\; {\rm GeV,}\; {\rm 500}\; {\rm GeV}$, the relative probabilities take the following values: $H_{11\left(\max \right)}^{\left(f\right)} \approx 1.7\times 10^{4} ;\; 2.4\times 10^{7} ;\; 9.0\times 10^{10} $ and  $H_{11\left(\max \right)}^{\left(\pm \right)} \approx 1.3\times 10^{4} ;\; 3.0\times 10^{9} ;\; 1.1\times 10^{15} $. From this it can be seen that for small values of the Breit-Wheeler parameter $\left(\varepsilon _{iBW} =2.5;\; E_{i} =50\; {\rm MeV}\; \right)$, the corresponding relative probabilities are of the same order of magnitude $\left(H_{11\left(\max \right)}^{\left(f\right)} {\rm \mathop{>}\limits_\sim }H_{11\left(\max \right)}^{\left(\pm \right)} \sim 10^{4} \right)$. However, with an increase in the Breit-Wheeler quantum parameter $\left(\varepsilon _{iBW} >>1\; \right)$ (with an increase in the energy of the initial electrons), the relative probability $H_{11\left(\max \right)}^{\left(\pm \right)} >>H_{11\left(\max \right)}^{\left(f\right)} $ (in accordance with the relation \eqref{ZEqnNum389568}).

\begin{figure}[h]
	\includegraphics[width=0.45\textwidth]{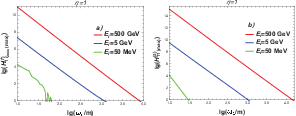}
	\caption{\label{fig 14} Dependence of the maximum relative probabilities $H_{1l_{2\min } \left(\max \right)}^{\left(f\right)} $ \eqref{ZEqnNum668878} (case a) and $H_{11\left(\max \right)}^{\left(\pm \right)} $ \eqref{ZEqnNum352099} (case b) on the characteristic energy of the Compton effect $\omega _{C} $\eqref{ZEqnNum291894} for fixed values of the initial electron energy.}
\end{figure} 

It should also be noted that along with high-probability processes of outgoing of final particles at minimum angles, there are also high-probability processes with outgoing of final particles at angles far from minimum(see peaks of relative probability distributions $H_{21}^{\left(f\right)} $ and $H_{12}^{\left(\pm \right)} $ in Figures 7 and 10). Indeed, from Figures 7 and 10, as well as Tables 2 and 4, it follows that for $\varepsilon _{iBW} =2.41$ and $E_{i} =50\; {\rm MeV}$, the corresponding relative probabilities take the following values: $H_{21\left(\max \right)}^{\left(f\right)} \approx 1.62\times 10^{4} $ (at the outgoing angles of final particles $\delta _{f\left(d\right)}^{2*} =34.95$ and $\delta _{\pm \left(u\right)}^{2*} =5.74$) and $H_{12\left(\max \right)}^{\left(\pm \right)} \approx 1.57\times 10^{4} $ (at the outgoing angles of final particles $\delta _{f\left(u\right)}^{2*} =2.79$ and $\delta _{\pm \left(d\right)}^{2*} =8.35$). Thus, for small electron energies and values of the Breit-Wheeler quantum parameter $\left(E_{i} =50\; {\rm MeV,\; }\varepsilon _{iBW} =2.41\right)$ in strong X-ray fields, relative probabilities $\left(H_{12\left(\max \right)}^{\left(\pm \right)} \sim H_{21\left(\max \right)}^{\left(f\right)} \right)\sim \left(H_{11\left(\max \right)}^{\left(\pm \right)} \sim H_{11\left(\max \right)}^{\left(f\right)} \right)\sim 10^{4} $ are obtained. At the same time, for large energies and values of the Breit-Wheeler quantum parameter $\left(E_{i} =500\; {\rm GeV,\; }\varepsilon _{iBW} =10^{5} \right)$ in strong X-ray fields , the relative probabilities $H_{21\left(\max \right)}^{\left(f\right)} \approx 2.3\times 10^{11} $ are (at the outgoing angles of final particles $\delta _{f\left(d\right)}^{2*} \approx 3.7\times 10^{9} $ and $\delta _{\pm \left(u\right)}^{2*} \approx 10^{5} $) and $H_{12\left(\max \right)}^{\left(\pm \right)} \approx 2.8\times 10^{15} $ (at the outgoing angles of final particles $\delta _{f\left(u\right)}^{2*} \approx 2.23$ and $\delta _{\pm \left(d\right)}^{2*} \approx 9.2\times 10^{8} $). Thus, $H_{21\left(\max \right)}^{\left(f\right)} \sim H_{11\left(\max \right)}^{\left(f\right)} \sim 10^{11} $ and $H_{12\left(\max \right)}^{\left(\pm \right)} \sim H_{11\left(\max \right)}^{\left(\pm \right)} \sim 10^{15} $.

It is important to note that Oleinik resonances occur not only in the field of a plane monochromatic wave, but also in the field of a plane pulsed wave, provided that the pulse $\tau $ time significantly exceeds the wave oscillation period $\tau >>\omega ^{-1} $ \cite{31}. However, for very short pulses $\tau \sim \omega ^{-1} $, Oleinik resonances may not appear. In this paper, an idealized case of a plane monochromatic electromagnetic wave was considered. In a real experiment, as well as in the vicinity of pulsars and magnetars, the electromagnetic wave is inhomogeneous in space and time. The study of Oleinik resonances in such fields is a rather complex independent task, which can be performed only by numerically solving the corresponding mathematical problem. The solution of the resonant problem in the field of a plane monochromatic wave, however, allows us to solve several important problems. First, determine the basic physical parameters of the problem (the characteristic energy of the process, the corresponding quantum parameters), which determine the resonant energy of final particles, as well as the value of the resonant probability of the process. Secondly, it allowed us to obtain analytical expressions for the resonant differential probability of the process. Note that all this is very important for the subsequent numerical analysis of this resonant process in an inhomogeneous electromagnetic field.

We also note that observations of the RTPP process in modern experimental facilities require fluxes of ultrarelativistic electrons of sufficiently high energies $\left(E_{i} {\rm \mathop{>}\limits_\sim }250{\rm \; GeV}\right)$. However, in the universe, in particular, near pulsars and magnetars, high-energy electron fluxes colliding with a strong X-ray wave are possible. At the same time, near such objects in strong X-ray fields, cascades of resonant QED processes are possible, such as: the resonant spontaneous bremsstrahlung radiation during scattering of ultrarelativistic electrons by nuclei \cite{66}, the Bethe-Heitler resonant process \cite{67}, the resonant Breit-Wheeler process \cite{68}, and the resonant Compton effect \cite{69}, etc. These processes are interconnected and can generate fluxes of high-energy gamma-quanta and ultrarelativistic electrons and positrons. Thus, the results obtained can be used to explain narrow fluxes of high-energy gamma-quanta near neutron stars, such as double X-ray systems operating on accretion \cite{109,110}, X-ray/gamma pulsars operating on rotation \cite{111,112} and magnetars operating on a magnetic field \cite{113,114}.

In conclusion, we formulate the main results:

\begin{enumerate}
	\item  Under resonant conditions, the intermediate virtual gamma-quantum becomes real. As a result, the initial second-order process by the fine structure constant in the wave field effectively splits into two first-order processes. In the first vertex, the Compton effect stimulated by an external field takes place, and in the second vertex, the Breit-Wheeler process stimulated by an external field takes place. 
	
	\item  Due to the laws of conservation of energy and momentum, the RTPP process can take place in two cases. In the first case, the energies of final particles and their outgoing angles are determined by the Compton effect stimulated by an external field. In the second case, the energies of final particles and their outgoing angles are determined by the Breit-Wheeler process stimulated by an external field. It is important to note that in the RTPP process, electron and positron pairs are born with equal energies. 
	
	\item  In the first case, the outgoing angle of the final electron (ultrarelativistic parameter $\delta _{f\left(d\right)}^{2} $), as well as the quantum parameter of the Compton effect $\left(\varepsilon _{iC} \right)$ and the number of absorbed gamma-quanta of the wave at the first vertex $\left(l_{1} \right)$, determine the outgoing angle of the electron-positron pair (ultrarelativistic parameter $\delta _{\pm \left(u\right)}^{2} $), as well as the energies of the electron (on the lower branch), electron-positron pair (on the upper branch) and the number of absorbed gamma-quanta of the wave at the second vertex $\left(l_{2} \right)$. In this case, if the quantum parameter $\varepsilon _{iC} >>1$, then the energy of the initial electrons, basically, passes into the energy of the electron-positron pair. 
	
	\item  In the second case, the outgoing angle of the electron-positron pair (an ultrarelativistic parameter $\delta _{\pm \left(d\right)}^{2} $), as well as the quantum parameter of the Breit-Willer process $\left(\varepsilon _{iBW} \right)$ and the number of absorbed gamma-quanta of the wave at the second vertex $\left(l_{2} \right)$, determine the outgoing angle of the electron (an ultrarelativistic parameter $\delta _{f\left(u\right)}^{2} $), as well as the energies of the electron-positron pair (on the lower branch) and the electron (on the upper branches). In this case, if the quantum parameter $\varepsilon _{iBW} >>1$, then the energy of the initial electrons, basically, goes to the energy of the final electron. 
	
	\item  The value of the differential probability of the RTPP process essentially depends on the quantum parameters of the Compton effect $\left(\varepsilon _{iC} \right)$ and Breit-Wheeler process $\left(\varepsilon _{iBW} \right)$, the characteristic energy of the Compton effect $\left(\omega _{C} \right)$, and the number of gamma-quanta absorbed at the first  and second vertices of the wave.
	
	\item  In the region of optical frequencies and small electromagnetic wave intensities $\left(\omega \sim 1{\rm \; eV,\; I}{\rm \mathop{<}\limits_\sim 10}^{20} {\rm \; Wcm}^{{\rm -2}} \right)$, the relative differential probability of the RTPP process (in units of the total probability of the Compton effect stimulated by an external field) exceeds unity for sufficiently large initial electron energies $\left(E_{i} {\rm \mathop{>}\limits_\sim }250{\rm \; GeV}\right)$. And for very high initial electron energies $\left(E_{i} {\rm }=87{\rm \; TeV}\right)$, the RTPP process becomes dominant $\left(H_{11\left(\max \right)}^{\left(\pm \right)} \approx 4.2\times 10^{3} \right)$ (see Figures 12 and 13, as well as Table 6).
	
	\item   In a sufficiently wide range of frequencies and intensities of a strong X-ray wave from $\omega \approx 18{\rm \; eV,}\; I\approx 5.85\times 10^{20} {\rm \; Wcm}^{-2} $ to $\omega \approx 26{\rm \; keV,}\; I\approx 1.3\times 10^{27} {\rm \; Wcm}^{-2} $, as well as in a wide range of initial electron energies $\left(E_{i} {\rm \mathop{>}\limits_\sim }50{\rm \; MeV}\right)$, the RTPP process becomes dominant and can exceed the total probability of the Compton effect stimulated by an external field by many orders of magnitude (see Figure 14). 
	
	\item  In  the RTPP process, quantum entanglement of final particles states occurs, when the measurement of the electron (electron-positron pair) outgoing angle uniquely determines the energies of final particles and the electron-positron pair (electron) outgoing angle. Moreover, this effect can significantly (by many orders of magnitude) exceed the corresponding Compton effect stimulated by an external field.
\end{enumerate}
\nocite{*}
\bibliography{prabiblio}

\begin{thebibliography}{114}%
\makeatletter
\providecommand \@ifxundefined [1]{%
 \@ifx{#1\undefined}
}%
\providecommand \@ifnum [1]{%
 \ifnum #1\expandafter \@firstoftwo
 \else \expandafter \@secondoftwo
 \fi
}%
\providecommand \@ifx [1]{%
 \ifx #1\expandafter \@firstoftwo
 \else \expandafter \@secondoftwo
 \fi
}%
\providecommand \natexlab [1]{#1}%
\providecommand \enquote  [1]{``#1''}%
\providecommand \bibnamefont  [1]{#1}%
\providecommand \bibfnamefont [1]{#1}%
\providecommand \citenamefont [1]{#1}%
\providecommand \href@noop [0]{\@secondoftwo}%
\providecommand \href [0]{\begingroup \@sanitize@url \@href}%
\providecommand \@href[1]{\@@startlink{#1}\@@href}%
\providecommand \@@href[1]{\endgroup#1\@@endlink}%
\providecommand \@sanitize@url [0]{\catcode `\\12\catcode `\$12\catcode
  `\&12\catcode `\#12\catcode `\^12\catcode `\_12\catcode `\%12\relax}%
\providecommand \@@startlink[1]{}%
\providecommand \@@endlink[0]{}%
\providecommand \url  [0]{\begingroup\@sanitize@url \@url }%
\providecommand \@url [1]{\endgroup\@href {#1}{\urlprefix }}%
\providecommand \urlprefix  [0]{URL }%
\providecommand \Eprint [0]{\href }%
\providecommand \doibase [0]{http://dx.doi.org/}%
\providecommand \selectlanguage [0]{\@gobble}%
\providecommand \bibinfo  [0]{\@secondoftwo}%
\providecommand \bibfield  [0]{\@secondoftwo}%
\providecommand \translation [1]{[#1]}%
\providecommand \BibitemOpen [0]{}%
\providecommand \bibitemStop [0]{}%
\providecommand \bibitemNoStop [0]{.\EOS\space}%
\providecommand \EOS [0]{\spacefactor3000\relax}%
\providecommand \BibitemShut  [1]{\csname bibitem#1\endcsname}%
\let\auto@bib@innerbib\@empty
%</preamble>
\bibitem [{\citenamefont {Bula}\ \emph {et~al.}(1996)\citenamefont {Bula},
  \citenamefont {McDonald}, \citenamefont {Prebys}, \citenamefont {Bamber},
  \citenamefont {Boege}, \citenamefont {Kotseroglou}, \citenamefont
  {Melissinos}, \citenamefont {Meyerhofer}, \citenamefont {Ragg}, \citenamefont
  {Burke}, \citenamefont {Field}, \citenamefont {Horton-Smith}, \citenamefont
  {Odian}, \citenamefont {Spencer}, \citenamefont {Walz}, \citenamefont
  {Berridge}, \citenamefont {Bugg}, \citenamefont {Shmakov},\ and\
  \citenamefont {Weidemann}}]{1}%
  \BibitemOpen
  \bibfield  {author} {\bibinfo {author} {\bibfnamefont {C.}~\bibnamefont
  {Bula}}, \bibinfo {author} {\bibfnamefont {K.~T.}\ \bibnamefont {McDonald}},
  \bibinfo {author} {\bibfnamefont {E.~J.}\ \bibnamefont {Prebys}}, \bibinfo
  {author} {\bibfnamefont {C.}~\bibnamefont {Bamber}}, \bibinfo {author}
  {\bibfnamefont {S.}~\bibnamefont {Boege}}, \bibinfo {author} {\bibfnamefont
  {T.}~\bibnamefont {Kotseroglou}}, \bibinfo {author} {\bibfnamefont {A.~C.}\
  \bibnamefont {Melissinos}}, \bibinfo {author} {\bibfnamefont {D.~D.}\
  \bibnamefont {Meyerhofer}}, \bibinfo {author} {\bibfnamefont
  {W.}~\bibnamefont {Ragg}}, \bibinfo {author} {\bibfnamefont {D.~L.}\
  \bibnamefont {Burke}}, \bibinfo {author} {\bibfnamefont {R.~C.}\ \bibnamefont
  {Field}}, \bibinfo {author} {\bibfnamefont {G.}~\bibnamefont {Horton-Smith}},
  \bibinfo {author} {\bibfnamefont {A.~C.}\ \bibnamefont {Odian}}, \bibinfo
  {author} {\bibfnamefont {J.~E.}\ \bibnamefont {Spencer}}, \bibinfo {author}
  {\bibfnamefont {D.}~\bibnamefont {Walz}}, \bibinfo {author} {\bibfnamefont
  {S.~C.}\ \bibnamefont {Berridge}}, \bibinfo {author} {\bibfnamefont {W.~M.}\
  \bibnamefont {Bugg}}, \bibinfo {author} {\bibfnamefont {K.}~\bibnamefont
  {Shmakov}}, \ and\ \bibinfo {author} {\bibfnamefont {A.~W.}\ \bibnamefont
  {Weidemann}},\ }\href {\doibase 10.1103/physrevlett.76.3116} {\bibfield
  {journal} {\bibinfo  {journal} {Physical Review Letters}\ }\textbf {\bibinfo
  {volume} {76}},\ \bibinfo {pages} {3116} (\bibinfo {year}
  {1996})}\BibitemShut {NoStop}%
\bibitem [{\citenamefont {Turcu}\ \emph {et~al.}(2015)\citenamefont {Turcu},
  \citenamefont {Balascuta}, \citenamefont {Negoita}, \citenamefont
  {Jaroszynski},\ and\ \citenamefont {McKenna}}]{2}%
  \BibitemOpen
  \bibfield  {author} {\bibinfo {author} {\bibfnamefont {E.}~\bibnamefont
  {Turcu}}, \bibinfo {author} {\bibfnamefont {S.}~\bibnamefont {Balascuta}},
  \bibinfo {author} {\bibfnamefont {F.}~\bibnamefont {Negoita}}, \bibinfo
  {author} {\bibfnamefont {D.}~\bibnamefont {Jaroszynski}}, \ and\ \bibinfo
  {author} {\bibfnamefont {P.}~\bibnamefont {McKenna}}\ }(\bibinfo {year}
  {2015})\ pp.\ \bibinfo {pages} {416--420}\BibitemShut {NoStop}%
\bibitem [{\citenamefont {Popovici}\ \emph {et~al.}(2017)\citenamefont
  {Popovici}, \citenamefont {Mitu}, \citenamefont {Cata-Danil}, \citenamefont
  {Negoita},\ and\ \citenamefont {Ivan}}]{3}%
  \BibitemOpen
  \bibfield  {author} {\bibinfo {author} {\bibfnamefont {M.-A.}\ \bibnamefont
  {Popovici}}, \bibinfo {author} {\bibfnamefont {I.-O.}\ \bibnamefont {Mitu}},
  \bibinfo {author} {\bibfnamefont {G.}~\bibnamefont {Cata-Danil}}, \bibinfo
  {author} {\bibfnamefont {F.}~\bibnamefont {Negoita}}, \ and\ \bibinfo
  {author} {\bibfnamefont {C.}~\bibnamefont {Ivan}},\ }\href {\doibase
  10.1088/1361-6498/37/1/176} {\bibfield  {journal} {\bibinfo  {journal}
  {Journal of Radiological Protection}\ }\textbf {\bibinfo {volume} {37
  (2017)}},\ \bibinfo {pages} {176 } (\bibinfo {year} {2017})}\BibitemShut
  {NoStop}%
\bibitem [{\citenamefont {Danson}\ \emph {et~al.}(2019)\citenamefont {Danson},
  \citenamefont {Haefner}, \citenamefont {Bromage}, \citenamefont {Butcher},
  \citenamefont {Chanteloup}, \citenamefont {Chowdhury}, \citenamefont
  {Galvanauskas}, \citenamefont {Gizzi}, \citenamefont {Hein}, \citenamefont
  {Hillier}, \citenamefont {Hopps}, \citenamefont {Kato}, \citenamefont
  {Khazanov}, \citenamefont {Kodama}, \citenamefont {Korn}, \citenamefont {Li},
  \citenamefont {Li}, \citenamefont {Limpert}, \citenamefont {Ma},
  \citenamefont {Nam}, \citenamefont {Neely}, \citenamefont {Papadopoulos},
  \citenamefont {Penman}, \citenamefont {Qian}, \citenamefont {Rocca},
  \citenamefont {Shaykin}, \citenamefont {Siders}, \citenamefont {Spindloe},
  \citenamefont {Szatmari}, \citenamefont {Trines}, \citenamefont {Zhu},
  \citenamefont {Zhu},\ and\ \citenamefont {Zuegel}}]{4}%
  \BibitemOpen
  \bibfield  {author} {\bibinfo {author} {\bibfnamefont {C.~N.}\ \bibnamefont
  {Danson}}, \bibinfo {author} {\bibfnamefont {C.}~\bibnamefont {Haefner}},
  \bibinfo {author} {\bibfnamefont {J.}~\bibnamefont {Bromage}}, \bibinfo
  {author} {\bibfnamefont {T.}~\bibnamefont {Butcher}}, \bibinfo {author}
  {\bibfnamefont {J.-C.~F.}\ \bibnamefont {Chanteloup}}, \bibinfo {author}
  {\bibfnamefont {E.~A.}\ \bibnamefont {Chowdhury}}, \bibinfo {author}
  {\bibfnamefont {A.}~\bibnamefont {Galvanauskas}}, \bibinfo {author}
  {\bibfnamefont {L.~A.}\ \bibnamefont {Gizzi}}, \bibinfo {author}
  {\bibfnamefont {J.}~\bibnamefont {Hein}}, \bibinfo {author} {\bibfnamefont
  {D.~I.}\ \bibnamefont {Hillier}}, \bibinfo {author} {\bibfnamefont {N.~W.}\
  \bibnamefont {Hopps}}, \bibinfo {author} {\bibfnamefont {Y.}~\bibnamefont
  {Kato}}, \bibinfo {author} {\bibfnamefont {E.~A.}\ \bibnamefont {Khazanov}},
  \bibinfo {author} {\bibfnamefont {R.}~\bibnamefont {Kodama}}, \bibinfo
  {author} {\bibfnamefont {G.}~\bibnamefont {Korn}}, \bibinfo {author}
  {\bibfnamefont {R.}~\bibnamefont {Li}}, \bibinfo {author} {\bibfnamefont
  {Y.}~\bibnamefont {Li}}, \bibinfo {author} {\bibfnamefont {J.}~\bibnamefont
  {Limpert}}, \bibinfo {author} {\bibfnamefont {J.}~\bibnamefont {Ma}},
  \bibinfo {author} {\bibfnamefont {C.~H.}\ \bibnamefont {Nam}}, \bibinfo
  {author} {\bibfnamefont {D.}~\bibnamefont {Neely}}, \bibinfo {author}
  {\bibfnamefont {D.}~\bibnamefont {Papadopoulos}}, \bibinfo {author}
  {\bibfnamefont {R.~R.}\ \bibnamefont {Penman}}, \bibinfo {author}
  {\bibfnamefont {L.}~\bibnamefont {Qian}}, \bibinfo {author} {\bibfnamefont
  {J.~J.}\ \bibnamefont {Rocca}}, \bibinfo {author} {\bibfnamefont {A.~A.}\
  \bibnamefont {Shaykin}}, \bibinfo {author} {\bibfnamefont {C.~W.}\
  \bibnamefont {Siders}}, \bibinfo {author} {\bibfnamefont {C.}~\bibnamefont
  {Spindloe}}, \bibinfo {author} {\bibfnamefont {S.}~\bibnamefont {Szatmari}},
  \bibinfo {author} {\bibfnamefont {R.~M. G.~M.}\ \bibnamefont {Trines}},
  \bibinfo {author} {\bibfnamefont {J.}~\bibnamefont {Zhu}}, \bibinfo {author}
  {\bibfnamefont {P.}~\bibnamefont {Zhu}}, \ and\ \bibinfo {author}
  {\bibfnamefont {J.~D.}\ \bibnamefont {Zuegel}},\ }\href {\doibase
  10.1017/hpl.2019.36} {\bibfield  {journal} {\bibinfo  {journal} {High Power
  Laser Science and Engineering}\ }\textbf {\bibinfo {volume} {7}} (\bibinfo
  {year} {2019}),\ 10.1017/hpl.2019.36}\BibitemShut {NoStop}%
\bibitem [{\citenamefont {Yoon}\ \emph {et~al.}(2021)\citenamefont {Yoon},
  \citenamefont {Kim}, \citenamefont {Choi}, \citenamefont {Sung},
  \citenamefont {Lee}, \citenamefont {Lee},\ and\ \citenamefont {Nam}}]{5}%
  \BibitemOpen
  \bibfield  {author} {\bibinfo {author} {\bibfnamefont {J.~W.}\ \bibnamefont
  {Yoon}}, \bibinfo {author} {\bibfnamefont {Y.~G.}\ \bibnamefont {Kim}},
  \bibinfo {author} {\bibfnamefont {I.~W.}\ \bibnamefont {Choi}}, \bibinfo
  {author} {\bibfnamefont {J.~H.}\ \bibnamefont {Sung}}, \bibinfo {author}
  {\bibfnamefont {H.~W.}\ \bibnamefont {Lee}}, \bibinfo {author} {\bibfnamefont
  {S.~K.}\ \bibnamefont {Lee}}, \ and\ \bibinfo {author} {\bibfnamefont
  {C.~H.}\ \bibnamefont {Nam}},\ }\href {\doibase 10.1364/optica.420520}
  {\bibfield  {journal} {\bibinfo  {journal} {Optica}\ }\textbf {\bibinfo
  {volume} {8}},\ \bibinfo {pages} {630} (\bibinfo {year} {2021})}\BibitemShut
  {NoStop}%
\bibitem [{\citenamefont {Turcu}\ \emph {et~al.}(2019)\citenamefont {Turcu},
  \citenamefont {Shen}, \citenamefont {Neely}, \citenamefont {Sarri},
  \citenamefont {Tanaka}, \citenamefont {McKenna}, \citenamefont {Mangles},
  \citenamefont {Yu}, \citenamefont {Luo}, \citenamefont {Zhu},\ and\
  \citenamefont {Yin}}]{6}%
  \BibitemOpen
  \bibfield  {author} {\bibinfo {author} {\bibfnamefont {I.~C.~E.}\
  \bibnamefont {Turcu}}, \bibinfo {author} {\bibfnamefont {B.}~\bibnamefont
  {Shen}}, \bibinfo {author} {\bibfnamefont {D.}~\bibnamefont {Neely}},
  \bibinfo {author} {\bibfnamefont {G.}~\bibnamefont {Sarri}}, \bibinfo
  {author} {\bibfnamefont {K.~A.}\ \bibnamefont {Tanaka}}, \bibinfo {author}
  {\bibfnamefont {P.}~\bibnamefont {McKenna}}, \bibinfo {author} {\bibfnamefont
  {S.~P.~D.}\ \bibnamefont {Mangles}}, \bibinfo {author} {\bibfnamefont
  {T.-P.}\ \bibnamefont {Yu}}, \bibinfo {author} {\bibfnamefont
  {W.}~\bibnamefont {Luo}}, \bibinfo {author} {\bibfnamefont {X.-L.}\
  \bibnamefont {Zhu}}, \ and\ \bibinfo {author} {\bibfnamefont
  {Y.}~\bibnamefont {Yin}},\ }\href {\doibase 10.1017/hpl.2018.66} {\bibfield
  {journal} {\bibinfo  {journal} {High Power Laser Science and Engineering}\
  }\textbf {\bibinfo {volume} {7}} (\bibinfo {year} {2019}),\
  10.1017/hpl.2018.66}\BibitemShut {NoStop}%
\bibitem [{\citenamefont {Tanaka}\ \emph {et~al.}(2020)\citenamefont {Tanaka},
  \citenamefont {Spohr}, \citenamefont {Balabanski}, \citenamefont {Balascuta},
  \citenamefont {Capponi}, \citenamefont {Cernaianu}, \citenamefont {Cuciuc},
  \citenamefont {Cucoanes}, \citenamefont {Dancus}, \citenamefont {Dhal},
  \citenamefont {Doria}, \citenamefont {Ghenuche}, \citenamefont {Ghita},
  \citenamefont {Kisyov}, \citenamefont {Nastasa}, \citenamefont {Ong},
  \citenamefont {Rotaru}, \citenamefont {Sangwan},\ and\ \citenamefont
  {Zamfir}}]{7}%
  \BibitemOpen
  \bibfield  {author} {\bibinfo {author} {\bibfnamefont {K.}~\bibnamefont
  {Tanaka}}, \bibinfo {author} {\bibfnamefont {K.}~\bibnamefont {Spohr}},
  \bibinfo {author} {\bibfnamefont {D.}~\bibnamefont {Balabanski}}, \bibinfo
  {author} {\bibfnamefont {S.}~\bibnamefont {Balascuta}}, \bibinfo {author}
  {\bibfnamefont {L.}~\bibnamefont {Capponi}}, \bibinfo {author} {\bibfnamefont
  {M.}~\bibnamefont {Cernaianu}}, \bibinfo {author} {\bibfnamefont
  {M.}~\bibnamefont {Cuciuc}}, \bibinfo {author} {\bibfnamefont
  {A.}~\bibnamefont {Cucoanes}}, \bibinfo {author} {\bibfnamefont
  {I.}~\bibnamefont {Dancus}}, \bibinfo {author} {\bibfnamefont
  {A.}~\bibnamefont {Dhal}}, \bibinfo {author} {\bibfnamefont {D.}~\bibnamefont
  {Doria}}, \bibinfo {author} {\bibfnamefont {P.}~\bibnamefont {Ghenuche}},
  \bibinfo {author} {\bibfnamefont {D.~G.}\ \bibnamefont {Ghita}}, \bibinfo
  {author} {\bibfnamefont {S.}~\bibnamefont {Kisyov}}, \bibinfo {author}
  {\bibfnamefont {V.}~\bibnamefont {Nastasa}}, \bibinfo {author} {\bibfnamefont
  {J.~F.}\ \bibnamefont {Ong}}, \bibinfo {author} {\bibfnamefont
  {F.}~\bibnamefont {Rotaru}}, \bibinfo {author} {\bibfnamefont
  {D.}~\bibnamefont {Sangwan}}, \ and\ \bibinfo {author} {\bibfnamefont
  {N.}~\bibnamefont {Zamfir}},\ }\href {\doibase 10.1063/1.5093535} {\bibfield
  {journal} {\bibinfo  {journal} {Matter and Radiation at Extremes}\ }\textbf
  {\bibinfo {volume} {5}},\ \bibinfo {pages} {024402} (\bibinfo {year}
  {2020})}\BibitemShut {NoStop}%
\bibitem [{\citenamefont {Weber}\ \emph {et~al.}(2017)\citenamefont {Weber},
  \citenamefont {Bechet}, \citenamefont {Borneis}, \citenamefont {Brabec},
  \citenamefont {Bucka}, \citenamefont {Chacon-Golcher}, \citenamefont
  {Ciappina}, \citenamefont {DeMarco}, \citenamefont {Fajstavr}, \citenamefont
  {Falk}, \citenamefont {Rodreguez~Garcia}, \citenamefont {Grosz},
  \citenamefont {Gu}, \citenamefont {Hernandez}, \citenamefont {Holec},
  \citenamefont {Janecka}, \citenamefont {Jantac}, \citenamefont {Jirka},
  \citenamefont {Kadlecova},\ and\ \citenamefont {Yu}}]{8}%
  \BibitemOpen
  \bibfield  {author} {\bibinfo {author} {\bibfnamefont {S.}~\bibnamefont
  {Weber}}, \bibinfo {author} {\bibfnamefont {S.}~\bibnamefont {Bechet}},
  \bibinfo {author} {\bibfnamefont {S.}~\bibnamefont {Borneis}}, \bibinfo
  {author} {\bibfnamefont {L.}~\bibnamefont {Brabec}}, \bibinfo {author}
  {\bibfnamefont {M.}~\bibnamefont {Bucka}}, \bibinfo {author} {\bibfnamefont
  {E.}~\bibnamefont {Chacon-Golcher}}, \bibinfo {author} {\bibfnamefont
  {M.}~\bibnamefont {Ciappina}}, \bibinfo {author} {\bibfnamefont
  {M.}~\bibnamefont {DeMarco}}, \bibinfo {author} {\bibfnamefont
  {A.}~\bibnamefont {Fajstavr}}, \bibinfo {author} {\bibfnamefont
  {K.}~\bibnamefont {Falk}}, \bibinfo {author} {\bibfnamefont {E.}~\bibnamefont
  {Rodreguez~Garcia}}, \bibinfo {author} {\bibfnamefont {J.}~\bibnamefont
  {Grosz}}, \bibinfo {author} {\bibfnamefont {Y.-J.}\ \bibnamefont {Gu}},
  \bibinfo {author} {\bibfnamefont {J.-C.}\ \bibnamefont {Hernandez}}, \bibinfo
  {author} {\bibfnamefont {M.}~\bibnamefont {Holec}}, \bibinfo {author}
  {\bibfnamefont {P.}~\bibnamefont {Janecka}}, \bibinfo {author} {\bibfnamefont
  {M.}~\bibnamefont {Jantac}}, \bibinfo {author} {\bibfnamefont
  {M.}~\bibnamefont {Jirka}}, \bibinfo {author} {\bibfnamefont
  {H.}~\bibnamefont {Kadlecova}}, \ and\ \bibinfo {author} {\bibfnamefont
  {Q.}~\bibnamefont {Yu}},\ }\href {\doibase 10.1016/j.mre.2017.03.003}
  {\bibfield  {journal} {\bibinfo  {journal} {Matter and Radiation at
  Extremes}\ }\textbf {\bibinfo {volume} {2}} (\bibinfo {year} {2017}),\
  10.1016/j.mre.2017.03.003}\BibitemShut {NoStop}%
\bibitem [{\citenamefont {Papadopoulos}\ \emph {et~al.}(2016)\citenamefont
  {Papadopoulos}, \citenamefont {Zou}, \citenamefont {Le~Blanc}, \citenamefont
  {Cheriaux}, \citenamefont {Georges}, \citenamefont {Druon}, \citenamefont
  {Mennerat}, \citenamefont {Ramirez}, \citenamefont {Martin}, \citenamefont
  {Freneaux}, \citenamefont {Beluze}, \citenamefont {Lebas}, \citenamefont
  {Monot}, \citenamefont {Mathieu},\ and\ \citenamefont {Audebert}}]{9}%
  \BibitemOpen
  \bibfield  {author} {\bibinfo {author} {\bibfnamefont {D.}~\bibnamefont
  {Papadopoulos}}, \bibinfo {author} {\bibfnamefont {J.}~\bibnamefont {Zou}},
  \bibinfo {author} {\bibfnamefont {C.}~\bibnamefont {Le~Blanc}}, \bibinfo
  {author} {\bibfnamefont {G.}~\bibnamefont {Cheriaux}}, \bibinfo {author}
  {\bibfnamefont {P.}~\bibnamefont {Georges}}, \bibinfo {author} {\bibfnamefont
  {F.}~\bibnamefont {Druon}}, \bibinfo {author} {\bibfnamefont
  {G.}~\bibnamefont {Mennerat}}, \bibinfo {author} {\bibfnamefont
  {P.}~\bibnamefont {Ramirez}}, \bibinfo {author} {\bibfnamefont
  {L.}~\bibnamefont {Martin}}, \bibinfo {author} {\bibfnamefont
  {A.}~\bibnamefont {Freneaux}}, \bibinfo {author} {\bibfnamefont
  {A.}~\bibnamefont {Beluze}}, \bibinfo {author} {\bibfnamefont
  {N.}~\bibnamefont {Lebas}}, \bibinfo {author} {\bibfnamefont
  {P.}~\bibnamefont {Monot}}, \bibinfo {author} {\bibfnamefont
  {F.}~\bibnamefont {Mathieu}}, \ and\ \bibinfo {author} {\bibfnamefont
  {P.}~\bibnamefont {Audebert}},\ }\href {\doibase 10.1017/hpl.2016.34}
  {\bibfield  {journal} {\bibinfo  {journal} {High Power Laser Science and
  Engineering}\ }\textbf {\bibinfo {volume} {4}} (\bibinfo {year} {2016}),\
  10.1017/hpl.2016.34}\BibitemShut {NoStop}%
\bibitem [{\citenamefont {Bromage}\ \emph {et~al.}(2019)\citenamefont
  {Bromage}, \citenamefont {Bahk}, \citenamefont {Begishev}, \citenamefont
  {Dorrer}, \citenamefont {Guardalben}, \citenamefont {Hoffman}, \citenamefont
  {Oliver}, \citenamefont {Roides}, \citenamefont {Schiesser}, \citenamefont
  {Shoup}, \citenamefont {Spilatro}, \citenamefont {Webb}, \citenamefont
  {Weiner},\ and\ \citenamefont {Zuegel}}]{10}%
  \BibitemOpen
  \bibfield  {author} {\bibinfo {author} {\bibfnamefont {J.}~\bibnamefont
  {Bromage}}, \bibinfo {author} {\bibfnamefont {S.-W.}\ \bibnamefont {Bahk}},
  \bibinfo {author} {\bibfnamefont {I.}~\bibnamefont {Begishev}}, \bibinfo
  {author} {\bibfnamefont {C.}~\bibnamefont {Dorrer}}, \bibinfo {author}
  {\bibfnamefont {M.}~\bibnamefont {Guardalben}}, \bibinfo {author}
  {\bibfnamefont {B.}~\bibnamefont {Hoffman}}, \bibinfo {author} {\bibfnamefont
  {J.}~\bibnamefont {Oliver}}, \bibinfo {author} {\bibfnamefont
  {R.}~\bibnamefont {Roides}}, \bibinfo {author} {\bibfnamefont
  {E.}~\bibnamefont {Schiesser}}, \bibinfo {author} {\bibfnamefont
  {M.}~\bibnamefont {Shoup}}, \bibinfo {author} {\bibfnamefont
  {M.}~\bibnamefont {Spilatro}}, \bibinfo {author} {\bibfnamefont
  {B.}~\bibnamefont {Webb}}, \bibinfo {author} {\bibfnamefont {D.}~\bibnamefont
  {Weiner}}, \ and\ \bibinfo {author} {\bibfnamefont {J.}~\bibnamefont
  {Zuegel}},\ }\href {\doibase 10.1017/hpl.2018.64} {\bibfield  {journal}
  {\bibinfo  {journal} {High Power Laser Science and Engineering}\ }\textbf
  {\bibinfo {volume} {7}} (\bibinfo {year} {2019}),\
  10.1017/hpl.2018.64}\BibitemShut {NoStop}%
\bibitem [{\citenamefont {Rossbach}\ \emph {et~al.}(2019)\citenamefont
  {Rossbach}, \citenamefont {Schneider},\ and\ \citenamefont {Wurth}}]{11}%
  \BibitemOpen
  \bibfield  {author} {\bibinfo {author} {\bibfnamefont {J.}~\bibnamefont
  {Rossbach}}, \bibinfo {author} {\bibfnamefont {J.}~\bibnamefont {Schneider}},
  \ and\ \bibinfo {author} {\bibfnamefont {W.}~\bibnamefont {Wurth}},\ }\href
  {\doibase 10.1016/j.physrep.2019.02.002} {\bibfield  {journal} {\bibinfo
  {journal} {Physics Reports}\ }\textbf {\bibinfo {volume} {808}} (\bibinfo
  {year} {2019}),\ 10.1016/j.physrep.2019.02.002}\BibitemShut {NoStop}%
\bibitem [{\citenamefont {Gonoskov}\ \emph {et~al.}(2017)\citenamefont
  {Gonoskov}, \citenamefont {Bashinov}, \citenamefont {Bastrakov},
  \citenamefont {Efimenko}, \citenamefont {Ilderton}, \citenamefont {Kim},
  \citenamefont {Marklund}, \citenamefont {Meyerov}, \citenamefont {Muraviev},\
  and\ \citenamefont {Sergeev}}]{12}%
  \BibitemOpen
  \bibfield  {author} {\bibinfo {author} {\bibfnamefont {A.}~\bibnamefont
  {Gonoskov}}, \bibinfo {author} {\bibfnamefont {A.}~\bibnamefont {Bashinov}},
  \bibinfo {author} {\bibfnamefont {S.}~\bibnamefont {Bastrakov}}, \bibinfo
  {author} {\bibfnamefont {E.}~\bibnamefont {Efimenko}}, \bibinfo {author}
  {\bibfnamefont {A.}~\bibnamefont {Ilderton}}, \bibinfo {author}
  {\bibfnamefont {A.}~\bibnamefont {Kim}}, \bibinfo {author} {\bibfnamefont
  {M.}~\bibnamefont {Marklund}}, \bibinfo {author} {\bibfnamefont
  {I.}~\bibnamefont {Meyerov}}, \bibinfo {author} {\bibfnamefont
  {A.}~\bibnamefont {Muraviev}}, \ and\ \bibinfo {author} {\bibfnamefont
  {A.}~\bibnamefont {Sergeev}},\ }\href {\doibase 10.1103/PhysRevX.7.041003}
  {\bibfield  {journal} {\bibinfo  {journal} {Physical Review X}\ }\textbf
  {\bibinfo {volume} {7}},\ \bibinfo {pages} {041003} (\bibinfo {year}
  {2017})}\BibitemShut {NoStop}%
\bibitem [{\citenamefont {Magnusson}\ \emph {et~al.}(2019)\citenamefont
  {Magnusson}, \citenamefont {Gonoskov}, \citenamefont {Marklund},
  \citenamefont {Esirkepov}, \citenamefont {Koga}, \citenamefont {Kondo},
  \citenamefont {Bulanov}, \citenamefont {Korn},\ and\ \citenamefont
  {Bulanov}}]{13}%
  \BibitemOpen
  \bibfield  {author} {\bibinfo {author} {\bibfnamefont {J.}~\bibnamefont
  {Magnusson}}, \bibinfo {author} {\bibfnamefont {A.}~\bibnamefont {Gonoskov}},
  \bibinfo {author} {\bibfnamefont {M.}~\bibnamefont {Marklund}}, \bibinfo
  {author} {\bibfnamefont {T.~Z.}\ \bibnamefont {Esirkepov}}, \bibinfo {author}
  {\bibfnamefont {J.~K.}\ \bibnamefont {Koga}}, \bibinfo {author}
  {\bibfnamefont {K.}~\bibnamefont {Kondo}}, \bibinfo {author} {\bibfnamefont
  {S.}~\bibnamefont {Bulanov}}, \bibinfo {author} {\bibfnamefont
  {G.}~\bibnamefont {Korn}}, \ and\ \bibinfo {author} {\bibfnamefont {S.~S.}\
  \bibnamefont {Bulanov}},\ }\href {\doibase 10.1103/PhysRevLett.122.254801}
  {\bibfield  {journal} {\bibinfo  {journal} {Physical Review Letters}\
  }\textbf {\bibinfo {volume} {122}} (\bibinfo {year} {2019}),\
  10.1103/PhysRevLett.122.254801}\BibitemShut {NoStop}%
\bibitem [{\citenamefont {Zhu}\ \emph {et~al.}(2018)\citenamefont {Zhu},
  \citenamefont {Yu}, \citenamefont {Chen}, \citenamefont {Weng},\ and\
  \citenamefont {Sheng}}]{14}%
  \BibitemOpen
  \bibfield  {author} {\bibinfo {author} {\bibfnamefont {X.-L.}\ \bibnamefont
  {Zhu}}, \bibinfo {author} {\bibfnamefont {T.-P.}\ \bibnamefont {Yu}},
  \bibinfo {author} {\bibfnamefont {M.}~\bibnamefont {Chen}}, \bibinfo {author}
  {\bibfnamefont {S.-M.}\ \bibnamefont {Weng}}, \ and\ \bibinfo {author}
  {\bibfnamefont {Z.-M.}\ \bibnamefont {Sheng}},\ }\href {\doibase
  10.1088/1367-2630/aad71a} {\bibfield  {journal} {\bibinfo  {journal} {New
  Journal of Physics}\ }\textbf {\bibinfo {volume} {20}},\ \bibinfo {pages}
  {083013} (\bibinfo {year} {2018})}\BibitemShut {NoStop}%
\bibitem [{\citenamefont {Xu}\ \emph {et~al.}(2021)\citenamefont {Xu},
  \citenamefont {Padilla}, \citenamefont {Wang},\ and\ \citenamefont
  {Li}}]{15}%
  \BibitemOpen
  \bibfield  {author} {\bibinfo {author} {\bibfnamefont {H.}~\bibnamefont
  {Xu}}, \bibinfo {author} {\bibfnamefont {O.}~\bibnamefont {Padilla}},
  \bibinfo {author} {\bibfnamefont {D.}~\bibnamefont {Wang}}, \ and\ \bibinfo
  {author} {\bibfnamefont {M.}~\bibnamefont {Li}},\ }\href {\doibase
  10.32614/cran.package.changepoints} {\enquote {\bibinfo {title}
  {Changepoints: A collection of change-point detection methods},}\ } (\bibinfo
  {year} {2021})\BibitemShut {NoStop}%
\bibitem [{\citenamefont {Alejo}\ \emph {et~al.}(2019)\citenamefont {Alejo},
  \citenamefont {Samarin}, \citenamefont {Warwick},\ and\ \citenamefont
  {Sarri}}]{16}%
  \BibitemOpen
  \bibfield  {author} {\bibinfo {author} {\bibfnamefont {A.}~\bibnamefont
  {Alejo}}, \bibinfo {author} {\bibfnamefont {G.~M.}\ \bibnamefont {Samarin}},
  \bibinfo {author} {\bibfnamefont {J.~R.}\ \bibnamefont {Warwick}}, \ and\
  \bibinfo {author} {\bibfnamefont {G.}~\bibnamefont {Sarri}},\ }\href
  {\doibase 10.3389/fphy.2019.00049} {\bibfield  {journal} {\bibinfo  {journal}
  {Frontiers in Physics}\ }\textbf {\bibinfo {volume} {7}} (\bibinfo {year}
  {2019}),\ 10.3389/fphy.2019.00049}\BibitemShut {NoStop}%
\bibitem [{\citenamefont {Long}\ \emph {et~al.}(2019)\citenamefont {Long},
  \citenamefont {Zhou}, \citenamefont {Huang}, \citenamefont {Jiang},
  \citenamefont {Ju}, \citenamefont {Zhang}, \citenamefont {Cai}, \citenamefont
  {Yu}, \citenamefont {Qiao}, \citenamefont {Ruan},\ and\ \citenamefont
  {He}}]{17}%
  \BibitemOpen
  \bibfield  {author} {\bibinfo {author} {\bibfnamefont {T.~Y.}\ \bibnamefont
  {Long}}, \bibinfo {author} {\bibfnamefont {C.~T.}\ \bibnamefont {Zhou}},
  \bibinfo {author} {\bibfnamefont {T.~W.}\ \bibnamefont {Huang}}, \bibinfo
  {author} {\bibfnamefont {K.}~\bibnamefont {Jiang}}, \bibinfo {author}
  {\bibfnamefont {L.~B.}\ \bibnamefont {Ju}}, \bibinfo {author} {\bibfnamefont
  {H.}~\bibnamefont {Zhang}}, \bibinfo {author} {\bibfnamefont {T.~X.}\
  \bibnamefont {Cai}}, \bibinfo {author} {\bibfnamefont {M.~Y.}\ \bibnamefont
  {Yu}}, \bibinfo {author} {\bibfnamefont {B.}~\bibnamefont {Qiao}}, \bibinfo
  {author} {\bibfnamefont {S.~C.}\ \bibnamefont {Ruan}}, \ and\ \bibinfo
  {author} {\bibfnamefont {X.~T.}\ \bibnamefont {He}},\ }\href {\doibase
  10.1088/1361-6587/ab210c} {\bibfield  {journal} {\bibinfo  {journal} {Plasma
  Physics and Controlled Fusion}\ }\textbf {\bibinfo {volume} {61}},\ \bibinfo
  {pages} {085002} (\bibinfo {year} {2019})}\BibitemShut {NoStop}%
\bibitem [{\citenamefont {Terzic}\ \emph {et~al.}(2021)\citenamefont {Terzic},
  \citenamefont {Mckaig}, \citenamefont {Johnson}, \citenamefont
  {Dharanikota},\ and\ \citenamefont {Krafft}}]{18}%
  \BibitemOpen
  \bibfield  {author} {\bibinfo {author} {\bibfnamefont {B.}~\bibnamefont
  {Terzic}}, \bibinfo {author} {\bibfnamefont {J.}~\bibnamefont {Mckaig}},
  \bibinfo {author} {\bibfnamefont {E.}~\bibnamefont {Johnson}}, \bibinfo
  {author} {\bibfnamefont {T.}~\bibnamefont {Dharanikota}}, \ and\ \bibinfo
  {author} {\bibfnamefont {G.}~\bibnamefont {Krafft}},\ }\href {\doibase
  10.1103/PhysRevAccelBeams.24.094401} {\bibfield  {journal} {\bibinfo
  {journal} {Physical Review Accelerators and Beams}\ }\textbf {\bibinfo
  {volume} {94}},\ \bibinfo {pages} {094401} (\bibinfo {year}
  {2021})}\BibitemShut {NoStop}%
\bibitem [{\citenamefont {Gunther}(2023)}]{19}%
  \BibitemOpen
  \bibfield  {author} {\bibinfo {author} {\bibfnamefont {B.}~\bibnamefont
  {Gunther}},\ }\enquote {\bibinfo {title} {Overview on inverse compton x-ray
  sources},}\ \ (\bibinfo {year} {2023})\ pp.\ \bibinfo {pages}
  {117--147}\BibitemShut {NoStop}%
\bibitem [{\citenamefont {Nielsen}\ \emph {et~al.}(2023)\citenamefont
  {Nielsen}, \citenamefont {Holtzapple}, \citenamefont {Lund}, \citenamefont
  {Surrow}, \citenamefont {S\o{}rensen}, \citenamefont {S\o{}rensen},\ and\
  \citenamefont {Uggerh\o{}j}}]{20}%
  \BibitemOpen
  \bibfield  {author} {\bibinfo {author} {\bibfnamefont {C.~F.}\ \bibnamefont
  {Nielsen}}, \bibinfo {author} {\bibfnamefont {R.}~\bibnamefont {Holtzapple}},
  \bibinfo {author} {\bibfnamefont {M.~M.}\ \bibnamefont {Lund}}, \bibinfo
  {author} {\bibfnamefont {J.~H.}\ \bibnamefont {Surrow}}, \bibinfo {author}
  {\bibfnamefont {A.~H.}\ \bibnamefont {S\o{}rensen}}, \bibinfo {author}
  {\bibfnamefont {M.~B.}\ \bibnamefont {S\o{}rensen}}, \ and\ \bibinfo {author}
  {\bibfnamefont {U.~I.}\ \bibnamefont {Uggerh\o{}j}} (\bibinfo {collaboration}
  {CERN NA63 Collaboration}),\ }\href {\doibase 10.1103/PhysRevD.108.052013}
  {\bibfield  {journal} {\bibinfo  {journal} {Phys. Rev. D}\ }\textbf {\bibinfo
  {volume} {108}},\ \bibinfo {pages} {052013} (\bibinfo {year}
  {2023})}\BibitemShut {NoStop}%
\bibitem [{\citenamefont {Macleod}\ \emph {et~al.}(2023)\citenamefont
  {Macleod}, \citenamefont {Hadjisolomou}, \citenamefont {Jeong},\ and\
  \citenamefont {Bulanov}}]{21}%
  \BibitemOpen
  \bibfield  {author} {\bibinfo {author} {\bibfnamefont {A.}~\bibnamefont
  {Macleod}}, \bibinfo {author} {\bibfnamefont {P.}~\bibnamefont
  {Hadjisolomou}}, \bibinfo {author} {\bibfnamefont {T.~M.}\ \bibnamefont
  {Jeong}}, \ and\ \bibinfo {author} {\bibfnamefont {S.~V.}\ \bibnamefont
  {Bulanov}},\ }in\ \href {\doibase 10.1117/12.2665777} {\emph {\bibinfo
  {booktitle} {Research Using Extreme Light: Entering New Frontiers with
  Petawatt-Class Lasers V}}},\ \bibinfo {editor} {edited by\ \bibinfo {editor}
  {\bibfnamefont {S.~V.}\ \bibnamefont {Bulanov}}\ and\ \bibinfo {editor}
  {\bibfnamefont {L.~O.}\ \bibnamefont {Silva}}}\ (\bibinfo  {publisher}
  {SPIE},\ \bibinfo {year} {2023})\ p.~\bibinfo {pages} {2}\BibitemShut
  {NoStop}%
\bibitem [{\citenamefont {Powell}\ \emph {et~al.}(2024)\citenamefont {Powell},
  \citenamefont {Jolly}, \citenamefont {Vallieres}, \citenamefont
  {Fillion-Gourdeau}, \citenamefont {Payeur}, \citenamefont {Fourmaux},
  \citenamefont {Lytova}, \citenamefont {Piche}, \citenamefont {Ibrahim},
  \citenamefont {MacLean},\ and\ \citenamefont {Legare}}]{22}%
  \BibitemOpen
  \bibfield  {author} {\bibinfo {author} {\bibfnamefont {J.}~\bibnamefont
  {Powell}}, \bibinfo {author} {\bibfnamefont {S.}~\bibnamefont {Jolly}},
  \bibinfo {author} {\bibfnamefont {S.}~\bibnamefont {Vallieres}}, \bibinfo
  {author} {\bibfnamefont {F.}~\bibnamefont {Fillion-Gourdeau}}, \bibinfo
  {author} {\bibfnamefont {S.}~\bibnamefont {Payeur}}, \bibinfo {author}
  {\bibfnamefont {S.}~\bibnamefont {Fourmaux}}, \bibinfo {author}
  {\bibfnamefont {M.}~\bibnamefont {Lytova}}, \bibinfo {author} {\bibfnamefont
  {M.}~\bibnamefont {Piche}}, \bibinfo {author} {\bibfnamefont
  {H.}~\bibnamefont {Ibrahim}}, \bibinfo {author} {\bibfnamefont
  {S.}~\bibnamefont {MacLean}}, \ and\ \bibinfo {author} {\bibfnamefont
  {F.}~\bibnamefont {Legare}},\ }\href {\doibase
  10.1103/PhysRevLett.133.155001} {\bibfield  {journal} {\bibinfo  {journal}
  {Physical Review Letters}\ }\textbf {\bibinfo {volume} {133}} (\bibinfo
  {year} {2024}),\ 10.1103/PhysRevLett.133.155001}\BibitemShut {NoStop}%
\bibitem [{\citenamefont {Nikishov}\ and\ \citenamefont {Ritus}(1970)}]{23}%
  \BibitemOpen
  \bibfield  {author} {\bibinfo {author} {\bibfnamefont {A.~I.}\ \bibnamefont
  {Nikishov}}\ and\ \bibinfo {author} {\bibfnamefont {V.~I.}\ \bibnamefont
  {Ritus}},\ }\href {\doibase 10.1070/PU1970v013n02ABEH004234} {\bibfield
  {journal} {\bibinfo  {journal} {Phys. Usp.}\ }\textbf {\bibinfo {volume}
  {13}},\ \bibinfo {pages} {303} (\bibinfo {year} {1970})}\BibitemShut
  {NoStop}%
\bibitem [{\citenamefont {Nikishov}\ and\ \citenamefont {Ritus}(1979)}]{24}%
  \BibitemOpen
  \bibfield  {author} {\bibinfo {author} {\bibfnamefont {A.~I.}\ \bibnamefont
  {Nikishov}}\ and\ \bibinfo {author} {\bibfnamefont {V.~I.}\ \bibnamefont
  {Ritus}},\ }\href@noop {} {\bibfield  {journal} {\bibinfo  {journal} {Nauka}\
  } (\bibinfo {year} {1979})}\BibitemShut {NoStop}%
\bibitem [{\citenamefont {Nikishov}(1987)}]{25}%
  \BibitemOpen
  \bibfield  {author} {\bibinfo {author} {\bibfnamefont {A.~I.}\ \bibnamefont
  {Nikishov}},\ }\href {\doibase 10.1070/PU1987v030n06ABEH002865} {\bibfield
  {journal} {\bibinfo  {journal} {Phys. Usp.}\ }\textbf {\bibinfo {volume}
  {30}},\ \bibinfo {pages} {551} (\bibinfo {year} {1987})}\BibitemShut
  {NoStop}%
\bibitem [{\citenamefont {Roshchupkin}(1996)}]{26}%
  \BibitemOpen
  \bibfield  {author} {\bibinfo {author} {\bibfnamefont {S.~P.}\ \bibnamefont
  {Roshchupkin}},\ }\href@noop {} {\bibfield  {journal} {\bibinfo  {journal}
  {Laser Physics}\ }\textbf {\bibinfo {volume} {6}},\ \bibinfo {pages} {837}
  (\bibinfo {year} {1996})}\BibitemShut {NoStop}%
\bibitem [{\citenamefont {Roshchupkin}\ \emph {et~al.}(2000)\citenamefont
  {Roshchupkin}, \citenamefont {Tsybul'nik},\ and\ \citenamefont
  {Chmirev}}]{27}%
  \BibitemOpen
  \bibfield  {author} {\bibinfo {author} {\bibfnamefont {S.}~\bibnamefont
  {Roshchupkin}}, \bibinfo {author} {\bibfnamefont {V.}~\bibnamefont
  {Tsybul'nik}}, \ and\ \bibinfo {author} {\bibfnamefont {A.}~\bibnamefont
  {Chmirev}},\ }\href@noop {} {\bibfield  {journal} {\bibinfo  {journal} {Laser
  Physics}\ }\textbf {\bibinfo {volume} {10}},\ \bibinfo {pages} {1256}
  (\bibinfo {year} {2000})}\BibitemShut {NoStop}%
\bibitem [{\citenamefont {Ehlotzky}\ \emph {et~al.}(2009)\citenamefont
  {Ehlotzky}, \citenamefont {Krajewska},\ and\ \citenamefont {Kaminski}}]{28}%
  \BibitemOpen
  \bibfield  {author} {\bibinfo {author} {\bibfnamefont {F.}~\bibnamefont
  {Ehlotzky}}, \bibinfo {author} {\bibfnamefont {K.}~\bibnamefont {Krajewska}},
  \ and\ \bibinfo {author} {\bibfnamefont {J.~Z.}\ \bibnamefont {Kaminski}},\
  }\href {https://api.semanticscholar.org/CorpusID:120788094} {\bibfield
  {journal} {\bibinfo  {journal} {Reports on Progress in Physics}\ }\textbf
  {\bibinfo {volume} {72}},\ \bibinfo {pages} {046401} (\bibinfo {year}
  {2009})}\BibitemShut {NoStop}%
\bibitem [{\citenamefont {Ruffini}\ \emph {et~al.}(2009)\citenamefont
  {Ruffini}, \citenamefont {Vereshchagin},\ and\ \citenamefont {Xue}}]{29}%
  \BibitemOpen
  \bibfield  {author} {\bibinfo {author} {\bibfnamefont {R.}~\bibnamefont
  {Ruffini}}, \bibinfo {author} {\bibfnamefont {G.}~\bibnamefont
  {Vereshchagin}}, \ and\ \bibinfo {author} {\bibfnamefont {S.-S.}\
  \bibnamefont {Xue}},\ }\href {\doibase 10.1016/j.physrep.2009.10.004}
  {\bibfield  {journal} {\bibinfo  {journal} {Physics Reports}\ }\textbf
  {\bibinfo {volume} {487}} (\bibinfo {year} {2009}),\
  10.1016/j.physrep.2009.10.004}\BibitemShut {NoStop}%
\bibitem [{\citenamefont {Piazza}\ \emph {et~al.}(2011)\citenamefont {Piazza},
  \citenamefont {Muller}, \citenamefont {Hatsagortsyan},\ and\ \citenamefont
  {Keitel}}]{30}%
  \BibitemOpen
  \bibfield  {author} {\bibinfo {author} {\bibfnamefont {A.}~\bibnamefont
  {Piazza}}, \bibinfo {author} {\bibfnamefont {C.}~\bibnamefont {Muller}},
  \bibinfo {author} {\bibfnamefont {K.}~\bibnamefont {Hatsagortsyan}}, \ and\
  \bibinfo {author} {\bibfnamefont {C.}~\bibnamefont {Keitel}},\ }\href
  {\doibase 10.1103/RevModPhys.84.1177} {\bibfield  {journal} {\bibinfo
  {journal} {Reviews of Modern Physics}\ }\textbf {\bibinfo {volume} {84}}
  (\bibinfo {year} {2011}),\ 10.1103/RevModPhys.84.1177}\BibitemShut {NoStop}%
\bibitem [{\citenamefont {Roshchupkin}\ \emph {et~al.}(2012)\citenamefont
  {Roshchupkin}, \citenamefont {Lebedev}, \citenamefont {Padusenko},\ and\
  \citenamefont {Voroshilo}}]{31}%
  \BibitemOpen
  \bibfield  {author} {\bibinfo {author} {\bibfnamefont {S.~P.}\ \bibnamefont
  {Roshchupkin}}, \bibinfo {author} {\bibfnamefont {A.~A.}\ \bibnamefont
  {Lebedev}}, \bibinfo {author} {\bibfnamefont {E.~A.}\ \bibnamefont
  {Padusenko}}, \ and\ \bibinfo {author} {\bibfnamefont {A.~I.}\ \bibnamefont
  {Voroshilo}},\ }\href {\doibase 10.1134/s1054660x12060084} {\bibfield
  {journal} {\bibinfo  {journal} {Laser Physics}\ }\textbf {\bibinfo {volume}
  {22}},\ \bibinfo {pages} {1113} (\bibinfo {year} {2012})}\BibitemShut
  {NoStop}%
\bibitem [{\citenamefont {Hartin}(2018)}]{32}%
  \BibitemOpen
  \bibfield  {author} {\bibinfo {author} {\bibfnamefont {A.}~\bibnamefont
  {Hartin}},\ }\href {\doibase 10.1142/s0217751x18300119} {\bibfield  {journal}
  {\bibinfo  {journal} {International Journal of Modern Physics A}\ }\textbf
  {\bibinfo {volume} {33}},\ \bibinfo {pages} {1830011} (\bibinfo {year}
  {2018})}\BibitemShut {NoStop}%
\bibitem [{\citenamefont {Mironov}\ \emph {et~al.}(2020)\citenamefont
  {Mironov}, \citenamefont {Meuren},\ and\ \citenamefont {Fedotov}}]{33}%
  \BibitemOpen
  \bibfield  {author} {\bibinfo {author} {\bibfnamefont {A.}~\bibnamefont
  {Mironov}}, \bibinfo {author} {\bibfnamefont {S.}~\bibnamefont {Meuren}}, \
  and\ \bibinfo {author} {\bibfnamefont {A.}~\bibnamefont {Fedotov}},\ }\href
  {\doibase 10.1103/PhysRevD.102.053005} {\bibfield  {journal} {\bibinfo
  {journal} {Physical Review D}\ }\textbf {\bibinfo {volume} {102}} (\bibinfo
  {year} {2020}),\ 10.1103/PhysRevD.102.053005}\BibitemShut {NoStop}%
\bibitem [{\citenamefont {Gonoskov}\ \emph {et~al.}(2022)\citenamefont
  {Gonoskov}, \citenamefont {Blackburn}, \citenamefont {Marklund},\ and\
  \citenamefont {Bulanov}}]{34}%
  \BibitemOpen
  \bibfield  {author} {\bibinfo {author} {\bibfnamefont {A.}~\bibnamefont
  {Gonoskov}}, \bibinfo {author} {\bibfnamefont {T.~G.}\ \bibnamefont
  {Blackburn}}, \bibinfo {author} {\bibfnamefont {M.}~\bibnamefont {Marklund}},
  \ and\ \bibinfo {author} {\bibfnamefont {S.~S.}\ \bibnamefont {Bulanov}},\
  }\href {\doibase 10.1103/RevModPhys.94.045001} {\bibfield  {journal}
  {\bibinfo  {journal} {Rev. Mod. Phys.}\ }\textbf {\bibinfo {volume} {94}},\
  \bibinfo {pages} {045001} (\bibinfo {year} {2022})}\BibitemShut {NoStop}%
\bibitem [{\citenamefont {Fedotov}\ \emph {et~al.}(2023)\citenamefont
  {Fedotov}, \citenamefont {Ilderton}, \citenamefont {Karbstein}, \citenamefont
  {King}, \citenamefont {Seipt}, \citenamefont {Taya},\ and\ \citenamefont
  {Torgrimsson}}]{35}%
  \BibitemOpen
  \bibfield  {author} {\bibinfo {author} {\bibfnamefont {A.}~\bibnamefont
  {Fedotov}}, \bibinfo {author} {\bibfnamefont {A.}~\bibnamefont {Ilderton}},
  \bibinfo {author} {\bibfnamefont {F.}~\bibnamefont {Karbstein}}, \bibinfo
  {author} {\bibfnamefont {B.}~\bibnamefont {King}}, \bibinfo {author}
  {\bibfnamefont {D.}~\bibnamefont {Seipt}}, \bibinfo {author} {\bibfnamefont
  {H.}~\bibnamefont {Taya}}, \ and\ \bibinfo {author} {\bibfnamefont
  {G.}~\bibnamefont {Torgrimsson}},\ }\href {\doibase
  10.1016/j.physrep.2023.01.003} {\bibfield  {journal} {\bibinfo  {journal}
  {Physics Reports}\ }\textbf {\bibinfo {volume} {1010}},\ \bibinfo {pages} {1}
  (\bibinfo {year} {2023})}\BibitemShut {NoStop}%
\bibitem [{\citenamefont {Roshchupkin}\ and\ \citenamefont
  {Voroshilo}(2008)}]{36}%
  \BibitemOpen
  \bibfield  {author} {\bibinfo {author} {\bibfnamefont {S.~P.}\ \bibnamefont
  {Roshchupkin}}\ and\ \bibinfo {author} {\bibfnamefont {A.}~\bibnamefont
  {Voroshilo}},\ }\href {\doibase 10.13140/2.1.4071.2328} {\emph {\bibinfo
  {title} {Resonant and Coherent Effects of Quantum Electrodynamics in the
  Light Field}}}\ (\bibinfo  {publisher} {Naukova Dumka, Kiev},\ \bibinfo
  {year} {2008})\BibitemShut {NoStop}%
\bibitem [{\citenamefont {Roshchupkin}\ and\ \citenamefont
  {Lebed'}(2013)}]{37}%
  \BibitemOpen
  \bibfield  {author} {\bibinfo {author} {\bibfnamefont {S.~P.}\ \bibnamefont
  {Roshchupkin}}\ and\ \bibinfo {author} {\bibfnamefont {A.}~\bibnamefont
  {Lebed'}},\ }\href@noop {} {\emph {\bibinfo {title} {Effects of Quantum
  Electrodynamics in strong impulse laser fieldsd}}}\ (\bibinfo  {publisher}
  {Naukova Dumka, Kiev},\ \bibinfo {year} {2013})\BibitemShut {NoStop}%
\bibitem [{\citenamefont {Greiner}\ \emph {et~al.}(2012)\citenamefont
  {Greiner}, \citenamefont {M{\"u}ller},\ and\ \citenamefont {Rafelski}}]{38}%
  \BibitemOpen
  \bibfield  {author} {\bibinfo {author} {\bibfnamefont {W.}~\bibnamefont
  {Greiner}}, \bibinfo {author} {\bibfnamefont {B.}~\bibnamefont {M{\"u}ller}},
  \ and\ \bibinfo {author} {\bibfnamefont {J.}~\bibnamefont {Rafelski}},\
  }\href {https://books.google.ru/books?id=Wh3-CAAAQBAJ} {\emph {\bibinfo
  {title} {Quantum Electrodynamics of Strong Fields: With an Introduction into
  Modern Relativistic Quantum Mechanics}}},\ Theoretical and Mathematical
  Physics\ (\bibinfo  {publisher} {Springer Berlin Heidelberg},\ \bibinfo
  {year} {2012})\BibitemShut {NoStop}%
\bibitem [{\citenamefont {Narozhny}\ and\ \citenamefont {Fofanov}(1996)}]{39}%
  \BibitemOpen
  \bibfield  {author} {\bibinfo {author} {\bibfnamefont {N.~B.}\ \bibnamefont
  {Narozhny}}\ and\ \bibinfo {author} {\bibfnamefont {M.~S.}\ \bibnamefont
  {Fofanov}},\ }\enquote {\bibinfo {title} {Photon emission by an electron
  colliding with a short focused laser pulse},}\ in\ \href {\doibase
  10.1007/978-94-009-0261-9_39} {\emph {\bibinfo {booktitle} {Super-Intense
  Laser-Atom Physics IV}}}\ (\bibinfo  {publisher} {Springer Netherlands},\
  \bibinfo {year} {1996})\ pp.\ \bibinfo {pages} {411--420}\BibitemShut
  {NoStop}%
\bibitem [{\citenamefont {Narozhny}\ and\ \citenamefont {Fofanov}(2000)}]{40}%
  \BibitemOpen
  \bibfield  {author} {\bibinfo {author} {\bibfnamefont {N.~B.}\ \bibnamefont
  {Narozhny}}\ and\ \bibinfo {author} {\bibfnamefont {M.~S.}\ \bibnamefont
  {Fofanov}},\ }\href {\doibase 10.1134/1.559160} {\bibfield  {journal}
  {\bibinfo  {journal} {Journal of Experimental and Theoretical Physics}\
  }\textbf {\bibinfo {volume} {90}},\ \bibinfo {pages} {753} (\bibinfo {year}
  {2000})}\BibitemShut {NoStop}%
\bibitem [{\citenamefont {Harvey}\ \emph {et~al.}(2009)\citenamefont {Harvey},
  \citenamefont {Heinzl},\ and\ \citenamefont {Ilderton}}]{41}%
  \BibitemOpen
  \bibfield  {author} {\bibinfo {author} {\bibfnamefont {C.}~\bibnamefont
  {Harvey}}, \bibinfo {author} {\bibfnamefont {T.}~\bibnamefont {Heinzl}}, \
  and\ \bibinfo {author} {\bibfnamefont {A.}~\bibnamefont {Ilderton}},\ }\href
  {\doibase 10.1103/physreva.79.063407} {\bibfield  {journal} {\bibinfo
  {journal} {Physical Review A}\ }\textbf {\bibinfo {volume} {79}} (\bibinfo
  {year} {2009}),\ 10.1103/physreva.79.063407}\BibitemShut {NoStop}%
\bibitem [{\citenamefont {Boca}\ and\ \citenamefont {Florescu}(2009)}]{42}%
  \BibitemOpen
  \bibfield  {author} {\bibinfo {author} {\bibfnamefont {M.}~\bibnamefont
  {Boca}}\ and\ \bibinfo {author} {\bibfnamefont {V.}~\bibnamefont
  {Florescu}},\ }\href {\doibase 10.1103/physreva.80.053403} {\bibfield
  {journal} {\bibinfo  {journal} {Physical Review A}\ }\textbf {\bibinfo
  {volume} {80}} (\bibinfo {year} {2009}),\
  10.1103/physreva.80.053403}\BibitemShut {NoStop}%
\bibitem [{\citenamefont {Mackenroth}\ and\ \citenamefont
  {Di~Piazza}(2011)}]{43}%
  \BibitemOpen
  \bibfield  {author} {\bibinfo {author} {\bibfnamefont {F.}~\bibnamefont
  {Mackenroth}}\ and\ \bibinfo {author} {\bibfnamefont {A.}~\bibnamefont
  {Di~Piazza}},\ }\href {\doibase 10.1103/physreva.83.032106} {\bibfield
  {journal} {\bibinfo  {journal} {Physical Review A}\ }\textbf {\bibinfo
  {volume} {83}} (\bibinfo {year} {2011}),\
  10.1103/physreva.83.032106}\BibitemShut {NoStop}%
\bibitem [{\citenamefont {Seipt}\ and\ \citenamefont {Kampfer}(2011)}]{44}%
  \BibitemOpen
  \bibfield  {author} {\bibinfo {author} {\bibfnamefont {D.}~\bibnamefont
  {Seipt}}\ and\ \bibinfo {author} {\bibfnamefont {B.}~\bibnamefont
  {Kampfer}},\ }\href {\doibase 10.1103/physreva.83.022101} {\bibfield
  {journal} {\bibinfo  {journal} {Physical Review A}\ }\textbf {\bibinfo
  {volume} {83}} (\bibinfo {year} {2011}),\
  10.1103/physreva.83.022101}\BibitemShut {NoStop}%
\bibitem [{\citenamefont {Mackenroth}\ and\ \citenamefont
  {Di~Piazza}(2013)}]{45}%
  \BibitemOpen
  \bibfield  {author} {\bibinfo {author} {\bibfnamefont {F.}~\bibnamefont
  {Mackenroth}}\ and\ \bibinfo {author} {\bibfnamefont {A.}~\bibnamefont
  {Di~Piazza}},\ }\href {\doibase 10.1103/physrevlett.110.070402} {\bibfield
  {journal} {\bibinfo  {journal} {Physical Review Letters}\ }\textbf {\bibinfo
  {volume} {110}} (\bibinfo {year} {2013}),\
  10.1103/physrevlett.110.070402}\BibitemShut {NoStop}%
\bibitem [{\citenamefont {Seipt}\ and\ \citenamefont {Kampfer}(2012)}]{46}%
  \BibitemOpen
  \bibfield  {author} {\bibinfo {author} {\bibfnamefont {D.}~\bibnamefont
  {Seipt}}\ and\ \bibinfo {author} {\bibfnamefont {B.}~\bibnamefont
  {Kampfer}},\ }\href {\doibase 10.1103/physrevd.85.101701} {\bibfield
  {journal} {\bibinfo  {journal} {Physical Review D}\ }\textbf {\bibinfo
  {volume} {85}} (\bibinfo {year} {2012}),\
  10.1103/physrevd.85.101701}\BibitemShut {NoStop}%
\bibitem [{\citenamefont {Boca}\ \emph {et~al.}(2012)\citenamefont {Boca},
  \citenamefont {Dinu},\ and\ \citenamefont {Florescu}}]{47}%
  \BibitemOpen
  \bibfield  {author} {\bibinfo {author} {\bibfnamefont {M.}~\bibnamefont
  {Boca}}, \bibinfo {author} {\bibfnamefont {V.}~\bibnamefont {Dinu}}, \ and\
  \bibinfo {author} {\bibfnamefont {V.}~\bibnamefont {Florescu}},\ }\href
  {\doibase 10.1103/physreva.86.013414} {\bibfield  {journal} {\bibinfo
  {journal} {Physical Review A}\ }\textbf {\bibinfo {volume} {86}} (\bibinfo
  {year} {2012}),\ 10.1103/physreva.86.013414}\BibitemShut {NoStop}%
\bibitem [{\citenamefont {Dai}\ \emph {et~al.}(2023)\citenamefont {Dai},
  \citenamefont {Jiang}, \citenamefont {Jiang}, \citenamefont {Shaisultanov},\
  and\ \citenamefont {Chen}}]{48}%
  \BibitemOpen
  \bibfield  {author} {\bibinfo {author} {\bibfnamefont {Y.-N.}\ \bibnamefont
  {Dai}}, \bibinfo {author} {\bibfnamefont {J.-J.}\ \bibnamefont {Jiang}},
  \bibinfo {author} {\bibfnamefont {Y.-H.}\ \bibnamefont {Jiang}}, \bibinfo
  {author} {\bibfnamefont {R.}~\bibnamefont {Shaisultanov}}, \ and\ \bibinfo
  {author} {\bibfnamefont {Y.-Y.}\ \bibnamefont {Chen}},\ }\href {\doibase
  10.1103/physrevd.108.056025} {\bibfield  {journal} {\bibinfo  {journal}
  {Physical Review D}\ }\textbf {\bibinfo {volume} {108}} (\bibinfo {year}
  {2023}),\ 10.1103/physrevd.108.056025}\BibitemShut {NoStop}%
\bibitem [{\citenamefont {Khalaf}\ and\ \citenamefont {Kaminer}(2023)}]{49}%
  \BibitemOpen
  \bibfield  {author} {\bibinfo {author} {\bibfnamefont {M.}~\bibnamefont
  {Khalaf}}\ and\ \bibinfo {author} {\bibfnamefont {I.}~\bibnamefont
  {Kaminer}},\ }\href {\doibase 10.1126/sciadv.ade0932} {\bibfield  {journal}
  {\bibinfo  {journal} {Science Advances}\ }\textbf {\bibinfo {volume} {9}}
  (\bibinfo {year} {2023}),\ 10.1126/sciadv.ade0932}\BibitemShut {NoStop}%
\bibitem [{\citenamefont {Nishiura}\ and\ \citenamefont {Ioka}(2024)}]{50}%
  \BibitemOpen
  \bibfield  {author} {\bibinfo {author} {\bibfnamefont {R.}~\bibnamefont
  {Nishiura}}\ and\ \bibinfo {author} {\bibfnamefont {K.}~\bibnamefont
  {Ioka}},\ }\href {\doibase 10.1103/physrevd.109.043048} {\bibfield  {journal}
  {\bibinfo  {journal} {Physical Review D}\ }\textbf {\bibinfo {volume} {109}}
  (\bibinfo {year} {2024}),\ 10.1103/physrevd.109.043048}\BibitemShut {NoStop}%
\bibitem [{\citenamefont {Podszus}\ \emph {et~al.}(2023)\citenamefont
  {Podszus}, \citenamefont {Dinu},\ and\ \citenamefont {Di~Piazza}}]{51}%
  \BibitemOpen
  \bibfield  {author} {\bibinfo {author} {\bibfnamefont {T.}~\bibnamefont
  {Podszus}}, \bibinfo {author} {\bibfnamefont {V.}~\bibnamefont {Dinu}}, \
  and\ \bibinfo {author} {\bibfnamefont {A.}~\bibnamefont {Di~Piazza}},\ }in\
  \href {\doibase 10.1117/12.2671101} {\emph {\bibinfo {booktitle} {Research
  Using Extreme Light: Entering New Frontiers with Petawatt-Class Lasers V}}},\
  \bibinfo {editor} {edited by\ \bibinfo {editor} {\bibfnamefont {S.~V.}\
  \bibnamefont {Bulanov}}\ and\ \bibinfo {editor} {\bibfnamefont {L.~O.}\
  \bibnamefont {Silva}}}\ (\bibinfo  {publisher} {SPIE},\ \bibinfo {year}
  {2023})\ p.~\bibinfo {pages} {3}\BibitemShut {NoStop}%
\bibitem [{\citenamefont {Titov}(2024)}]{52}%
  \BibitemOpen
  \bibfield  {author} {\bibinfo {author} {\bibfnamefont {A.~I.}\ \bibnamefont
  {Titov}},\ }\href {\doibase 10.1134/s1063779624700588} {\bibfield  {journal}
  {\bibinfo  {journal} {Physics of Particles and Nuclei}\ }\textbf {\bibinfo
  {volume} {55}},\ \bibinfo {pages} {920} (\bibinfo {year} {2024})}\BibitemShut
  {NoStop}%
\bibitem [{\citenamefont {Li}\ \emph {et~al.}(2023)\citenamefont {Li},
  \citenamefont {Chen}, \citenamefont {Hatsagortsyan}, \citenamefont
  {Di~Piazza}, \citenamefont {Tamburini},\ and\ \citenamefont {Keitel}}]{53}%
  \BibitemOpen
  \bibfield  {author} {\bibinfo {author} {\bibfnamefont {Y.-F.}\ \bibnamefont
  {Li}}, \bibinfo {author} {\bibfnamefont {Y.-Y.}\ \bibnamefont {Chen}},
  \bibinfo {author} {\bibfnamefont {K.~Z.}\ \bibnamefont {Hatsagortsyan}},
  \bibinfo {author} {\bibfnamefont {A.}~\bibnamefont {Di~Piazza}}, \bibinfo
  {author} {\bibfnamefont {M.}~\bibnamefont {Tamburini}}, \ and\ \bibinfo
  {author} {\bibfnamefont {C.~H.}\ \bibnamefont {Keitel}},\ }\href {\doibase
  10.1103/physrevd.107.116020} {\bibfield  {journal} {\bibinfo  {journal}
  {Physical Review D}\ }\textbf {\bibinfo {volume} {107}} (\bibinfo {year}
  {2023}),\ 10.1103/physrevd.107.116020}\BibitemShut {NoStop}%
\bibitem [{\citenamefont {Gelfer}\ \emph {et~al.}(2022)\citenamefont {Gelfer},
  \citenamefont {Fedotov}, \citenamefont {Mironov},\ and\ \citenamefont
  {Weber}}]{54}%
  \BibitemOpen
  \bibfield  {author} {\bibinfo {author} {\bibfnamefont {E.~G.}\ \bibnamefont
  {Gelfer}}, \bibinfo {author} {\bibfnamefont {A.~M.}\ \bibnamefont {Fedotov}},
  \bibinfo {author} {\bibfnamefont {A.~A.}\ \bibnamefont {Mironov}}, \ and\
  \bibinfo {author} {\bibfnamefont {S.}~\bibnamefont {Weber}},\ }\href
  {\doibase 10.1103/physrevd.106.056013} {\bibfield  {journal} {\bibinfo
  {journal} {Physical Review D}\ }\textbf {\bibinfo {volume} {106}} (\bibinfo
  {year} {2022}),\ 10.1103/physrevd.106.056013}\BibitemShut {NoStop}%
\bibitem [{\citenamefont {Song}\ \emph {et~al.}(2024)\citenamefont {Song},
  \citenamefont {Wang}, \citenamefont {Chen},\ and\ \citenamefont
  {Sheng}}]{55}%
  \BibitemOpen
  \bibfield  {author} {\bibinfo {author} {\bibfnamefont {H.-H.}\ \bibnamefont
  {Song}}, \bibinfo {author} {\bibfnamefont {W.-M.}\ \bibnamefont {Wang}},
  \bibinfo {author} {\bibfnamefont {M.}~\bibnamefont {Chen}}, \ and\ \bibinfo
  {author} {\bibfnamefont {Z.-M.}\ \bibnamefont {Sheng}},\ }\href {\doibase
  10.1103/physreve.109.035204} {\bibfield  {journal} {\bibinfo  {journal}
  {Physical Review E}\ }\textbf {\bibinfo {volume} {109}} (\bibinfo {year}
  {2024}),\ 10.1103/physreve.109.035204}\BibitemShut {NoStop}%
\bibitem [{\citenamefont {Gao}\ and\ \citenamefont {Tang}(2022)}]{56}%
  \BibitemOpen
  \bibfield  {author} {\bibinfo {author} {\bibfnamefont {Y.}~\bibnamefont
  {Gao}}\ and\ \bibinfo {author} {\bibfnamefont {S.}~\bibnamefont {Tang}},\
  }\href {\doibase 10.1103/physrevd.106.056003} {\bibfield  {journal} {\bibinfo
   {journal} {Physical Review D}\ }\textbf {\bibinfo {volume} {106}} (\bibinfo
  {year} {2022}),\ 10.1103/physrevd.106.056003}\BibitemShut {NoStop}%
\bibitem [{\citenamefont {Golub}\ \emph {et~al.}(2021)\citenamefont {Golub},
  \citenamefont {Villalba-Chavez},\ and\ \citenamefont {Muller}}]{57}%
  \BibitemOpen
  \bibfield  {author} {\bibinfo {author} {\bibfnamefont {A.}~\bibnamefont
  {Golub}}, \bibinfo {author} {\bibfnamefont {S.}~\bibnamefont
  {Villalba-Chavez}}, \ and\ \bibinfo {author} {\bibfnamefont {C.}~\bibnamefont
  {Muller}},\ }\href {\doibase 10.1103/PhysRevD.103.096002} {\bibfield
  {journal} {\bibinfo  {journal} {Physical Review D}\ }\textbf {\bibinfo
  {volume} {103}} (\bibinfo {year} {2021}),\
  10.1103/PhysRevD.103.096002}\BibitemShut {NoStop}%
\bibitem [{\citenamefont {Tang}\ and\ \citenamefont {King}(2021)}]{58}%
  \BibitemOpen
  \bibfield  {author} {\bibinfo {author} {\bibfnamefont {S.}~\bibnamefont
  {Tang}}\ and\ \bibinfo {author} {\bibfnamefont {B.}~\bibnamefont {King}},\
  }\href {\doibase 10.1103/physrevd.104.096019} {\bibfield  {journal} {\bibinfo
   {journal} {Physical Review D}\ }\textbf {\bibinfo {volume} {104}} (\bibinfo
  {year} {2021}),\ 10.1103/physrevd.104.096019}\BibitemShut {NoStop}%
\bibitem [{\citenamefont {Degli~Esposti}\ and\ \citenamefont
  {Torgrimsson}(2022)}]{59}%
  \BibitemOpen
  \bibfield  {author} {\bibinfo {author} {\bibfnamefont {G.}~\bibnamefont
  {Degli~Esposti}}\ and\ \bibinfo {author} {\bibfnamefont {G.}~\bibnamefont
  {Torgrimsson}},\ }\href {\doibase 10.1103/physrevd.105.096036} {\bibfield
  {journal} {\bibinfo  {journal} {Physical Review D}\ }\textbf {\bibinfo
  {volume} {105}} (\bibinfo {year} {2022}),\
  10.1103/physrevd.105.096036}\BibitemShut {NoStop}%
\bibitem [{\citenamefont {Blackburn}\ and\ \citenamefont {King}(2022)}]{60}%
  \BibitemOpen
  \bibfield  {author} {\bibinfo {author} {\bibfnamefont {T.~G.}\ \bibnamefont
  {Blackburn}}\ and\ \bibinfo {author} {\bibfnamefont {B.}~\bibnamefont
  {King}},\ }\href {\doibase 10.1140/epjc/s10052-021-09955-3} {\bibfield
  {journal} {\bibinfo  {journal} {The European Physical Journal C}\ }\textbf
  {\bibinfo {volume} {82}} (\bibinfo {year} {2022}),\
  10.1140/epjc/s10052-021-09955-3}\BibitemShut {NoStop}%
\bibitem [{\citenamefont {Dai}\ \emph {et~al.}(2021)\citenamefont {Dai},
  \citenamefont {Shen}, \citenamefont {Li}, \citenamefont {Shaisultanov},
  \citenamefont {Hatsagortsyan}, \citenamefont {Keitel},\ and\ \citenamefont
  {Chen}}]{61}%
  \BibitemOpen
  \bibfield  {author} {\bibinfo {author} {\bibfnamefont {Y.-N.}\ \bibnamefont
  {Dai}}, \bibinfo {author} {\bibfnamefont {B.-F.}\ \bibnamefont {Shen}},
  \bibinfo {author} {\bibfnamefont {J.-X.}\ \bibnamefont {Li}}, \bibinfo
  {author} {\bibfnamefont {R.}~\bibnamefont {Shaisultanov}}, \bibinfo {author}
  {\bibfnamefont {K.~Z.}\ \bibnamefont {Hatsagortsyan}}, \bibinfo {author}
  {\bibfnamefont {C.~H.}\ \bibnamefont {Keitel}}, \ and\ \bibinfo {author}
  {\bibfnamefont {Y.-Y.}\ \bibnamefont {Chen}},\ }\href {\doibase
  10.1063/5.0063633} {\bibfield  {journal} {\bibinfo  {journal} {Matter and
  Radiation at Extremes}\ }\textbf {\bibinfo {volume} {7}} (\bibinfo {year}
  {2021}),\ 10.1063/5.0063633}\BibitemShut {NoStop}%
\bibitem [{\citenamefont {Hafizi}\ \emph {et~al.}(2019)\citenamefont {Hafizi},
  \citenamefont {Gordon},\ and\ \citenamefont {Kaganovich}}]{62}%
  \BibitemOpen
  \bibfield  {author} {\bibinfo {author} {\bibfnamefont {B.}~\bibnamefont
  {Hafizi}}, \bibinfo {author} {\bibfnamefont {D.~F.}\ \bibnamefont {Gordon}},
  \ and\ \bibinfo {author} {\bibfnamefont {D.}~\bibnamefont {Kaganovich}},\
  }\href {\doibase 10.1103/physrevlett.122.233201} {\bibfield  {journal}
  {\bibinfo  {journal} {Physical Review Letters}\ }\textbf {\bibinfo {volume}
  {122}} (\bibinfo {year} {2019}),\ 10.1103/physrevlett.122.233201}\BibitemShut
  {NoStop}%
\bibitem [{\citenamefont {Oleinik}(1967)}]{63}%
  \BibitemOpen
  \bibfield  {author} {\bibinfo {author} {\bibfnamefont {V.~P.}\ \bibnamefont
  {Oleinik}},\ }\href {https://api.semanticscholar.org/CorpusID:117337427}
  {\bibfield  {journal} {\bibinfo  {journal} {Journal of Experimental and
  Theoretical Physics}\ }\textbf {\bibinfo {volume} {25}},\ \bibinfo {pages}
  {697} (\bibinfo {year} {1967})}\BibitemShut {NoStop}%
\bibitem [{\citenamefont {Oleinik}(1968)}]{64}%
  \BibitemOpen
  \bibfield  {author} {\bibinfo {author} {\bibfnamefont {V.~P.}\ \bibnamefont
  {Oleinik}},\ }\href {https://api.semanticscholar.org/CorpusID:117337427}
  {\bibfield  {journal} {\bibinfo  {journal} {Journal of Experimental and
  Theoretical Physics}\ }\textbf {\bibinfo {volume} {26}},\ \bibinfo {pages}
  {1132} (\bibinfo {year} {1968})}\BibitemShut {NoStop}%
\bibitem [{\citenamefont {Sizykh}\ \emph {et~al.}(2021)\citenamefont {Sizykh},
  \citenamefont {Roshchupkin},\ and\ \citenamefont {Dubov}}]{65}%
  \BibitemOpen
  \bibfield  {author} {\bibinfo {author} {\bibfnamefont {G.~K.}\ \bibnamefont
  {Sizykh}}, \bibinfo {author} {\bibfnamefont {S.~P.}\ \bibnamefont
  {Roshchupkin}}, \ and\ \bibinfo {author} {\bibfnamefont {V.~V.}\ \bibnamefont
  {Dubov}},\ }\href {\doibase 10.3390/universe7070210} {\bibfield  {journal}
  {\bibinfo  {journal} {Universe}\ }\textbf {\bibinfo {volume} {7}},\ \bibinfo
  {pages} {210} (\bibinfo {year} {2021})}\BibitemShut {NoStop}%
\bibitem [{\citenamefont {Roshchupkin}\ \emph {et~al.}(2021)\citenamefont
  {Roshchupkin}, \citenamefont {Larin},\ and\ \citenamefont {Dubov}}]{66}%
  \BibitemOpen
  \bibfield  {author} {\bibinfo {author} {\bibfnamefont {S.~P.}\ \bibnamefont
  {Roshchupkin}}, \bibinfo {author} {\bibfnamefont {N.~R.}\ \bibnamefont
  {Larin}}, \ and\ \bibinfo {author} {\bibfnamefont {V.~V.}\ \bibnamefont
  {Dubov}},\ }\href {\doibase 10.1103/physrevd.104.116011} {\bibfield
  {journal} {\bibinfo  {journal} {Physical Review D}\ }\textbf {\bibinfo
  {volume} {104}} (\bibinfo {year} {2021}),\
  10.1103/physrevd.104.116011}\BibitemShut {NoStop}%
\bibitem [{\citenamefont {Roshchupkin}\ \emph {et~al.}(2022)\citenamefont
  {Roshchupkin}, \citenamefont {Dubov}, \citenamefont {Dubov},\ and\
  \citenamefont {Starodub}}]{67}%
  \BibitemOpen
  \bibfield  {author} {\bibinfo {author} {\bibfnamefont {S.~P.}\ \bibnamefont
  {Roshchupkin}}, \bibinfo {author} {\bibfnamefont {A.~V.}\ \bibnamefont
  {Dubov}}, \bibinfo {author} {\bibfnamefont {V.~V.}\ \bibnamefont {Dubov}}, \
  and\ \bibinfo {author} {\bibfnamefont {S.~S.}\ \bibnamefont {Starodub}},\
  }\href {\doibase 10.1088/1367-2630/ac46e3} {\bibfield  {journal} {\bibinfo
  {journal} {New Journal of Physics}\ }\textbf {\bibinfo {volume} {24}},\
  \bibinfo {pages} {013020} (\bibinfo {year} {2022})}\BibitemShut {NoStop}%
\bibitem [{\citenamefont {Roshchupkin}\ \emph {et~al.}(2023)\citenamefont
  {Roshchupkin}, \citenamefont {Serov},\ and\ \citenamefont {Dubov}}]{68}%
  \BibitemOpen
  \bibfield  {author} {\bibinfo {author} {\bibfnamefont {S.~P.}\ \bibnamefont
  {Roshchupkin}}, \bibinfo {author} {\bibfnamefont {V.~D.}\ \bibnamefont
  {Serov}}, \ and\ \bibinfo {author} {\bibfnamefont {V.~V.}\ \bibnamefont
  {Dubov}},\ }\href {\doibase 10.3390/photonics10080949} {\bibfield  {journal}
  {\bibinfo  {journal} {Photonics}\ }\textbf {\bibinfo {volume} {10}},\
  \bibinfo {pages} {949} (\bibinfo {year} {2023})}\BibitemShut {NoStop}%
\bibitem [{\citenamefont {Roshchupkin}\ and\ \citenamefont
  {Makarov}(2024)}]{69}%
  \BibitemOpen
  \bibfield  {author} {\bibinfo {author} {\bibfnamefont {S.~P.}\ \bibnamefont
  {Roshchupkin}}\ and\ \bibinfo {author} {\bibfnamefont {S.~B.}\ \bibnamefont
  {Makarov}},\ }\href {\doibase 10.3390/photonics11070597} {\bibfield
  {journal} {\bibinfo  {journal} {Photonics}\ }\textbf {\bibinfo {volume}
  {11}},\ \bibinfo {pages} {597} (\bibinfo {year} {2024})}\BibitemShut
  {NoStop}%
\bibitem [{\citenamefont {King}\ and\ \citenamefont {Ruhl}(2013)}]{70}%
  \BibitemOpen
  \bibfield  {author} {\bibinfo {author} {\bibfnamefont {B.}~\bibnamefont
  {King}}\ and\ \bibinfo {author} {\bibfnamefont {H.}~\bibnamefont {Ruhl}},\
  }\href {\doibase 10.1103/physrevd.88.013005} {\bibfield  {journal} {\bibinfo
  {journal} {Physical Review D}\ }\textbf {\bibinfo {volume} {88}} (\bibinfo
  {year} {2013}),\ 10.1103/physrevd.88.013005}\BibitemShut {NoStop}%
\bibitem [{\citenamefont {Tang}\ and\ \citenamefont {King}(2023)}]{71}%
  \BibitemOpen
  \bibfield  {author} {\bibinfo {author} {\bibfnamefont {S.}~\bibnamefont
  {Tang}}\ and\ \bibinfo {author} {\bibfnamefont {B.}~\bibnamefont {King}},\
  }\href {\doibase 10.1103/physrevd.107.096004} {\bibfield  {journal} {\bibinfo
   {journal} {Physical Review D}\ }\textbf {\bibinfo {volume} {107}} (\bibinfo
  {year} {2023}),\ 10.1103/physrevd.107.096004}\BibitemShut {NoStop}%
\bibitem [{\citenamefont {Hu}\ \emph {et~al.}(2010)\citenamefont {Hu},
  \citenamefont {Muller},\ and\ \citenamefont {Keitel}}]{72}%
  \BibitemOpen
  \bibfield  {author} {\bibinfo {author} {\bibfnamefont {H.}~\bibnamefont
  {Hu}}, \bibinfo {author} {\bibfnamefont {C.}~\bibnamefont {Muller}}, \ and\
  \bibinfo {author} {\bibfnamefont {C.~H.}\ \bibnamefont {Keitel}},\ }\href
  {\doibase 10.1103/physrevlett.105.080401} {\bibfield  {journal} {\bibinfo
  {journal} {Physical Review Letters}\ }\textbf {\bibinfo {volume} {105}}
  (\bibinfo {year} {2010}),\ 10.1103/physrevlett.105.080401}\BibitemShut
  {NoStop}%
\bibitem [{\citenamefont {Dinu}\ and\ \citenamefont {Torgrimsson}(2018)}]{73}%
  \BibitemOpen
  \bibfield  {author} {\bibinfo {author} {\bibfnamefont {V.}~\bibnamefont
  {Dinu}}\ and\ \bibinfo {author} {\bibfnamefont {G.}~\bibnamefont
  {Torgrimsson}},\ }\href {\doibase 10.1103/physrevd.97.036021} {\bibfield
  {journal} {\bibinfo  {journal} {Physical Review D}\ }\textbf {\bibinfo
  {volume} {97}} (\bibinfo {year} {2018}),\
  10.1103/physrevd.97.036021}\BibitemShut {NoStop}%
\bibitem [{\citenamefont {Ilderton}(2011)}]{74}%
  \BibitemOpen
  \bibfield  {author} {\bibinfo {author} {\bibfnamefont {A.}~\bibnamefont
  {Ilderton}},\ }\href {\doibase 10.1103/physrevlett.106.020404} {\bibfield
  {journal} {\bibinfo  {journal} {Physical Review Letters}\ }\textbf {\bibinfo
  {volume} {106}} (\bibinfo {year} {2011}),\
  10.1103/physrevlett.106.020404}\BibitemShut {NoStop}%
\bibitem [{\citenamefont {King}\ and\ \citenamefont {Fedotov}(2018)}]{75}%
  \BibitemOpen
  \bibfield  {author} {\bibinfo {author} {\bibfnamefont {B.}~\bibnamefont
  {King}}\ and\ \bibinfo {author} {\bibfnamefont {A.~M.}\ \bibnamefont
  {Fedotov}},\ }\href {\doibase 10.1103/physrevd.98.016005} {\bibfield
  {journal} {\bibinfo  {journal} {Physical Review D}\ }\textbf {\bibinfo
  {volume} {98}} (\bibinfo {year} {2018}),\
  10.1103/physrevd.98.016005}\BibitemShut {NoStop}%
\bibitem [{\citenamefont {Mackenroth}\ and\ \citenamefont
  {Di~Piazza}(2018)}]{76}%
  \BibitemOpen
  \bibfield  {author} {\bibinfo {author} {\bibfnamefont {F.}~\bibnamefont
  {Mackenroth}}\ and\ \bibinfo {author} {\bibfnamefont {A.}~\bibnamefont
  {Di~Piazza}},\ }\href {\doibase 10.1103/physrevd.98.116002} {\bibfield
  {journal} {\bibinfo  {journal} {Physical Review D}\ }\textbf {\bibinfo
  {volume} {98}} (\bibinfo {year} {2018}),\
  10.1103/physrevd.98.116002}\BibitemShut {NoStop}%
\bibitem [{\citenamefont {Yaghjian}(2021)}]{77}%
  \BibitemOpen
  \bibfield  {author} {\bibinfo {author} {\bibfnamefont {A.~D.}\ \bibnamefont
  {Yaghjian}},\ }\href {\doibase 10.1103/physrevaccelbeams.24.114002}
  {\bibfield  {journal} {\bibinfo  {journal} {Physical Review Accelerators and
  Beams}\ }\textbf {\bibinfo {volume} {24}} (\bibinfo {year} {2021}),\
  10.1103/physrevaccelbeams.24.114002}\BibitemShut {NoStop}%
\bibitem [{\citenamefont {Lv}\ \emph {et~al.}(2022)\citenamefont {Lv},
  \citenamefont {Raicher}, \citenamefont {Keitel},\ and\ \citenamefont
  {Hatsagortsyan}}]{78}%
  \BibitemOpen
  \bibfield  {author} {\bibinfo {author} {\bibfnamefont {Q.~Z.}\ \bibnamefont
  {Lv}}, \bibinfo {author} {\bibfnamefont {E.}~\bibnamefont {Raicher}},
  \bibinfo {author} {\bibfnamefont {C.}~\bibnamefont {Keitel}}, \ and\ \bibinfo
  {author} {\bibfnamefont {K.~Z.}\ \bibnamefont {Hatsagortsyan}},\ }\href
  {\doibase 10.1103/physrevlett.128.024801} {\bibfield  {journal} {\bibinfo
  {journal} {Physical Review Letters}\ }\textbf {\bibinfo {volume} {128}}
  (\bibinfo {year} {2022}),\ 10.1103/physrevlett.128.024801}\BibitemShut
  {NoStop}%
\bibitem [{\citenamefont {Schulze}\ \emph {et~al.}(2022)\citenamefont
  {Schulze}, \citenamefont {Grabiger}, \citenamefont {Loetzsch}, \citenamefont
  {Marx-Glowna}, \citenamefont {Schmitt}, \citenamefont {Garcia}, \citenamefont
  {Hippler}, \citenamefont {Huang}, \citenamefont {Karbstein}, \citenamefont
  {Kon\^opkov\'a}, \citenamefont {Schlenvoigt}, \citenamefont {Schwinkendorf},
  \citenamefont {Strohm}, \citenamefont {Toncian}, \citenamefont {Uschmann},
  \citenamefont {Wille}, \citenamefont {Zastrau}, \citenamefont
  {R\"ohlsberger}, \citenamefont {St\"ohlker}, \citenamefont {Cowan},\ and\
  \citenamefont {Paulus}}]{79}%
  \BibitemOpen
  \bibfield  {author} {\bibinfo {author} {\bibfnamefont {K.~S.}\ \bibnamefont
  {Schulze}}, \bibinfo {author} {\bibfnamefont {B.}~\bibnamefont {Grabiger}},
  \bibinfo {author} {\bibfnamefont {R.}~\bibnamefont {Loetzsch}}, \bibinfo
  {author} {\bibfnamefont {B.}~\bibnamefont {Marx-Glowna}}, \bibinfo {author}
  {\bibfnamefont {A.~T.}\ \bibnamefont {Schmitt}}, \bibinfo {author}
  {\bibfnamefont {A.~L.}\ \bibnamefont {Garcia}}, \bibinfo {author}
  {\bibfnamefont {W.}~\bibnamefont {Hippler}}, \bibinfo {author} {\bibfnamefont
  {L.}~\bibnamefont {Huang}}, \bibinfo {author} {\bibfnamefont
  {F.}~\bibnamefont {Karbstein}}, \bibinfo {author} {\bibfnamefont
  {Z.}~\bibnamefont {Kon\^opkov\'a}}, \bibinfo {author} {\bibfnamefont {H.-P.}\
  \bibnamefont {Schlenvoigt}}, \bibinfo {author} {\bibfnamefont {J.-P.}\
  \bibnamefont {Schwinkendorf}}, \bibinfo {author} {\bibfnamefont
  {C.}~\bibnamefont {Strohm}}, \bibinfo {author} {\bibfnamefont
  {T.}~\bibnamefont {Toncian}}, \bibinfo {author} {\bibfnamefont
  {I.}~\bibnamefont {Uschmann}}, \bibinfo {author} {\bibfnamefont {H.-C.}\
  \bibnamefont {Wille}}, \bibinfo {author} {\bibfnamefont {U.}~\bibnamefont
  {Zastrau}}, \bibinfo {author} {\bibfnamefont {R.}~\bibnamefont
  {R\"ohlsberger}}, \bibinfo {author} {\bibfnamefont {T.}~\bibnamefont
  {St\"ohlker}}, \bibinfo {author} {\bibfnamefont {T.~E.}\ \bibnamefont
  {Cowan}}, \ and\ \bibinfo {author} {\bibfnamefont {G.~G.}\ \bibnamefont
  {Paulus}},\ }\href {\doibase 10.1103/PhysRevResearch.4.013220} {\bibfield
  {journal} {\bibinfo  {journal} {Phys. Rev. Res.}\ }\textbf {\bibinfo {volume}
  {4}},\ \bibinfo {pages} {013220} (\bibinfo {year} {2022})}\BibitemShut
  {NoStop}%
\bibitem [{\citenamefont {Salgado}\ \emph {et~al.}(2022)\citenamefont
  {Salgado}, \citenamefont {Cavanagh}, \citenamefont {Tamburini}, \citenamefont
  {Storey}, \citenamefont {Beyer}, \citenamefont {Bucksbaum}, \citenamefont
  {Chen}, \citenamefont {Piazza}, \citenamefont {Gerstmayr}, \citenamefont
  {Harsh}, \citenamefont {Isele}, \citenamefont {Junghans}, \citenamefont
  {Keitel}, \citenamefont {Kuschel}, \citenamefont {Nielsen}, \citenamefont
  {Reis}, \citenamefont {Roedel}, \citenamefont {Sarri}, \citenamefont
  {Seidel},\ and\ \citenamefont {Zepf}}]{80}%
  \BibitemOpen
  \bibfield  {author} {\bibinfo {author} {\bibfnamefont {F.}~\bibnamefont
  {Salgado}}, \bibinfo {author} {\bibfnamefont {N.}~\bibnamefont {Cavanagh}},
  \bibinfo {author} {\bibfnamefont {M.}~\bibnamefont {Tamburini}}, \bibinfo
  {author} {\bibfnamefont {D.}~\bibnamefont {Storey}}, \bibinfo {author}
  {\bibfnamefont {R.}~\bibnamefont {Beyer}}, \bibinfo {author} {\bibfnamefont
  {P.}~\bibnamefont {Bucksbaum}}, \bibinfo {author} {\bibfnamefont
  {Z.}~\bibnamefont {Chen}}, \bibinfo {author} {\bibfnamefont {A.}~\bibnamefont
  {Piazza}}, \bibinfo {author} {\bibfnamefont {E.}~\bibnamefont {Gerstmayr}},
  \bibinfo {author} {\bibnamefont {Harsh}}, \bibinfo {author} {\bibfnamefont
  {E.}~\bibnamefont {Isele}}, \bibinfo {author} {\bibfnamefont
  {A.}~\bibnamefont {Junghans}}, \bibinfo {author} {\bibfnamefont
  {C.}~\bibnamefont {Keitel}}, \bibinfo {author} {\bibfnamefont
  {S.}~\bibnamefont {Kuschel}}, \bibinfo {author} {\bibfnamefont
  {C.}~\bibnamefont {Nielsen}}, \bibinfo {author} {\bibfnamefont
  {D.}~\bibnamefont {Reis}}, \bibinfo {author} {\bibfnamefont {C.}~\bibnamefont
  {Roedel}}, \bibinfo {author} {\bibfnamefont {G.}~\bibnamefont {Sarri}},
  \bibinfo {author} {\bibfnamefont {A.}~\bibnamefont {Seidel}}, \ and\ \bibinfo
  {author} {\bibfnamefont {M.}~\bibnamefont {Zepf}},\ }\href {\doibase
  10.1088/1367-2630/ac4283} {\bibfield  {journal} {\bibinfo  {journal} {New
  Journal of Physics}\ }\textbf {\bibinfo {volume} {24}} (\bibinfo {year}
  {2022}),\ 10.1088/1367-2630/ac4283}\BibitemShut {NoStop}%
\bibitem [{\citenamefont {Borneis}\ \emph {et~al.}(2021)\citenamefont
  {Borneis}, \citenamefont {Lastovicka}, \citenamefont {Sokol}, \citenamefont
  {Jeong}, \citenamefont {Condamine}, \citenamefont {Renner}, \citenamefont
  {Bohlin}, \citenamefont {Fajstavr}, \citenamefont {Hernandez}, \citenamefont
  {Jourdain}, \citenamefont {Kumar}, \citenamefont {Modransky}, \citenamefont
  {Pokorny}, \citenamefont {Wolf}, \citenamefont {Zhai}, \citenamefont {Korn},\
  and\ \citenamefont {Weber}}]{81}%
  \BibitemOpen
  \bibfield  {author} {\bibinfo {author} {\bibfnamefont {S.}~\bibnamefont
  {Borneis}}, \bibinfo {author} {\bibfnamefont {T.}~\bibnamefont {Lastovicka}},
  \bibinfo {author} {\bibfnamefont {M.}~\bibnamefont {Sokol}}, \bibinfo
  {author} {\bibfnamefont {T.-M.}\ \bibnamefont {Jeong}}, \bibinfo {author}
  {\bibfnamefont {F.}~\bibnamefont {Condamine}}, \bibinfo {author}
  {\bibfnamefont {O.}~\bibnamefont {Renner}}, \bibinfo {author} {\bibfnamefont
  {H.}~\bibnamefont {Bohlin}}, \bibinfo {author} {\bibfnamefont
  {A.}~\bibnamefont {Fajstavr}}, \bibinfo {author} {\bibfnamefont {J.-C.}\
  \bibnamefont {Hernandez}}, \bibinfo {author} {\bibfnamefont {N.}~\bibnamefont
  {Jourdain}}, \bibinfo {author} {\bibfnamefont {D.}~\bibnamefont {Kumar}},
  \bibinfo {author} {\bibfnamefont {D.}~\bibnamefont {Modransky}}, \bibinfo
  {author} {\bibfnamefont {A.}~\bibnamefont {Pokorny}}, \bibinfo {author}
  {\bibfnamefont {A.}~\bibnamefont {Wolf}}, \bibinfo {author} {\bibfnamefont
  {S.}~\bibnamefont {Zhai}}, \bibinfo {author} {\bibfnamefont {G.}~\bibnamefont
  {Korn}}, \ and\ \bibinfo {author} {\bibfnamefont {S.}~\bibnamefont {Weber}},\
  }\href {\doibase 10.1017/hpl.2021.16} {\bibfield  {journal} {\bibinfo
  {journal} {High Power Laser Science and Engineering}\ }\textbf {\bibinfo
  {volume} {9}} (\bibinfo {year} {2021}),\ 10.1017/hpl.2021.16}\BibitemShut
  {NoStop}%
\bibitem [{\citenamefont {Di~Piazza}\ and\ \citenamefont
  {Fronimos}(2022)}]{82}%
  \BibitemOpen
  \bibfield  {author} {\bibinfo {author} {\bibfnamefont {A.}~\bibnamefont
  {Di~Piazza}}\ and\ \bibinfo {author} {\bibfnamefont {F.~P.}\ \bibnamefont
  {Fronimos}},\ }\href {\doibase 10.1103/physrevd.105.116019} {\bibfield
  {journal} {\bibinfo  {journal} {Physical Review D}\ }\textbf {\bibinfo
  {volume} {105}} (\bibinfo {year} {2022}),\
  10.1103/physrevd.105.116019}\BibitemShut {NoStop}%
\bibitem [{\citenamefont {Jirka}\ \emph {et~al.}(2022)\citenamefont {Jirka},
  \citenamefont {Sasorov},\ and\ \citenamefont {Bulanov}}]{83}%
  \BibitemOpen
  \bibfield  {author} {\bibinfo {author} {\bibfnamefont {M.}~\bibnamefont
  {Jirka}}, \bibinfo {author} {\bibfnamefont {P.}~\bibnamefont {Sasorov}}, \
  and\ \bibinfo {author} {\bibfnamefont {S.~V.}\ \bibnamefont {Bulanov}},\
  }\href {\doibase 10.1103/physrevd.105.113004} {\bibfield  {journal} {\bibinfo
   {journal} {Physical Review D}\ }\textbf {\bibinfo {volume} {105}} (\bibinfo
  {year} {2022}),\ 10.1103/physrevd.105.113004}\BibitemShut {NoStop}%
\bibitem [{\citenamefont {Di~Piazza}\ and\ \citenamefont
  {Patuleanu}(2021)}]{84}%
  \BibitemOpen
  \bibfield  {author} {\bibinfo {author} {\bibfnamefont {A.}~\bibnamefont
  {Di~Piazza}}\ and\ \bibinfo {author} {\bibfnamefont {T.}~\bibnamefont
  {Patuleanu}},\ }\href {\doibase 10.1103/physrevd.104.076003} {\bibfield
  {journal} {\bibinfo  {journal} {Physical Review D}\ }\textbf {\bibinfo
  {volume} {104}} (\bibinfo {year} {2021}),\
  10.1103/physrevd.104.076003}\BibitemShut {NoStop}%
\bibitem [{\citenamefont {Muller}\ \emph {et~al.}(2003)\citenamefont {Muller},
  \citenamefont {Voitkiv},\ and\ \citenamefont {Grun}}]{85}%
  \BibitemOpen
  \bibfield  {author} {\bibinfo {author} {\bibfnamefont {C.}~\bibnamefont
  {Muller}}, \bibinfo {author} {\bibfnamefont {A.~B.}\ \bibnamefont {Voitkiv}},
  \ and\ \bibinfo {author} {\bibfnamefont {N.}~\bibnamefont {Grun}},\ }\href
  {\doibase 10.1103/physreva.67.063407} {\bibfield  {journal} {\bibinfo
  {journal} {Physical Review A}\ }\textbf {\bibinfo {volume} {67}} (\bibinfo
  {year} {2003}),\ 10.1103/physreva.67.063407}\BibitemShut {NoStop}%
\bibitem [{\citenamefont {Krachkov}\ \emph {et~al.}(2019)\citenamefont
  {Krachkov}, \citenamefont {Di~Piazza},\ and\ \citenamefont {Milstein}}]{86}%
  \BibitemOpen
  \bibfield  {author} {\bibinfo {author} {\bibfnamefont {P.}~\bibnamefont
  {Krachkov}}, \bibinfo {author} {\bibfnamefont {A.}~\bibnamefont {Di~Piazza}},
  \ and\ \bibinfo {author} {\bibfnamefont {A.}~\bibnamefont {Milstein}},\
  }\href {\doibase 10.1016/j.physletb.2019.134814} {\bibfield  {journal}
  {\bibinfo  {journal} {Physics Letters B}\ }\textbf {\bibinfo {volume}
  {797}},\ \bibinfo {pages} {134814} (\bibinfo {year} {2019})}\BibitemShut
  {NoStop}%
\bibitem [{\citenamefont {Habibi}\ \emph {et~al.}(2023)\citenamefont {Habibi},
  \citenamefont {Arefiev},\ and\ \citenamefont {Toncian}}]{87}%
  \BibitemOpen
  \bibfield  {author} {\bibinfo {author} {\bibfnamefont {M.}~\bibnamefont
  {Habibi}}, \bibinfo {author} {\bibfnamefont {A.}~\bibnamefont {Arefiev}}, \
  and\ \bibinfo {author} {\bibfnamefont {T.}~\bibnamefont {Toncian}},\ }\href
  {\doibase 10.1063/5.0167288} {\bibfield  {journal} {\bibinfo  {journal}
  {Physics of Plasmas}\ }\textbf {\bibinfo {volume} {30}} (\bibinfo {year}
  {2023}),\ 10.1063/5.0167288}\BibitemShut {NoStop}%
\bibitem [{\citenamefont {Gong}\ \emph {et~al.}(2023)\citenamefont {Gong},
  \citenamefont {Hatsagortsyan},\ and\ \citenamefont {Keitel}}]{88}%
  \BibitemOpen
  \bibfield  {author} {\bibinfo {author} {\bibfnamefont {Z.}~\bibnamefont
  {Gong}}, \bibinfo {author} {\bibfnamefont {K.~Z.}\ \bibnamefont
  {Hatsagortsyan}}, \ and\ \bibinfo {author} {\bibfnamefont {C.~H.}\
  \bibnamefont {Keitel}},\ }\href {\doibase 10.1103/physrevlett.130.015101}
  {\bibfield  {journal} {\bibinfo  {journal} {Physical Review Letters}\
  }\textbf {\bibinfo {volume} {130}} (\bibinfo {year} {2023}),\
  10.1103/physrevlett.130.015101}\BibitemShut {NoStop}%
\bibitem [{\citenamefont {Zhang}\ \emph {et~al.}(2023)\citenamefont {Zhang},
  \citenamefont {Zhang},\ and\ \citenamefont {Zhou}}]{89}%
  \BibitemOpen
  \bibfield  {author} {\bibinfo {author} {\bibfnamefont {B.}~\bibnamefont
  {Zhang}}, \bibinfo {author} {\bibfnamefont {Z.-M.}\ \bibnamefont {Zhang}}, \
  and\ \bibinfo {author} {\bibfnamefont {W.-M.}\ \bibnamefont {Zhou}},\ }\href
  {\doibase 10.1063/5.0157663} {\bibfield  {journal} {\bibinfo  {journal}
  {Matter and Radiation at Extremes}\ }\textbf {\bibinfo {volume} {8}}
  (\bibinfo {year} {2023}),\ 10.1063/5.0157663}\BibitemShut {NoStop}%
\bibitem [{\citenamefont {Chen}\ \emph {et~al.}(2022)\citenamefont {Chen},
  \citenamefont {Hatsagortsyan}, \citenamefont {Keitel},\ and\ \citenamefont
  {Shaisultanov}}]{90}%
  \BibitemOpen
  \bibfield  {author} {\bibinfo {author} {\bibfnamefont {Y.-Y.}\ \bibnamefont
  {Chen}}, \bibinfo {author} {\bibfnamefont {K.~Z.}\ \bibnamefont
  {Hatsagortsyan}}, \bibinfo {author} {\bibfnamefont {C.~H.}\ \bibnamefont
  {Keitel}}, \ and\ \bibinfo {author} {\bibfnamefont {R.}~\bibnamefont
  {Shaisultanov}},\ }\href {\doibase 10.1103/physrevd.105.116013} {\bibfield
  {journal} {\bibinfo  {journal} {Physical Review D}\ }\textbf {\bibinfo
  {volume} {105}} (\bibinfo {year} {2022}),\
  10.1103/physrevd.105.116013}\BibitemShut {NoStop}%
\bibitem [{\citenamefont {Adamo}\ \emph {et~al.}(2021)\citenamefont {Adamo},
  \citenamefont {Ilderton},\ and\ \citenamefont {MacLeod}}]{91}%
  \BibitemOpen
  \bibfield  {author} {\bibinfo {author} {\bibfnamefont {T.}~\bibnamefont
  {Adamo}}, \bibinfo {author} {\bibfnamefont {A.}~\bibnamefont {Ilderton}}, \
  and\ \bibinfo {author} {\bibfnamefont {A.~J.}\ \bibnamefont {MacLeod}},\
  }\href {\doibase 10.1103/physrevd.104.116013} {\bibfield  {journal} {\bibinfo
   {journal} {Physical Review D}\ }\textbf {\bibinfo {volume} {104}} (\bibinfo
  {year} {2021}),\ 10.1103/physrevd.104.116013}\BibitemShut {NoStop}%
\bibitem [{\citenamefont {MacLeod}\ and\ \citenamefont {King}(2024)}]{92}%
  \BibitemOpen
  \bibfield  {author} {\bibinfo {author} {\bibfnamefont {A.~J.}\ \bibnamefont
  {MacLeod}}\ and\ \bibinfo {author} {\bibfnamefont {B.}~\bibnamefont {King}},\
  }\href {\doibase 10.1103/physreva.110.032216} {\bibfield  {journal} {\bibinfo
   {journal} {Physical Review A}\ }\textbf {\bibinfo {volume} {110}} (\bibinfo
  {year} {2024}),\ 10.1103/physreva.110.032216}\BibitemShut {NoStop}%
\bibitem [{\citenamefont {Dahiri}\ \emph {et~al.}(2022)\citenamefont {Dahiri},
  \citenamefont {Baouahi}, \citenamefont {Jakha}, \citenamefont {Mouslih},
  \citenamefont {Manaut},\ and\ \citenamefont {Taj}}]{93}%
  \BibitemOpen
  \bibfield  {author} {\bibinfo {author} {\bibfnamefont {I.}~\bibnamefont
  {Dahiri}}, \bibinfo {author} {\bibfnamefont {M.}~\bibnamefont {Baouahi}},
  \bibinfo {author} {\bibfnamefont {M.}~\bibnamefont {Jakha}}, \bibinfo
  {author} {\bibfnamefont {S.}~\bibnamefont {Mouslih}}, \bibinfo {author}
  {\bibfnamefont {B.}~\bibnamefont {Manaut}}, \ and\ \bibinfo {author}
  {\bibfnamefont {S.}~\bibnamefont {Taj}},\ }\href {\doibase
  10.1016/j.cjph.2022.03.048} {\bibfield  {journal} {\bibinfo  {journal}
  {Chinese Journal of Physics}\ }\textbf {\bibinfo {volume} {77}},\ \bibinfo
  {pages} {1691–1700} (\bibinfo {year} {2022})}\BibitemShut {NoStop}%
\bibitem [{\citenamefont {Karlovets}\ \emph {et~al.}(2021)\citenamefont
  {Karlovets}, \citenamefont {Serbo},\ and\ \citenamefont {Surzhykov}}]{94}%
  \BibitemOpen
  \bibfield  {author} {\bibinfo {author} {\bibfnamefont {D.~V.}\ \bibnamefont
  {Karlovets}}, \bibinfo {author} {\bibfnamefont {V.~G.}\ \bibnamefont
  {Serbo}}, \ and\ \bibinfo {author} {\bibfnamefont {A.}~\bibnamefont
  {Surzhykov}},\ }\href {\doibase 10.1103/physreva.104.023101} {\bibfield
  {journal} {\bibinfo  {journal} {Physical Review A}\ }\textbf {\bibinfo
  {volume} {104}} (\bibinfo {year} {2021}),\
  10.1103/physreva.104.023101}\BibitemShut {NoStop}%
\bibitem [{\citenamefont {Mendonca}(2024)}]{95}%
  \BibitemOpen
  \bibfield  {author} {\bibinfo {author} {\bibfnamefont {J.~T.}\ \bibnamefont
  {Mendonca}},\ }\href {\doibase 10.3390/photonics11050448} {\bibfield
  {journal} {\bibinfo  {journal} {Photonics}\ }\textbf {\bibinfo {volume}
  {11}},\ \bibinfo {pages} {448} (\bibinfo {year} {2024})}\BibitemShut
  {NoStop}%
\bibitem [{\citenamefont {Gies}\ \emph {et~al.}(2022)\citenamefont {Gies},
  \citenamefont {Karbstein},\ and\ \citenamefont {Klar}}]{96}%
  \BibitemOpen
  \bibfield  {author} {\bibinfo {author} {\bibfnamefont {H.}~\bibnamefont
  {Gies}}, \bibinfo {author} {\bibfnamefont {F.}~\bibnamefont {Karbstein}}, \
  and\ \bibinfo {author} {\bibfnamefont {L.}~\bibnamefont {Klar}},\ }\href
  {\doibase 10.1103/physrevd.106.116005} {\bibfield  {journal} {\bibinfo
  {journal} {Physical Review D}\ }\textbf {\bibinfo {volume} {106}} (\bibinfo
  {year} {2022}),\ 10.1103/physrevd.106.116005}\BibitemShut {NoStop}%
\bibitem [{\citenamefont {Li}\ \emph {et~al.}(2021)\citenamefont {Li},
  \citenamefont {Gan}, \citenamefont {Wang}, \citenamefont {Jiao},
  \citenamefont {Jin}, \citenamefont {Zhuo}, \citenamefont {Zhou},
  \citenamefont {Zhu}, \citenamefont {He},\ and\ \citenamefont {Qiao}}]{97}%
  \BibitemOpen
  \bibfield  {author} {\bibinfo {author} {\bibfnamefont {X.~B.}\ \bibnamefont
  {Li}}, \bibinfo {author} {\bibfnamefont {L.~F.}\ \bibnamefont {Gan}},
  \bibinfo {author} {\bibfnamefont {J.}~\bibnamefont {Wang}}, \bibinfo {author}
  {\bibfnamefont {J.~L.}\ \bibnamefont {Jiao}}, \bibinfo {author}
  {\bibfnamefont {S.}~\bibnamefont {Jin}}, \bibinfo {author} {\bibfnamefont
  {H.~B.}\ \bibnamefont {Zhuo}}, \bibinfo {author} {\bibfnamefont {C.~T.}\
  \bibnamefont {Zhou}}, \bibinfo {author} {\bibfnamefont {S.~P.}\ \bibnamefont
  {Zhu}}, \bibinfo {author} {\bibfnamefont {X.~T.}\ \bibnamefont {He}}, \ and\
  \bibinfo {author} {\bibfnamefont {B.}~\bibnamefont {Qiao}},\ }\href {\doibase
  10.1088/1367-2630/ac4055} {\bibfield  {journal} {\bibinfo  {journal} {New
  Journal of Physics}\ }\textbf {\bibinfo {volume} {23}},\ \bibinfo {pages}
  {123043} (\bibinfo {year} {2021})}\BibitemShut {NoStop}%
\bibitem [{\citenamefont {Krajewska}\ \emph {et~al.}(2021)\citenamefont
  {Krajewska}, \citenamefont {Kaminski},\ and\ \citenamefont {Muller}}]{98}%
  \BibitemOpen
  \bibfield  {author} {\bibinfo {author} {\bibfnamefont {K.}~\bibnamefont
  {Krajewska}}, \bibinfo {author} {\bibfnamefont {J.~Z.}\ \bibnamefont
  {Kaminski}}, \ and\ \bibinfo {author} {\bibfnamefont {C.}~\bibnamefont
  {Muller}},\ }\href {\doibase 10.1088/1367-2630/ac231e} {\bibfield  {journal}
  {\bibinfo  {journal} {New Journal of Physics}\ }\textbf {\bibinfo {volume}
  {23}},\ \bibinfo {pages} {095012} (\bibinfo {year} {2021})}\BibitemShut
  {NoStop}%
\bibitem [{\citenamefont {Pastor}\ \emph {et~al.}(2024)\citenamefont {Pastor},
  \citenamefont {Alvarez-Estrada}, \citenamefont {Roso},\ and\ \citenamefont
  {Castejon}}]{99}%
  \BibitemOpen
  \bibfield  {author} {\bibinfo {author} {\bibfnamefont {I.}~\bibnamefont
  {Pastor}}, \bibinfo {author} {\bibfnamefont {R.~F.}\ \bibnamefont
  {Alvarez-Estrada}}, \bibinfo {author} {\bibfnamefont {L.}~\bibnamefont
  {Roso}}, \ and\ \bibinfo {author} {\bibfnamefont {F.}~\bibnamefont
  {Castejon}},\ }\href {\doibase 10.3390/photonics11020113} {\bibfield
  {journal} {\bibinfo  {journal} {Photonics}\ }\textbf {\bibinfo {volume}
  {11}},\ \bibinfo {pages} {113} (\bibinfo {year} {2024})}\BibitemShut
  {NoStop}%
\bibitem [{\citenamefont {El~Asri}\ \emph {et~al.}(2021)\citenamefont
  {El~Asri}, \citenamefont {Mouslih}, \citenamefont {Jakha}, \citenamefont
  {Manaut}, \citenamefont {Attaourti}, \citenamefont {Taj},\ and\ \citenamefont
  {Benbrik}}]{100}%
  \BibitemOpen
  \bibfield  {author} {\bibinfo {author} {\bibfnamefont {S.}~\bibnamefont
  {El~Asri}}, \bibinfo {author} {\bibfnamefont {S.}~\bibnamefont {Mouslih}},
  \bibinfo {author} {\bibfnamefont {M.}~\bibnamefont {Jakha}}, \bibinfo
  {author} {\bibfnamefont {B.}~\bibnamefont {Manaut}}, \bibinfo {author}
  {\bibfnamefont {Y.}~\bibnamefont {Attaourti}}, \bibinfo {author}
  {\bibfnamefont {S.}~\bibnamefont {Taj}}, \ and\ \bibinfo {author}
  {\bibfnamefont {R.}~\bibnamefont {Benbrik}},\ }\href {\doibase
  10.1103/physrevd.104.113001} {\bibfield  {journal} {\bibinfo  {journal}
  {Physical Review D}\ }\textbf {\bibinfo {volume} {104}} (\bibinfo {year}
  {2021}),\ 10.1103/physrevd.104.113001}\BibitemShut {NoStop}%
\bibitem [{\citenamefont {Ouali}\ \emph {et~al.}(2022)\citenamefont {Ouali},
  \citenamefont {Ouhammou}, \citenamefont {Taj}, \citenamefont {Benbrik},
  \citenamefont {Manaut},\ and\ \citenamefont {El~Idrissi}}]{101}%
  \BibitemOpen
  \bibfield  {author} {\bibinfo {author} {\bibfnamefont {M.}~\bibnamefont
  {Ouali}}, \bibinfo {author} {\bibfnamefont {M.}~\bibnamefont {Ouhammou}},
  \bibinfo {author} {\bibfnamefont {S.}~\bibnamefont {Taj}}, \bibinfo {author}
  {\bibfnamefont {R.}~\bibnamefont {Benbrik}}, \bibinfo {author} {\bibfnamefont
  {B.}~\bibnamefont {Manaut}}, \ and\ \bibinfo {author} {\bibfnamefont
  {M.}~\bibnamefont {El~Idrissi}},\ }\href {\doibase 10.1088/1555-6611/ac8fe8}
  {\bibfield  {journal} {\bibinfo  {journal} {Laser Physics}\ }\textbf
  {\bibinfo {volume} {32}},\ \bibinfo {pages} {106002} (\bibinfo {year}
  {2022})}\BibitemShut {NoStop}%
\bibitem [{\citenamefont {Dai}\ \emph {et~al.}(2024)\citenamefont {Dai},
  \citenamefont {Hatsagortsyan}, \citenamefont {Keitel},\ and\ \citenamefont
  {Chen}}]{102}%
  \BibitemOpen
  \bibfield  {author} {\bibinfo {author} {\bibfnamefont {Y.-N.}\ \bibnamefont
  {Dai}}, \bibinfo {author} {\bibfnamefont {K.~Z.}\ \bibnamefont
  {Hatsagortsyan}}, \bibinfo {author} {\bibfnamefont {C.~H.}\ \bibnamefont
  {Keitel}}, \ and\ \bibinfo {author} {\bibfnamefont {Y.-Y.}\ \bibnamefont
  {Chen}},\ }\href {\doibase 10.1103/physrevd.110.012008} {\bibfield  {journal}
  {\bibinfo  {journal} {Physical Review D}\ }\textbf {\bibinfo {volume} {110}}
  (\bibinfo {year} {2024}),\ 10.1103/physrevd.110.012008}\BibitemShut {NoStop}%
\bibitem [{\citenamefont {Olofsson}\ and\ \citenamefont
  {Gonoskov}(2022)}]{103}%
  \BibitemOpen
  \bibfield  {author} {\bibinfo {author} {\bibfnamefont {C.}~\bibnamefont
  {Olofsson}}\ and\ \bibinfo {author} {\bibfnamefont {A.}~\bibnamefont
  {Gonoskov}},\ }\href {\doibase 10.1103/physreva.106.063512} {\bibfield
  {journal} {\bibinfo  {journal} {Physical Review A}\ }\textbf {\bibinfo
  {volume} {106}} (\bibinfo {year} {2022}),\
  10.1103/physreva.106.063512}\BibitemShut {NoStop}%
\bibitem [{\citenamefont {Zaim}\ \emph {et~al.}(2024)\citenamefont {Zaim},
  \citenamefont {Sainte-Marie}, \citenamefont {Fedeli}, \citenamefont
  {Bartoli}, \citenamefont {Huebl}, \citenamefont {Leblanc}, \citenamefont
  {Vay},\ and\ \citenamefont {Vincenti}}]{104}%
  \BibitemOpen
  \bibfield  {author} {\bibinfo {author} {\bibfnamefont {N.}~\bibnamefont
  {Zaim}}, \bibinfo {author} {\bibfnamefont {A.}~\bibnamefont {Sainte-Marie}},
  \bibinfo {author} {\bibfnamefont {L.}~\bibnamefont {Fedeli}}, \bibinfo
  {author} {\bibfnamefont {P.}~\bibnamefont {Bartoli}}, \bibinfo {author}
  {\bibfnamefont {A.}~\bibnamefont {Huebl}}, \bibinfo {author} {\bibfnamefont
  {A.}~\bibnamefont {Leblanc}}, \bibinfo {author} {\bibfnamefont {J.-L.}\
  \bibnamefont {Vay}}, \ and\ \bibinfo {author} {\bibfnamefont
  {H.}~\bibnamefont {Vincenti}},\ }\href {\doibase
  10.1103/PhysRevLett.132.175002} {\bibfield  {journal} {\bibinfo  {journal}
  {Physical Review Letters}\ }\textbf {\bibinfo {volume} {132}} (\bibinfo
  {year} {2024}),\ 10.1103/PhysRevLett.132.175002}\BibitemShut {NoStop}%
\bibitem [{\citenamefont {Martinez}\ \emph {et~al.}(2023)\citenamefont
  {Martinez}, \citenamefont {Barbosa},\ and\ \citenamefont {Vranic}}]{105}%
  \BibitemOpen
  \bibfield  {author} {\bibinfo {author} {\bibfnamefont {B.}~\bibnamefont
  {Martinez}}, \bibinfo {author} {\bibfnamefont {B.}~\bibnamefont {Barbosa}}, \
  and\ \bibinfo {author} {\bibfnamefont {M.}~\bibnamefont {Vranic}},\ }\href
  {\doibase 10.1103/physrevaccelbeams.26.011301} {\bibfield  {journal}
  {\bibinfo  {journal} {Physical Review Accelerators and Beams}\ }\textbf
  {\bibinfo {volume} {26}} (\bibinfo {year} {2023}),\
  10.1103/physrevaccelbeams.26.011301}\BibitemShut {NoStop}%
\bibitem [{\citenamefont {Volkov}(1935)}]{106}%
  \BibitemOpen
  \bibfield  {author} {\bibinfo {author} {\bibfnamefont {D.~M.}\ \bibnamefont
  {Volkov}},\ }\href@noop {} {\bibfield  {journal} {\bibinfo  {journal} {Z.
  Phys.}\ }\textbf {\bibinfo {volume} {94}},\ \bibinfo {pages} {250} (\bibinfo
  {year} {1935})}\BibitemShut {NoStop}%
\bibitem [{\citenamefont {Berestetskii}\ \emph {et~al.}(1982)\citenamefont
  {Berestetskii}, \citenamefont {Lifshitz},\ and\ \citenamefont
  {Pitaevskii}}]{108}%
  \BibitemOpen
  \bibfield  {author} {\bibinfo {author} {\bibfnamefont {V.}~\bibnamefont
  {Berestetskii}}, \bibinfo {author} {\bibfnamefont {E.}~\bibnamefont
  {Lifshitz}}, \ and\ \bibinfo {author} {\bibfnamefont {L.}~\bibnamefont
  {Pitaevskii}},\ }\href {https://books.google.ru/books?id=YlwKR5JNWDgC} {\emph
  {\bibinfo {title} {Quantum Electrodynamics: Volume 4}}},\ Course of
  theoretical physics\ (\bibinfo  {publisher} {Elsevier Science},\ \bibinfo
  {year} {1982})\BibitemShut {NoStop}%
\bibitem [{\citenamefont {Breit}\ and\ \citenamefont {Wigner}(1936)}]{107}%
  \BibitemOpen
  \bibfield  {author} {\bibinfo {author} {\bibfnamefont {G.}~\bibnamefont
  {Breit}}\ and\ \bibinfo {author} {\bibfnamefont {E.}~\bibnamefont {Wigner}},\
  }\href {\doibase 10.1103/physrev.49.519} {\bibfield  {journal} {\bibinfo
  {journal} {Physical Review}\ }\textbf {\bibinfo {volume} {49}},\ \bibinfo
  {pages} {519} (\bibinfo {year} {1936})}\BibitemShut {NoStop}%
\bibitem [{\citenamefont {Deng}\ \emph {et~al.}(2020)\citenamefont {Deng},
  \citenamefont {Gao}, \citenamefont {Li},\ and\ \citenamefont {Shao}}]{109}%
  \BibitemOpen
  \bibfield  {author} {\bibinfo {author} {\bibfnamefont {Z.-L.}\ \bibnamefont
  {Deng}}, \bibinfo {author} {\bibfnamefont {Z.-F.}\ \bibnamefont {Gao}},
  \bibinfo {author} {\bibfnamefont {X.-D.}\ \bibnamefont {Li}}, \ and\ \bibinfo
  {author} {\bibfnamefont {Y.}~\bibnamefont {Shao}},\ }\href {\doibase
  10.3847/1538-4357/ab76c4} {\bibfield  {journal} {\bibinfo  {journal} {The
  Astrophysical Journal}\ }\textbf {\bibinfo {volume} {892}},\ \bibinfo {pages}
  {4} (\bibinfo {year} {2020})}\BibitemShut {NoStop}%
\bibitem [{\citenamefont {Deng}\ \emph {et~al.}(2021)\citenamefont {Deng},
  \citenamefont {Li}, \citenamefont {Gao},\ and\ \citenamefont {Shao}}]{110}%
  \BibitemOpen
  \bibfield  {author} {\bibinfo {author} {\bibfnamefont {Z.-L.}\ \bibnamefont
  {Deng}}, \bibinfo {author} {\bibfnamefont {X.-D.}\ \bibnamefont {Li}},
  \bibinfo {author} {\bibfnamefont {Z.-F.}\ \bibnamefont {Gao}}, \ and\
  \bibinfo {author} {\bibfnamefont {Y.}~\bibnamefont {Shao}},\ }\href {\doibase
  10.3847/1538-4357/abe0b2} {\bibfield  {journal} {\bibinfo  {journal} {The
  Astrophysical Journal}\ }\textbf {\bibinfo {volume} {909}},\ \bibinfo {pages}
  {174} (\bibinfo {year} {2021})}\BibitemShut {NoStop}%
\bibitem [{\citenamefont {Gao}\ \emph {et~al.}(2017)\citenamefont {Gao},
  \citenamefont {Wang}, \citenamefont {Shan}, \citenamefont {Li},\ and\
  \citenamefont {Wang}}]{111}%
  \BibitemOpen
  \bibfield  {author} {\bibinfo {author} {\bibfnamefont {Z.-F.}\ \bibnamefont
  {Gao}}, \bibinfo {author} {\bibfnamefont {N.}~\bibnamefont {Wang}}, \bibinfo
  {author} {\bibfnamefont {H.}~\bibnamefont {Shan}}, \bibinfo {author}
  {\bibfnamefont {X.-D.}\ \bibnamefont {Li}}, \ and\ \bibinfo {author}
  {\bibfnamefont {W.}~\bibnamefont {Wang}},\ }\href {\doibase
  10.3847/1538-4357/aa8f49} {\bibfield  {journal} {\bibinfo  {journal} {The
  Astrophysical Journal}\ }\textbf {\bibinfo {volume} {849}},\ \bibinfo {pages}
  {19} (\bibinfo {year} {2017})}\BibitemShut {NoStop}%
\bibitem [{\citenamefont {Wang}\ \emph {et~al.}(2020)\citenamefont {Wang},
  \citenamefont {Gao}, \citenamefont {Jia}, \citenamefont {Wang},\ and\
  \citenamefont {Li}}]{112}%
  \BibitemOpen
  \bibfield  {author} {\bibinfo {author} {\bibfnamefont {H.}~\bibnamefont
  {Wang}}, \bibinfo {author} {\bibfnamefont {Z.-F.}\ \bibnamefont {Gao}},
  \bibinfo {author} {\bibfnamefont {H.-Y.}\ \bibnamefont {Jia}}, \bibinfo
  {author} {\bibfnamefont {N.}~\bibnamefont {Wang}}, \ and\ \bibinfo {author}
  {\bibfnamefont {X.-D.}\ \bibnamefont {Li}},\ }\href {\doibase
  10.3390/universe6050063} {\bibfield  {journal} {\bibinfo  {journal}
  {Universe}\ }\textbf {\bibinfo {volume} {6}},\ \bibinfo {pages} {63}
  (\bibinfo {year} {2020})}\BibitemShut {NoStop}%
\bibitem [{\citenamefont {Gao}\ \emph {et~al.}(2015)\citenamefont {Gao},
  \citenamefont {Li}, \citenamefont {Wang}, \citenamefont {Yuan}, \citenamefont
  {Wang}, \citenamefont {Peng},\ and\ \citenamefont {Du}}]{113}%
  \BibitemOpen
  \bibfield  {author} {\bibinfo {author} {\bibfnamefont {Z.~F.}\ \bibnamefont
  {Gao}}, \bibinfo {author} {\bibfnamefont {X.-D.}\ \bibnamefont {Li}},
  \bibinfo {author} {\bibfnamefont {N.}~\bibnamefont {Wang}}, \bibinfo {author}
  {\bibfnamefont {J.~P.}\ \bibnamefont {Yuan}}, \bibinfo {author}
  {\bibfnamefont {P.}~\bibnamefont {Wang}}, \bibinfo {author} {\bibfnamefont
  {Q.~H.}\ \bibnamefont {Peng}}, \ and\ \bibinfo {author} {\bibfnamefont
  {Y.~J.}\ \bibnamefont {Du}},\ }\href {\doibase 10.1093/mnras/stv2465}
  {\bibfield  {journal} {\bibinfo  {journal} {Monthly Notices of the Royal
  Astronomical Society}\ }\textbf {\bibinfo {volume} {456}},\ \bibinfo {pages}
  {55} (\bibinfo {year} {2015})}\BibitemShut {NoStop}%
\bibitem [{\citenamefont {Yan}\ \emph {et~al.}(2021)\citenamefont {Yan},
  \citenamefont {Gao}, \citenamefont {Yang},\ and\ \citenamefont {Dong}}]{114}%
  \BibitemOpen
  \bibfield  {author} {\bibinfo {author} {\bibfnamefont {F.-Z.}\ \bibnamefont
  {Yan}}, \bibinfo {author} {\bibfnamefont {Z.-F.}\ \bibnamefont {Gao}},
  \bibinfo {author} {\bibfnamefont {W.-S.}\ \bibnamefont {Yang}}, \ and\
  \bibinfo {author} {\bibfnamefont {A.-J.}\ \bibnamefont {Dong}},\ }\href
  {\doibase 10.1002/asna.202113913} {\bibfield  {journal} {\bibinfo  {journal}
  {Astronomische Nachrichten}\ }\textbf {\bibinfo {volume} {342}},\ \bibinfo
  {pages} {249} (\bibinfo {year} {2021})}\BibitemShut {NoStop}%
\end{thebibliography}%

\end{document}